\documentclass[aps,prd,nofootinbib,twocolumn,superscriptaddress,preprintnumbers,balancelastpage,longbibliography]{revtex4-1}

\usepackage{placeins}
\usepackage{amsmath,amssymb,mathtools,bm}
\usepackage{lipsum}
\usepackage{comment}

\usepackage{graphicx, color, hepunits}
\usepackage[dvipsnames]{xcolor}
\usepackage{float}
\usepackage{filecontents}
\usepackage{multirow}
\usepackage{slashed}
\usepackage[colorlinks=true 
,urlcolor=purple 
,anchorcolor=blue
,citecolor=blue 
,filecolor=magenta  
,linkcolor=blue 
,menucolor=blue
,linktocpage=true
,pdfproducer=medialab
,pdfa=true
]{hyperref}

\usepackage[utf8]{inputenc}
\usepackage[english]{babel}
\usepackage{soul}
\let\newvec\vec
\let\vec\mathbf

\newcommand{\D}{{\rm d}}

\newcommand{\mpl}{M_{\rm pl}}

\newcommand{\es}[2] {\begin{equation} \label{#1} \begin{split} #2 \end{split} \end{equation}}
\DeclareMathOperator{\csch}{csch}

\makeatletter
\def\l@subsubsection#1#2{}
\makeatother

\begin{document}

\title{
The Cosmological Dynamics of String Theory Axion Strings
}

\author{Joshua N. Benabou}
\email{joshua\_benabou@berkeley.edu}
\affiliation{Berkeley Center for Theoretical Physics, University of California, Berkeley, CA 94720, U.S.A.}
\affiliation{Theoretical Physics Group, Lawrence Berkeley National Laboratory, Berkeley, CA 94720, U.S.A.}

\author{Quentin Bonnefoy}
\email{q.bonnefoy@berkeley.edu}
\affiliation{Berkeley Center for Theoretical Physics, University of California, Berkeley, CA 94720, U.S.A.}
\affiliation{Theoretical Physics Group, Lawrence Berkeley National Laboratory, Berkeley, CA 94720, U.S.A.}

\author{Malte~Buschmann}
\email{m.s.a.buschmann@uva.nl}
\affiliation{Department of Physics, Princeton University, Princeton, NJ 08544, USA}
\affiliation{GRAPPA Institute, Institute for Theoretical Physics Amsterdam,
University of Amsterdam, Science Park 904, 1098 XH Amsterdam, The Netherlands}

\author{Soubhik Kumar}
\email{soubhik.kumar@nyu.edu}
\affiliation{Center for Cosmology and Particle Physics, Department of Physics,
New York University, New York, NY 10003, U.S.A.}
\affiliation{Berkeley Center for Theoretical Physics, University of California, Berkeley, CA 94720, U.S.A.}
\affiliation{Theoretical Physics Group, Lawrence Berkeley National Laboratory, Berkeley, CA 94720, U.S.A.}

\author{Benjamin R. Safdi}
\email{brsafdi@berkeley.edu}
\affiliation{Berkeley Center for Theoretical Physics, University of California, Berkeley, CA 94720, U.S.A.}
\affiliation{Theoretical Physics Group, Lawrence Berkeley National Laboratory, Berkeley, CA 94720, U.S.A.}

\date{\today}

\begin{abstract}
The quantum chromodynamics (QCD) axion may solve the strong {\it CP} problem and explain the dark matter (DM) abundance of our Universe. 
The axion was originally proposed to arise as the pseudo-Nambu Goldstone boson of global $\mathrm{U}(1)_{\rm PQ}$ Peccei-Quinn (PQ) symmetry breaking, but axions also arise generically in string theory as zero modes of higher-dimensional gauge fields.
In this work we show that string theory axions behave fundamentally differently from field theory axions in the early Universe.  Field theory axions may form axion strings if the PQ phase transition takes place after inflation.
In contrast, we show that string theory axions do not generically form axion strings.  In special inflationary paradigms, such as D-brane inflation, string theory axion strings may form; however, their tension is parametrically larger than that of field theory axion strings. We then show that such QCD axion strings overproduce the DM abundance for all allowed QCD axion masses and are thus ruled out, except in scenarios with large warping. A loop-hole to this conclusion arises in the axiverse, where an axion string could be composed of multiple different axion mass eigenstates; a heavier eigenstate could collapse the network earlier, allowing for the QCD axion to produce the correct DM abundance and also generating observable gravitational wave signals.
\end{abstract}
\maketitle

\tableofcontents

\section{Introduction}

Axions emerge naturally in the context of string theory compactifications~\cite{Witten:1984dg,Choi:1985je,Barr:1985hk,Svrcek:2006yi,Arvanitaki:2009fg}. In these constructions the 4-dimensional axions arise as the zero modes of higher-dimensional gauge fields. 
Axions from string theory could include the quantum chromodynamics (QCD) axion, which may solve the strong {\it CP} problem related to the absence of a neutron electric dipole moment~\cite{Peccei:1977hh,Peccei:1977ur,Weinberg:1977ma,Wilczek:1977pj}.  The QCD axion may also explain the dark matter (DM) abundance in the Universe~\cite{Preskill:1982cy,Abbott:1982af,Dine:1982ah}.  String theory may also give rise to a large number of axion-like particles, with one being the QCD axion; this scenario is known as the axiverse~\cite{Svrcek:2006yi, Arvanitaki:2009fg}, and it has been realized recently in explicit string theory compactifications~\cite{Cicoli:2012sz,Demirtas:2018akl,Halverson:2019cmy,Demirtas:2021gsq}.

One advantage of string theory axions over field-theory axions, where the axion arises as the Goldstone mode of a spontaneously broken $\mathrm{U}(1)_{\rm PQ}$ symmetry called the Peccei–Quinn (PQ) symmetry~\cite{Peccei:1977hh,Peccei:1977ur}, is that string theory axions may more naturally evade the so-called PQ quality problem~\cite{Georgi:1981pu,Lazarides:1985bj,Kamionkowski:1992mf, Ghigna:1992iv, Barr:1992qq, Holman:1992us}.  In the field theory constructions we require a high-quality $\mathrm{U}(1)_{\rm PQ}$ symmetry such that QCD instantons provide the dominant source of the axion's potential, to the precision required by experimental measurements of the neutron electric dipole moment. However, global symmetries are expected to be violated in the context of quantum gravity
(see, {\it e.g.},~\cite{Harlow:2018tng, Reece:2023czb} and references therein).
Under the assumption that this breaking arises through Planck-suppressed operators, field theory models with high-quality PQ global symmetries have
been designed, but they require non-minimal structures
at the PQ scale to make the symmetry accidental at low energies (see,
{\it e.g.},~\cite{DiLuzio:2020wdo} for examples).
String theory constructions mitigate the PQ quality problem by protecting the axion mass by the higher-dimensional gauge invariance of the gauge field that gives rise to the axion at low energies. In such scenarios, only non-perturbative effects, whose strength depends on the compactification details, can give the axion a mass. Ref.~\cite{Demirtas:2021gsq}, for example, considered a large ensemble of orientifold compactifications in type IIB string theory on Calabi-Yau hypersurfaces and found that the strong-{\it CP} problem was solved to adequate precision in approximately 99.7\% of the different compactifications, with stringy instanton effects contributing sufficiently sub-dominant bare axion masses.    

The strong motivation for string theory axions sparked thorough studies of the various roles they can play in cosmology (see, {\it e.g.}, the recent string cosmology reviews~\cite{Flauger:2022hie,Cvetic:2022fnv,Green:2022hhj,Cicoli:2023opf}). 
However, there have been few studies of cosmic axion strings in the context of string theory, despite their possibly important roles~\cite{Witten:1985fp,Harvey:1988in,Dvali:2003zh,March-Russell:2021zfq}.  This work investigates the structure, formation mechanisms, dynamics, and phenomenology of axion strings in string theory and, more generally, extra dimension ultraviolet (UV) completions.

Cosmic axion strings are known to produce rich cosmological histories, and thus they have been extensively scrutinized in field theory models. On the one hand, the effective field theory (EFT) of string theory axions is identical to that of field theory axions (or their supersymmetric counterparts).  On the other hand, however, in certain cosmological scenarios axion strings may form, and at the cores of those strings the axion-only EFT is singular, with the UV completion being restored (see~\cite{Safdi:2022xkm} for a review).  This implies that the cosmology of axion strings is sensitive to the UV completion of the theory. 

String theory axions are associated to axion strings whose core is not smoothed out in any EFT description (see, {\it e.g.},~\cite{Dolan:2017vmn,Greene:1989ya,Reece:2018zvv,Lanza:2020qmt,Lanza:2021udy,March-Russell:2021zfq,Heidenreich:2021yda}),
and is instead resolved in terms of fundamental strings or wrapped branes in the string theory UV completion. These fundamental objects indeed have the appropriate (electric or magnetic) charges under the higher-dimensional gauge fields which give rise to the axion to be axion strings in 4D.
The cosmological evolution of a network of such fundamental one-dimensional objects significantly differs from the one of their field theory counterparts, due to their tension, possible instabilities, and different reconnection probabilities~\cite{Dvali:2003zj,Jones:2003da,Copeland:2003bj,Jackson:2004zg,Polchinski:2004ia,Hanany:2005bc,Hashimoto:2005hi,Polchinski:2005bg,Copeland:2009ga,Banks:2010zn}. Cosmic superstrings have also been extensively discussed recently as sources of gravitational waves (GWs)~\cite{Auclair:2019wcv,Sousa:2020sxs,LIGOScientific:2021nrg,Blasi:2020mfx,Blanco-Pillado:2021ygr,Boileau:2021gbr,Ellis:2023tsl,Ellis:2023oxs}. 
However, the role played by axions on the cosmological dynamics of the strings, and conversely, the role played by the strings on the axion cosmology has not been previously explored. 
In this work, we delve into the physics of axion strings in string theory compactifications, and we show that it is qualitatively different than that of field theory axions. 

Field theory axion strings always form as long as the reheating temperature $T_{\rm RH}\gtrsim f_a$, the axion decay constant\footnote{In this paper, the axion decay constant $f_a$ is defined to be the periodicity of the canonically-normalized axion, which is closely related to the PQ symmetry breaking scale in field theory models.}, for {\it any} inflationary mechanism.
The reason for this is straightforward: string formation requires a spontaneously-broken global symmetry according to the Kibble-Zurek mechanism~\cite{Kibble:1976sj,Zurek:1985qw}, which is the spontaneously-broken $\mathrm{U}(1)_{\rm PQ}$ symmetry in the field theory axion constructions. 
On the other hand, for string-theory axions, topological defects do not form even when $T_{\rm RH}\gtrsim f_a$. Above such temperatures, the 4D EFT breaks down but the PQ symmetry is not restored; instead, it is replaced by a higher-dimensional gauge invariance (thereby solving the aforementioned PQ quality problem). Consequently, as we argue in more detail below  (see also~\cite{Reece:2023czb,Cicoli:2022fzy}), string theory axion strings do not form in the standard thermal cosmology, {\it i.e.,} in scenarios when the last stages of inflation, reheating, and the subsequent thermal history are well described by an EFT; their formation requires additional ingredients.
 Such special circumstances in which string theory axion strings can form are actually known.
For example, D-brane annihilation after D-brane inflation~\cite{Dvali:1998pa} can produce fundamental (F-) and D-strings, or more generally ${\rm D}(1+p)$-branes wrapped on $p$-cycles~\cite{Jones:2002cv, Sarangi:2002yt}. 
Also, if $T_{\rm RH}$ was larger than the (warped down) string scale, a Hagedorn phase transition can take place as the Universe cools, producing F-strings~\cite{ENGLERT1988423, Polchinski:2004ia, Frey:2005jk}. Therefore, it is important to understand the properties and the evolution of the resulting axion string network. 

In this context, we show that the tension of string theory axion strings is parametrically larger than the tension of field theory axion strings unless the extra dimensions are strongly warped. (In the latter case, one recovers the field theory relation for the string tension, as expected from gauge-string dualities.) The expected range of string tensions $\mu_{\rm eff}$ is as follows. An upper bound has been conjectured in the context of the weak gravity conjecture (WGC) by~\cite{Hebecker:2017wsu,Hebecker:2017uix,
Harlow:2022ich},
\es{eq:upperBoundTension}{
\mu_{\rm eff} \lesssim \kappa f_a M_{\rm pl} \,,
}
with $\kappa$ a numerical factor of order unity. This is the so-called ``magnetic axion WGC.''
Meanwhile, the tension of the strings is bounded from below by
\es{eq:lowerBoundTension}{
\pi f_a^2 \log(m_r / H) \lesssim \mu_{\rm eff}
}
as a result of the axion configuration,
where $m_r$ is the mass of the modulus field that regulates the string core ($m_r \sim f_a$ being natural for field theory axions), and $H$ is the Hubble parameter. 
Axion strings saturating the lower limit emerge in field theory UV completions and for strongly-warped extra dimensions; we illustrate how the upper limit is saturated for axion strings arising in flat compactifications of string theory. 

A large axion string tension impacts how strings emit axions; for example, the density of strings, the amount of axions they radiate, as well as the balance between axion and GW emissions, are all affected. Combining all effects, we find that, if string theory axion strings do form in flat compactifications, the axion generated by the strings cannot be the QCD axion, as otherwise the relic DM abundance of those axions would over-close the universe, even allowing for extra entropy dilution by a possible period of early matter domination.  This effectively rules out axion string production of QCD axion DM in string theory UV completions in non-warped scenarios. 
However, we identify an exception to the previous claim in axiverse constructions: if the axion sourced by the string network is instead a linear combination of the QCD axion and other heavier axions, then we show that the QCD axion can constitute the DM. Furthermore, unlike in field theory constructions, we show that this scenario can give rise to observable GW signals.  

\section{Axion strings in 4D versus higher-dimensional field theory}\label{sec:axion_strings}

Axion strings are characterized by the property that, in traversing a circle enclosing an axion string core, the axion, which is a periodic field, undergoes a full field excursion. The axion-only picture of axion strings is clearly singular, since in that picture the axion field would have an infinite derivative at the core. 
In PQ UV completions, the radial mode, which is otherwise massive and frozen at its vacuum expectation value (VEV), is restored at the location of the string core and resolves the singularity by driving the full complex PQ field to zero. PQ axion strings have been the subject of extensive analytic and numerical work since they play important roles in determining the QCD axion DM abundance if the PQ symmetry is broken after inflation~\cite{Vilenkin:1982ks,Sikivie:1982qv,Davis:1986xc,Harari:1987ht,Shellard:1987bv,Davis:1989nj,Hagmann:1990mj,Battye:1993jv,Battye:1994au,Yamaguchi:1998gx,Klaer:2017ond,Gorghetto:2018myk,Vaquero:2018tib,Buschmann:2019icd,Gorghetto:2020qws,Dine:2020pds,Buschmann:2021sdq,OHare:2021zrq,Gorghetto:2022ikz}. 

 The PQ axion theory may be realized through a minimal scalar sector with Lagrangian
 \es{eq:minimalPQ}{
 {\mathcal L} = |\partial_\mu \Phi|^2 - \lambda_\Phi \left( |\Phi|^2 - {f_a^2 \over 2} \right)^2 \,,
 }
with $\Phi$ connected to the Standard Model (SM) through SM or new, vector-like fermions and possibly a Higgs coupling. At high temperatures the field $\Phi$ is in thermal equilibrium with the SM, and $\Phi$ has a symmetry-restoring thermal mass $m_{\rm therm} \sim \sqrt{\lambda_\Phi}T$.  Assuming the reheat temperature after inflation $T_{\rm RH}$ is above $f_a$, then when the Universe cools down below $f_a$ the field $\Phi$ undergoes $\mathrm{U}(1)_{\rm PQ}$ symmetry breaking, leading to the axion $a$ as a Goldstone mode and the radial mode $s$ as a heavy state with mass $m_s = \sqrt{2 \lambda_\Phi} f_a$:
 \es{}{
 \Phi = {\left(f_a + s \right) \over \sqrt{2}} e^{i a /f_a} \,.
 }
Let us consider an infinitely straight and static PQ axion string (see~\cite{Safdi:2022xkm} for a review).  For a string stretched in the ${\bf \hat z}$ direction, the solution is given by
 \es{}{
 \Phi(r,\theta,z) = {f_a \over \sqrt{2}} g(m_s r) e^{i \theta} \,,
 }
 where $r$ is the radial direction from the string core and $\theta$ is the polar angle. The dimensionless function $g(x)$ goes to zero as $x \to 0$ to regulate the singularity at the string core, while $g(x) \to 1$ for $x \gg 1$. When $r>0$, the axion field is well-defined and undergoes the expected winding as we go around the string, $a(r,\theta)/f_a=\theta \mod 2\pi$. The energy density in the string is given by 
 \es{}{
 \rho_{\rm str} = |\overrightarrow{\nabla} \Phi|^2 + \lambda_\Phi \left( |\Phi|^2 - {f_a^2 \over 2} \right)^2 \,,
 }
while the tension $\mu_{\rm eff}$\footnote{Note that we differentiate $\mu_{\rm eff}$ from $\mu$ by defining $\mu$ to be the near-core tension, with $\mu_{\rm eff}$ the IR-divergent quantity that is regulated by the finite distance to the next nearest string in a cosmological context.} is computed to be
 \es{}{
 \mu_{\rm eff} = \int_0^{2 \pi} \D \theta \int_0^{r_{\rm IR}} \D r \, r \rho_{\rm str} = \pi f_a^2  \log\left(\gamma\, m_s r_{\rm IR} \right) \,,
 }
where $\gamma$ is some constant of order unity. 
Note that $r_{\rm IR}$ is an IR cut-off, which regulates the logarithmically divergent contribution to the tension that arises from the gradient energy in the axion field. In practice, we expect $r_{\rm IR}$ to be the distance to the next string, which should be of the order of $1/H$, where $H$ is the Hubble parameter. The dimensionless constant $\gamma$ of order unity accounts for the UV and IR finite contribution to the tension from the radial mode, which is of order $\pi f_a^2$. In practice, for the QCD axion, we may take $m_s \sim f_a \sim 10^{11}$ GeV to obtain the correct relic DM abundance~\cite{Buschmann:2021sdq}. Then, near the QCD phase transition $H \sim T_{\rm QCD}^2 / M_{\rm pl}$, with $T_{\rm QCD} \sim 100$ MeV and $M_{\rm pl}$ the reduced Planck mass, such that $\log(m_s / H) \sim 10^2$ while $\log(\gamma) \sim 1$.  Thus, the tension of PQ axion strings is dominated by the surrounding axion field configuration.

\subsection{Flat extra dimension}
\label{sec:flat_extra_dim}

Let us now contrast the field theory axion string described above with a string theory axion string. As a proxy for a complete string theory compactification, we consider first a simpler extra-dimensional setup with an axion arising from a one-form gauge field~\cite{Arkani-Hamed:2003xts,Choi:2003wr}, which already captures the physics we wish to highlight. 
We first consider the case of a flat 5D theory with the fifth dimension compactified on a $S^1/\mathbb{Z}_2$ orbifold, while in the next subsection, we deal with the case of a warped extra dimension.

In the field theory setup, the singularity in the axion field configuration at the string core is regulated by the heavy radial mode. In the extra-dimensional UV completions, the singularity is also potentially regulated by a massive scalar mode (a modulus field), though the exact nature of that modulus field is model-dependent. In the construction below we illustrate the case in which the radion that controls the size of the fifth dimension is the modulus that regulates the string core, but we show later in this subsection that the dilaton -- or a linear combination of the radion and dilaton -- could also play that role.

We denote the 5D gauge field by  $A_M$ and study the coupled dynamics of $A_5$ and the radion $\rho$.
The latter appears in the parameterization of the 5D metric as~\cite{Appelquist:1983vs},
\es{eq:5D_metric}{
	\D s^2 = G_{MN} \D x^M \D x^N 
 =  \frac{b}{\rho(x)}g_{\mu\nu}(x)\D x^\mu \D x^\nu + \rho(x)^2 \D\phi^2,
}
and its potential stabilizes the size of the fifth dimension $\langle \rho \rangle=b$ via, {\it e.g.}, the Goldberger-Wise mechanism~\cite{Goldberger:1999uk,Goldberger:1999un}.  
Here, $\phi$ ranges from $0$ to $\pi$, with orbifold boundary conditions at $\phi=0$ and $\phi=\pi$ eliminating the zero mode of $A_\mu$ so that only the zero mode of $A_5$ survives in the low energy theory.
Substituting the metric~\eqref{eq:5D_metric} into the 5D general relativity action we find
\es{eq:5D_GR}{
    \int \D^4x \D\phi \sqrt{-G}\left(2M_5^3 R^{(5)}\right) &= \\ \frac{1}{2}\mpl^2
    \int \D^4x \sqrt{-g}  &\left( R^{(4)} -\frac{3}{2\rho^2} \partial_\mu\rho \partial^\mu \rho \right),
}
where $M_5$ is the 5D Planck scale and $R^{(5)}$ ($R^{(4)}$) is the 5D (4D) Ricci scalar.
Here the 4D reduced Planck scale $\mpl$ is determined by $M_5$ and the size of the extra dimension: $\mpl^2/4 = 2\pi b M_5^3$.
We also introduce a radion potential $V(\rho)$, such that ${\mathcal L}_\rho \supset -\int d^4x \sqrt{-g} V(\rho)$, whose form depends on the details of the stabilization mechanism and will be specified below.

To include $A_5$ in the analysis, we start with the action of a $\mathrm{U}(1)$ gauge theory,
\es{eq:5DF}{
    S_{\mathrm{U}(1)} =& -\frac{1}{4g_5^2}\int \D^4x \D\phi \sqrt{-G}F_{MN}F^{MN} \,,
    }
with $F_{MN}$ the 5D field strength tensor and $g_5$ the 5D gauge coupling. We then define the dimensionless 4D axion field $\vartheta(x)$ through the gauge-invariant Wilson loop (see, {\it e.g.},~\cite{Arkani-Hamed:2003xts}):
\es{eq:theta}{
{\vartheta(x)} = 2\int_0^{\pi} \D \phi \, {A_5} \,,
}
with $\vartheta(x)$ having periodicity $2 \pi$ as a result of large $U(1)$ gauge transformations and for a spectrum of integer charges. We then focus on the $\phi$-independent modes, since the $\phi$-dependent modes acquire masses through the Kaluza-Klein (KK) mechanism. Combining~\eqref{eq:5D_metric} and~\eqref{eq:5DF} with~\eqref{eq:theta} leads to the axion action 
\es{}{
S_\vartheta = -\int \D^4x \sqrt{-g} {1 \over 8 \pi^2 g_4^2 \rho^2} g^{\mu \nu} \partial_\mu \vartheta \partial_\nu \vartheta \,,
}
where we defined the effective 4D gauge coupling 
\es{eq:gaugec_4d}{
{1 \over g_4^2}  \equiv {2\pi b \over g_5^2}.
}
Then, it is convenient to define the axion as $a \equiv f_a \vartheta$, such that $a$ is canonically normalized and periodic with period $2 \pi f_a$, with the decay constant
\es{eq:fadef}{
f_a \equiv {1 \over 2 \pi b g_4} \,.
}
Thus, we obtain the following 4D action for the radion-axion system,
\es{eq:Srhoa}{
    S_{\rho-a} = \int \D^4x \sqrt{-g}&\left(-\frac{3 \mpl^2}{4\rho^2}(\partial_\mu\rho\partial^\mu\rho)-V(\rho) \right.\\ & \quad\left.- \frac{1}{2} \left( {b \over \rho} \right)^2\partial_\mu a\partial^\mu a \right) \,.
}
Note that this action is essentially identical to that which describes the coupling of a volume modulus to an axion in a 6D construction, as derived for instance in~\cite{March-Russell:2021zfq}.
More generally, although we focus on the case of the radion from a single extra dimension, the solutions to be discussed next capture qualitative features of several other scenarios. For instance, canonically normalizing the radion, $\D \varphi\equiv\sqrt{\frac{3}{2}}\mpl\frac{\D \rho}{\rho}=\mpl \, \D\log\rho^{\sqrt{\frac{3}{2}}}$ with a vanishing integration constant, leads to the action:
\es{eq:Sphia}{
    S_{\varphi-a} = \int \D^4x \sqrt{-g}&\left(-\frac{1}{2}(\partial_\mu\varphi)^2-V\left(e^{\sqrt{\frac{2}{3}}{\varphi\over \mpl}}\right) \right.\\ & \quad\left.- \frac{b^2e^{-\frac{2\sqrt 2}{\sqrt{3}}{\varphi\over \mpl}}}{2}\partial_\mu a\partial^\mu a \right) \,.
}
This is nothing but the action that couples an axion to a dilaton (for a specific axion-dilaton coupling). The axion strings in a theory of a radion, dilaton, and axion would also be described through a similar EFT, where now $\varphi$ corresponds to the (canonically-normalized) combination of the radion and dilaton which couples to the axion. 

Let us now consider infinitely straight, static string solutions extended in the ${\bf \hat z}$ direction. As in the PQ case, the azimuthal angle is denoted by $\theta$ and the radial distance away from the string core is $r$. String solutions with winding number unity have $a(r,\theta) = f_a \theta$, such that the axion field undergoes one full period of field excursion when traversing a spatial circle enclosing the string core. Let us then take the ansatz $\rho = \rho(r)$; the equation of motion for $\rho(r)$ obtained from~\eqref{eq:Srhoa} is then
\es{eq:rho_EOM}{
\rho'' - {\rho'^2 \over \rho} + {\rho' \over r} +{2 \over 3 (2 \pi g_4)^2 \mpl^2 r^2 \rho} - {2\rho^2 V'(\rho) \over 3 \mpl^2} = 0 \,.
}

We now compute $\rho(r)$ at small $r$ to understand the radion contribution to the string tension.  Note that at small $r$ we expect the extra dimension to decompactify ($\rho \to \infty$) for the axion string configuration to be non-singular.
This limit requires that the canonically normalized radion travels an infinite field space distance, in which case the Swampland Distance Conjecture (SDC)~\cite{Ooguri:2006in} states that the EFT should break down. In the present case, an irreducible source of breakdown is clearly associated with the appearance of light KK modes. In the cases where $\rho$ corresponds to a dilaton, the core of the axion string instead corresponds to a region where massive string theory modes become tensionless. A general treatment of infinite distance limits in the context of string theory axion strings can be found in~\cite{Lanza:2020qmt,Lanza:2021udy,Lanza:2022zyg,Grimm:2022sbl,Cota:2022yjw,Wiesner:2022qys,Martucci:2022krl}.

In the decompactification limit the stabilization potential $V(\rho)$ is expected to asymptote to a constant, which we take to be zero, $\lim_{\rho \to \infty} V(\rho) = 0$ (see, {\it e.g.},~\cite{Dine:1985he,Giddings:2003zw,Giddings:2004vr}). This can be checked explicitly using the stabilization potentials that we consider below. Correspondingly, the potential contribution in~\eqref{eq:rho_EOM} can be dropped for analyzing the small-$r$ behavior.\footnote{More precisely we require $\lim_{\rho \to \infty} V'(\rho) \to 0$ sufficiently quickly, as we discuss below in more details for the case of the Goldberger-Wise potential.}

At small $r$ we then find, to leading order in $1/r$, the solution
\es{eq:small_r_v1}{
\rho(r) \approx -{1 \over \pi \sqrt{6}} {1 \over g_4 \mpl} \log(c r/b) \,,
}
where $c$ is a constant prefactor that we determine below.
This solution is valid for $r \lesssim b$, as we see below from the numerical solution. Defining $z\equiv r e^{i\theta}$, the radion and axion profiles may be combined into
\es{BPSsolution}{
 \tau\equiv\sqrt{\frac{3}{2}}\frac{\mpl}{b}\rho+ia\approx -\frac{1}{2\pi b g_4}\log{c\bar z/b} \ ,
}
while in terms of $\tau$, the kinetic terms in the action read
\es{eq:Stau}{
    S_{\tau} \supset \int \D^4x \sqrt{-g} \left(- \frac{1}{2}\right)\left( \frac{2b}{\tau+\bar\tau} \right)^2|\partial\tau|^2 \,.
}
Therefore $\tau$ is a natural complex field to consider close to the string core and its profile is anti-holomorphic with respect to $z$. Such near-core solutions have been extensively studied in the context of supersymmetric theories~\cite{Greene:1989ya,March-Russell:2021zfq} and the swampland program~\cite{Lanza:2020qmt,Lanza:2021udy,Lanza:2022zyg,Grimm:2022sbl,Cota:2022yjw,Wiesner:2022qys,Martucci:2022krl}. (Actually, even if we do not deal with supersymmetry here, our 4D EFT precisely corresponds to the bosonic sector of a supersymmetric one with K\"ahler potential $\propto \log(\tau+\bar\tau)$.) 

The holomorphic or anti-holomorphic solutions are BPS~\cite{Bogomolny:1975de,Prasad:1975kr}: they saturate the following inequality on the (UV part of the) string tension $\mu$,
\es{BPSenergy}{
\mu = &\int \D^2x \, \frac{1}{2} \left( {b \over \rho} \right)^2|\newvec\partial \tau|^2\\
= &\ i\int \frac{1}{2}\left( {b \over \rho} \right)^2 \D\bar\tau \wedge \D\tau + \int \D^2x \, \frac{1}{2}\left( {b \over \rho} \right)^2\left|\partial_z\tau\right|^2\\
\geq & \ i\int \frac{1}{2}\left( {b \over \rho} \right)^2 \D\bar\tau \wedge \D\tau \ .
}
The resulting lower bound, saturated by the current solution, only depends on IR data:
\es{}{
i\int \frac{1}{2}\left( {b \over \rho} \right)^2 \D\bar\tau \wedge \D\tau= &- \sqrt{3\over 2}\mpl b\int {\D\rho \over \rho^2}\wedge \D a \\
= & \ \left(\frac{b}{\rho(r_{\rm IR})}\right)\sqrt{6}\pi f_a\mpl \,,
}
where $r_{\rm IR}$
is an IR cut-off for when the approximation in~\eqref{eq:small_r_v1} is no longer valid. This result agrees with the direct computation, as it should: 
\es{eq:mu_calc}{
\mu &\approx 2 \pi \int_0^{r_{\rm IR}} \D r \, r \left[ {3 \mpl^2 \over 4\rho^2} \rho'^2 + {1 \over 2} \left({b \over \rho}\right)^2 {f_a^2 \over r^2} \right] \\
&= \left(\frac{b}{\rho(r_{\rm IR})}\right)\sqrt{6}\pi f_a\mpl\leq \sqrt{6} \pi f_a \mpl \,.
}
Assuming that $\rho(r)$ falls monotonically with $r$, which we later confirm numerically, the smallest that $\rho(r_{\rm IR})$ can be is $b$, since this is the VEV asymptotically far from the string core.
Taking $\rho(r_{\rm IR}) \geq b$ leads to the inequality in the last line above. As we see below in the numerical calculations of the tension, it appears that the axion strings saturate this inequality.\footnote{The same result can be obtained by considering the BPS string magnetically charged under $A_M$ in 5D, which yields the correct description of the string near the core where the fifth dimension decompactifies. Our solution captures a smeared version of that string. (Forces on such BPS objects exactly balance, so that solutions can be superposed and even smeared. The resulting brane is again BPS in one dimension less, and it is charged under the (smeared) zero mode of the initial gauge field: in the present case this yields the axion.)} 
That is, we conjecture that the tension is given by 
\es{eq:mu_eff_flat}{
\mu_{\rm eff} = &~\sqrt{6} \pi f_a \mpl + \pi f_a^2 \log(m_r / H) + {\mathcal O}(f_a^2) \,,
}
where $m_r^2 \equiv \frac{2b^2}{3\mpl^2}V''(b)$ is the mass of the canonically-normalized radion at its minimum.  
The second term above is the axion contribution to the tension, which is cut off in the IR by the Hubble radius (we leave off sub-leading, non-logarithmically-divergent terms of order $f_a^2$).
This axion contribution, as in the field theory case, originates from the large-$r$ region, $r > 1/m_r$, for which the radion contribution is subdominant. As we comment on further, the result in~\eqref{eq:mu_eff_flat} is important because the term of order $f_a M_{\rm pl}$ dominates over the axion contribution to the tension for $f_a \lesssim 5 \times 10^{16}$ GeV.  This means that such strings will behave fundamentally differently from field theory axion strings in a cosmological context.  

For later purposes, we note the large-$r$ asymptotic:
\es{eq:rho_large_r_main_text}{
\rho = b\left[1 + {2 \over 3} \frac{f_a^2}{ \mpl^2}\left( {1 \over m_r r} \right)^2 + {\mathcal O}\left( {1 \over r^4} \right)\right].
}
This asymptotic form of the solution is valid for $m_r r \gg 1$.  To see this, one may compute the coefficient of the higher-order $1/r^4$ term in~\eqref{eq:rho_large_r_main_text} in the physical limit $\mpl \gg f_a$; in order for the $1/r^4$ term to be subdominant compared to the $1/r^2$ term, we need $m_r r \gg 1$.  

Below, we compute the tension numerically to provide evidence for the conjecture in~\eqref{eq:mu_eff_flat}.  Our numerical results support the scaling $\log(c) \sim \mpl/f_a$.
We also see that the dominant contribution to the string tension is at length scales exponentially smaller than $1/M_{\rm pl}$ from the string core for realistic values $f_a \sim 10^{11}$~GeV, as we quantify further below.
However, this brings the axion-radion EFT described above into question since that EFT is not valid below a length scale $\sim$$1/f_a$, when the KK excitations become dynamic (the KK scale could be even smaller if $2\pi g_4$ is small).
This suggests, in agreement with~\cite{Dolan:2017vmn,Greene:1989ya,Reece:2018zvv,Lanza:2020qmt,Lanza:2021udy,March-Russell:2021zfq,Heidenreich:2021yda}, that (i) we must go beyond the axion-radion EFT to accurately resolve the string core, and (ii) unlike in the field theory case the axion string core in the extra dimension case is really an object with no physical size.
Indeed, we show in the next section that in the context of string theory the axion string cores are fundamental strings or wrapped D-branes, which have no thickness in the transverse dimensions.

To study~\eqref{eq:rho_EOM} numerically, we must specify a radion potential. Goldberger-Wise stabilization provides a motivated choice, which may be obtained as follows, similar to the analysis in~\cite{Chacko:2002sb}. Consider a free 5D scalar field $\Psi$ with an action
\es{eq:bulk_scalar_action}{
	S_\Psi = \int \D^4x \D\phi \,\sqrt{-G}\left(-\frac{1}{2}\partial_M\Psi\partial^M\Psi-\frac{1}{2}M_\Psi^2\Psi^2-\Lambda_B\right) \,,
}
where we include a bulk cosmological constant term $\Lambda_B$.
This has a solution for the extra-dimensional profile of $\Psi$,
\es{eq:GW_soln}{
	\Psi(\phi) = A \exp(M_\Psi\rho\phi) +  B \exp(-M_\Psi\rho\phi) \,,
}
with the constants $A$ and $B$ determined by imposing boundary conditions $\Psi(0)=v_h$ and $\Psi(\pi)=v_v$.
Substituting $\Psi(\phi)$ back into~\eqref{eq:bulk_scalar_action} with the metric~\eqref{eq:5D_metric} and integrating over $\phi$ yields the Goldberger-Wise potential for the radion,
\es{eq:GW_pot}{
	V(\rho) = &\frac{M_\Psi b^2}{2\rho^2}\left(\alpha\coth(M_\Psi\pi\rho)-\beta\csch(M_\Psi\pi\rho)\right) \\
 &+ \frac{\pi b^2 \Lambda_B}{\rho}\,,
}
with $\alpha = v_h^2+v_v^2$ and $\beta = 2v_h v_v$. 
We impose $\beta >0$ and  $\Lambda_B <0$, such that the potential has a unique global minimum.

Defining $\tilde{r}=r/b$, $\tilde{\rho}=\rho/b$,~\eqref{eq:rho_EOM} can be written as the de-dimensionalized equation of motion 
\es{eq:rho_EOM_numerical_general}{
\tilde{\rho}'' - {\tilde{\rho}'^2 \over \tilde{\rho}} + {\tilde{\rho}' \over\tilde{r}} +\frac{2}{3} \left(\frac{f_a}{\mpl}\right)^2 \frac{1}{\tilde{r}^2 \tilde{\rho}} - \frac{2\tilde{\rho}^2b^3}{3\mpl^2}\frac{\partial V}{\partial \rho} = 0 \,,
}
where here the primed quantities are with respect to $\tilde{r}$. In particular for the Goldberger-Wise potential we have
\begin{widetext}
\es{eq:dV_GW}{
\frac{\partial V}{\partial \rho} = \frac{\pi^3 M_\Psi^5 (b M_\Psi)^2}{2 \hat{\rho}^3}\left[\hat{\alpha} \left (-2  \operatorname{coth}(\hat{\rho})- \hat{\rho} \operatorname{csch}^2(\hat{\rho})\right)+\hat{\beta}(\hat{\rho} \operatorname{coth}(\hat{\rho})+2) \operatorname{csch}(\hat{\rho})-2\hat{\Lambda}_B 
\hat{\rho}\right] \,,
}
\end{widetext}
where we define the dimensionless quantities $\hat{\rho} \equiv \pi M_\Psi  \rho$, $\hat{\alpha} = \alpha / M_\Psi^3$, $\hat{\beta} = \beta / M_\Psi^3$, $\hat{\Lambda}_B \equiv \Lambda_B/M_\Psi^5$. For a given set of order one parameters $(\hat{\alpha}, \hat{\beta}, \hat{\Lambda}_B)$, we numerically determine the unique root $\hat{\rho}_*$ of~\eqref{eq:dV_GW}, which allows us to evaluate the last term of~\eqref{eq:rho_EOM_numerical_general} as a function of $\tilde{\rho}$ using $\hat{\rho} = \tilde{\rho} \hat{\rho}_*$. By definition we have $\hat{\rho}_* = \pi M_\Psi b$, such that parametrically $b \sim M_\Psi^{-1}$. Therefore, $m_r \sim M_\Psi^2/\mpl$. We solve for the precise value of $m_r$ numerically by computing the second derivative of $V$ at $\hat{\rho}=\hat{\rho}_*$. Recalling~\eqref{eq:fadef}, it follows that $M_\mathrm{\Psi}\sim f_a$, such that the only dependence of~\eqref{eq:rho_EOM_numerical_general} on dimensionful parameters is from the fourth and last terms which are parametrically $(f_a/\mpl)^2$. 

\begin{figure}
    \centering
    \includegraphics[width=0.48\textwidth]{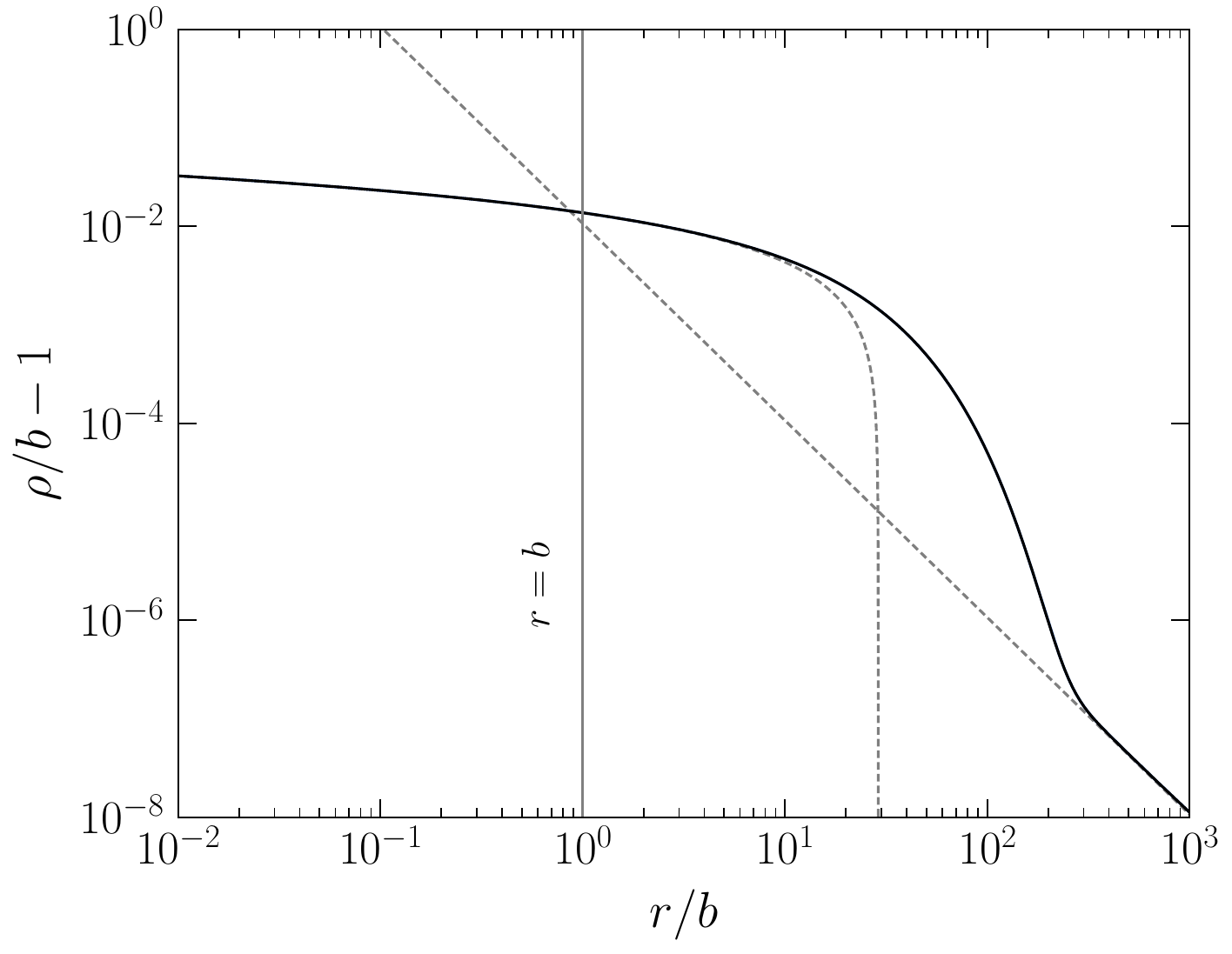}
    \caption{ Radion profile for a static, infinite string in an unwarped 5D construction (solid black). The small and large $r$ ansatzes in~\eqref{eq:small_r_v1} and~\eqref{eq:rho_large_r_main_text}, respectively, are illustrated in dashed grey. Here, $f_a/M_\mathrm{pl} = 5 \times 10^{-3}$.
    }
    \label{fig:radion_profile_1D_main_text}
\end{figure}

We use a 4$^\mathrm{th}$ order collocation method\footnote{This method is based on the algorithm of~\cite{Kierzenka2001ABS}, implemented in \texttt{scipy}'s~\cite{2020SciPy-NMeth} \texttt{scipy.integrate.solve\_bvp}.} to solve~\eqref{eq:rho_EOM_numerical_general} on a finite interval $[\tilde{r}_\mathrm{min}, \tilde{r}_\mathrm{max}]$ spanning 15 decades, with boundary conditions $\tilde{\rho}' = -\sqrt{\frac{2}{3}}\frac{f_a}{\mpl} \frac{1}{\tilde{r}}$ at $\tilde{r}=\tilde{r}_\mathrm{min} $ (from~\eqref{eq:small_r_v1}) and $\tilde{\rho} = 1 -\tilde{\rho}' \tilde{r}/2$ at $\tilde{r} = \tilde{r}_\mathrm{max}$ (from~\eqref{eq:rho_large_r_main_text}). 
Our fiducial choice of parameters entering in the potential is $(\hat{\alpha}, \hat{\beta}, \hat{\Lambda}_B) = (3, 1.9 , -0.4) $, and we also set $2\pi g_4=1$ so that $f_a=b^{-1}$. 
In Fig.~\ref{fig:radion_profile_1D_main_text} we illustrate the resulting radion profile for $f_a/\mpl = 5 \times 10^{-3}$, compared to the asymptotic form at small and large $r$ in~\eqref{eq:small_r_v1} and~\eqref{eq:rho_large_r_main_text}, respectively.

\begin{figure}
    \centering
    \includegraphics[width=0.48\textwidth]{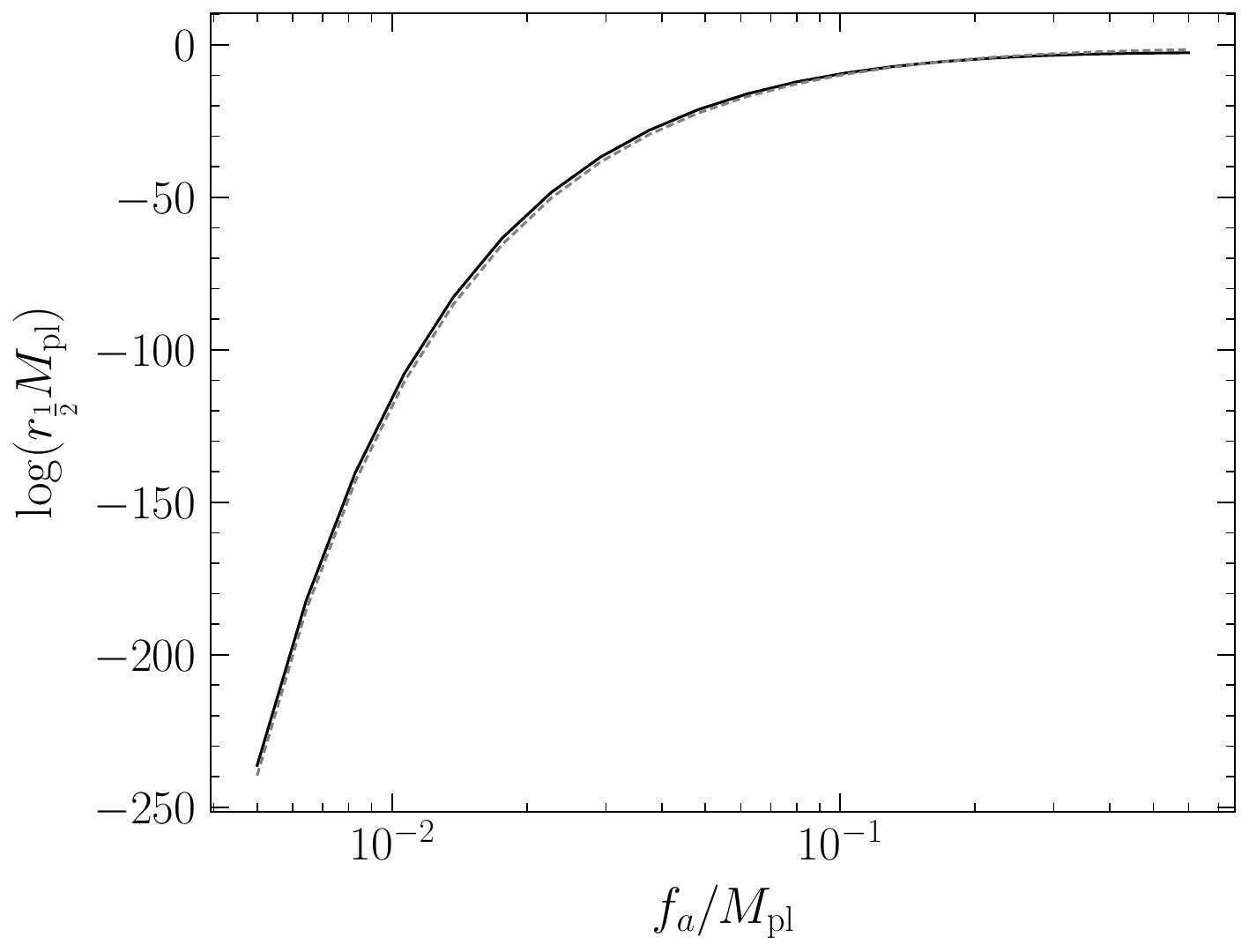}
    \caption{The string ``width" of the infinite string ansatz in units of the Planck length from the 5D EFT.  Note that the width is defined as the radial distance $r_{\frac{1}{2}}$ from the string core which corresponds to half of the UV part of the tension. The analytic estimate for the width in~\eqref{eq:small_width} is illustrated in dashed grey.  Interestingly, we find that the width of the strings is exponentially smaller than $M_{\rm pl}^{-1}$, suggesting that the cores of such strings are 1D objects with no physical width. 
    }
    \label{fig:string_width}
\end{figure}

We verify that for sufficiently low $f_a/\mpl$ the profile satisfies two key properties: (i) the radion approaches its VEV for $r \gtrsim b$, and (ii) the regime of validity of the small-$r$ ansatz extends up to $r \sim b$. Together with~\eqref{eq:mu_calc}, these properties confirm the conjecture~\eqref{eq:mu_eff_flat}.
Together with~\eqref{eq:small_r_v1}, they also imply that $ \log(c) \sim -\sqrt{\frac{3}{2}}\frac{\mpl}{f_a}$. Let us then define the ``width'' of the string as the distance $r_{\frac{1}{2}}$ from the string core that contains half of the UV part of the tension (note that for field theory strings we have $r_{\frac{1}{2}} \sim f_a^{-1}$). Then the above observations imply $\rho(r_{\frac{1}{2}}) \sim 2b$, giving $\log(cr_{\frac{1}{2}}/b)\sim -2 \sqrt{\frac{3}{2}}\frac{\mpl}{f_a}$. Hence, 
\es{eq:small_width}{\log\left(r_{\frac{1}{2}} \mpl\right) \sim -\sqrt{\frac{3}{2}}\frac{\mpl}{f_a} +\log\left(\frac{\mpl}{2\pi g_4 f_a}\right)\,,
}
meaning at least half of the tension is contained within a region much smaller than a Planck length from the string for $f_a \ll \mpl$. In Fig.~\ref{fig:string_width} we confirm this scaling holds to leading order by measuring $\log(c)$ from the numerical profiles.

Lastly, we point out that the above exponential dependence of the string ``width" on $\mpl/f_a$  seems to be largely independent of the choice of potential.   
Any potential falling at least as quickly as $1/\rho$ at large $\rho$ can be neglected at small $r$ in~\eqref{eq:rho_EOM}, such that the small $r$ behavior of $\rho$ is given by~\eqref{eq:small_r_v1}. Further,~\eqref{eq:rho_large_r_main_text} shows that the only dependence of the large $r$ behavior of $\rho$ on the potential is through $m_r$.  As a test, we repeat the logic of the preceding paragraphs for a potential of the form  $V(\rho) = -v^2 e^{-2\rho/b}\rho^2$, similar to that considered in {\it e.g.},~\cite{March-Russell:2021zfq}. We find that as long as $m_r$ is not exponentially suppressed relative to $f_a$, with an exponent of order $\mpl/f_a$, then \eqref{eq:small_width} holds to leading order.

\subsection{Warped extra dimension}\label{sec:EFTSolutionWarped}

We now consider the scenario of a warped extra dimension where the spacetime is a slice of a five-dimensional anti-de Sitter (AdS) geometry. 
We parameterize the metric as,
\es{eq:warp_metric}{
\D s^2 = e^{-2k\rho(x)\phi}g_{\mu\nu}\D x^\mu \D x^\nu + \rho(x)^2 \D\phi^2\, ,
}
where $\rho(x)$ is the radion and $k$ determines the AdS curvature scale.
As above, the extra dimensional coordinate $\phi$ ranges from $0$ to $\pi$ due to an orbifolding $\phi\leftrightarrow -\phi$ identification.
Substituting this ansatz in the 5D GR action and integrating over the extra dimension we find~\cite{Goldberger:1999un}
\es{}{
S_G =&~ \frac{2M_5^3}{k}\int \D^4x \sqrt{-g}\left(1-e^{-2k\pi\rho}\right)R^{(4)} \\
	-&~ \frac{12M_5^3}{k}\int \D^4x \sqrt{-g} \partial_\mu\left(e^{-k\pi\rho}\right)\partial^\mu\left(e^{-k\pi\rho}\right)\, .
	} 
To obtain the low energy effective $\mathrm{U}(1)$ action, we use an ansatz where $A_5 = A_5(x^\mu)$  and $A_\mu = A_\mu(\phi)$~\cite{Choi:2003wr}.
The equation of motion for $A_\mu$ can be written as,
\es{eq:Amu_equation}{
\partial_5\left(e^{-2k\rho\phi}F_{\mu 5}\right) = 0 \, .
}
Here the subscript $5$ refers to the coordinate $\phi$.
To solve~\eqref{eq:Amu_equation}, we can write $e^{-2k\rho\phi}F_{\mu 5} = f_\mu(x^\mu)$ for some general function $f_\mu(x^\mu)$.
By integrating over the extra dimension, with a boundary condition, $A_\mu(\phi=0)=A_\mu(\phi=\pi)=0$, we obtain
\es{}{
F_{\mu 5} = \frac{2k\pi \rho}{e^{2k\pi \rho}-1}\partial_\mu A_5 e^{2k\rho\phi} \, .
}
Substituting this back into the $\mathrm{U}(1)$ action, we arrive at
\es{eq:S_U1_warped}{
	S_{\mathrm{U}(1)} =&~-\frac{1}{4g_5^2}\int \D^5 x \sqrt{-G}G^{MN}G^{AB}F_{MA}F_{NB},\\
	=&~ -\frac{1}{g_5^2}\int \D^4x \sqrt{-g} \frac{2k\pi^2}{e^{2k\rho\pi}-1}g^{\mu\nu}\partial_\mu A_5 \partial_\nu A_5 \, .
}
We can rewrite this in terms of the canonically normalized radion field,\footnote{Note that $\varphi$ has a very different phenomenology than its flat space analog of~\eqref{eq:Sphia}. For instance, its interactions are stronger than gravitational in the warped case.} $\varphi/F = e^{-k\pi\rho}$ where $F^2 = 24M_5^3/k$:
\es{eq:A5_action_warp}{
S_{\mathrm{U}(1)} = -\frac{1}{g_5^2}\int \D^4x \sqrt{-g} \frac{2k\pi^2}{(F/\varphi)^2-1}g^{\mu\nu}\partial_\mu A_5 \partial_\nu A_5 \, .
}
Correspondingly, the gravity action gives
\es{eq:rad_action_warp}{
S_{G} =&~ \frac{2M_5^3}{k}\int \D^4x\sqrt{-g}\left(1-\left(\varphi\over F\right)^2\right)R^{(4)}\\
&-~ \frac{1}{2}\int \D^4x \sqrt{-g}\partial_\mu\varphi \partial^\mu\varphi \, .
}
We verify that in the limit $k\rightarrow 0$, the above action reduces to the corresponding action in flat space.\footnote{To show this, one can repeat the flat-extra-dimension analysis starting with a metric ansatz, $\D s^2 = g_{\mu\nu}(x)\D x^\mu \D x^\nu + \rho(x)^2 \D\phi^2$ which differs from~\eqref{eq:5D_metric} via a 4D Weyl rescaling factor $\langle\rho\rangle/\rho(x)$.} Note that canonically normalizing the radion requires an extra Weyl rescaling to obtain a pure Einstein-Hilbert term. This is performed in App.~\ref{app:warped_profiles}, where it is shown that at leading order in $\varphi/F$, one can simply neglect the first line of \eqref{eq:rad_action_warp} when one studies the radion-axion system.

The axion can be defined as before from the Wilson loop \eqref{eq:theta}.
Since $A_5$ is independent of $\phi$, we have $\vartheta = 2\pi A_5$, and we identify $\vartheta = a/f_a$ with the canonical axion field $a$ having a period of $2\pi f_a$.
Using this we can rewrite,
\es{eq:u1_warped}{
S_{\mathrm{U}(1)} = -\frac{1}{2}\int \D^4x \sqrt{-g}\frac{(F/\langle\varphi\rangle)^2-1}{(F/\varphi)^2-1} g^{\mu\nu}\partial_\mu a\partial_\nu a \,,
}
with
\es{eq:fa_warped}{
f_a^2 = \frac{k}{g_5^2\left[(F/\langle\varphi\rangle)^2-1\right]} \, .
}
In the limit of $k\rightarrow 0$, this reduces to $f_a = 1/(2\pi g_4\langle\rho\rangle)$ upon using~\eqref{eq:gaugec_4d}, as expected. 
We also see that the scale $f_a$ can be parametrically below the 5D UV scale $\sqrt{k}/g_5$ for a significant warp factor $\langle\varphi\rangle/F \ll 1$. Unlike the flat space, we note that $\varphi$ does not travel an infinite field space distance towards the core of the string so that the SDC does not allow us to predict a dramatic breakdown of the EFT at the core of the string. The absence of an infinite distance appears to be consistent with the holographic picture of our warped scenario, which is purely field-theoretical and does not require quantum gravity. To our knowledge, few studies of the SDC on warped geometries have been attempted so far~\cite{Blumenhagen:2022zzw,Seo:2023ssl,Etheredge:2023odp}, and none in the context of axion strings. This represents an interesting avenue for future work.

We now describe the axion string solution when we have an infinite string lying along the $\hat{\bf{z}}$ direction with $a(r,\theta) = f_a \theta$.  As before, $r$ is the radial direction away from the string core. Close to the string where $\rho \to \infty$ ($\varphi \ll F$) the approximate equation of motion for the radion is then given by
\es{eq:radion_eom_warp}{
\varphi''+ \frac{4(\varphi')^2}{F^2}+\frac{1}{r}\varphi'-\frac{k}{g_5^2}\frac{\varphi}{F^2r^2} - V'(\varphi)= 0 \,,
}
where we include the contribution from a stabilizing radion potential $V(\varphi)$.

The potential term may be neglected close to the string, and~\eqref{eq:radion_eom_warp} is then solved by an ansatz of the form $\varphi = C r^\alpha$ with $\alpha = \sqrt{k}/(F g_5)$.
Using this we can compute the contribution to the string tension from the inner core,
\es{eq:inner_core}{
\mu_{\rm in} = \pi \int_0^{r_I} \D r \,r \left[C^2\alpha^2 r^{2\alpha-2} + C^2\alpha^2 r^{2\alpha-2}\right],
}
where both the radion and the axion contribute equally.
The integral is dominated by the region near $r_I$ where we choose to cut off the integral:
\es{}{
\mu_{\rm in} = \pi \alpha \varphi(r_I)^2 \,.
}
Assuming $\varphi(r_I) \leq \langle \varphi \rangle$ and $\langle \varphi \rangle \ll F$, we may obtain an upper bound on $\mu_{\rm in}$:
\es{eq:warped_tension}{
\mu_{\rm in} \leq \pi \alpha \langle \varphi \rangle^2 = \pi f_a \langle \varphi\rangle = \sqrt{6}\pi f_a \mpl {\langle \varphi\rangle \over F} \, . 
}
Taking $k\sim M_5 \sim \mpl$, then 
 the above relation reduces to $\mu_{\rm in}\lesssim f_a^2$.
This implies that in a warped extra-dimensional scenario, the string theory axion string tension is similar to that of field theory axion strings; in particular, the tension is dominated by the axion contribution, given its logarithmic divergence. This is expected since both the axion and the radion are `composite' degrees of freedom in the 4D holographic theory. 
On the other hand, as the amount of warping decreases, the upper limit on the inner core tension in~\eqref{eq:warped_tension} reaches the flat extra dimension result in~\eqref{eq:mu_eff_flat}.

\begin{figure}
    \centering
    \includegraphics[width=0.48\textwidth]{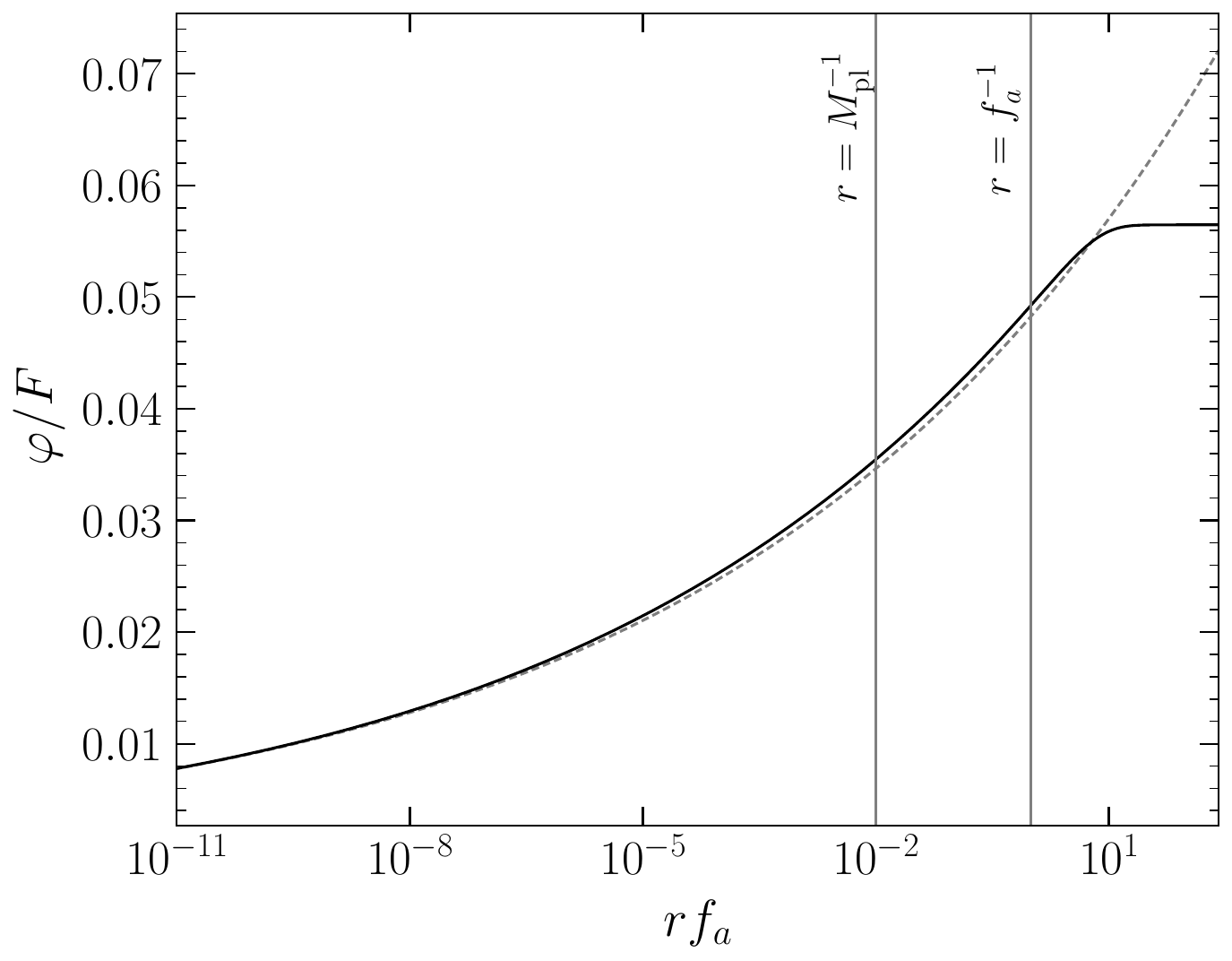}
    \caption{ Radion profile for an infinite string, as in Fig.~\ref{fig:radion_profile_1D_main_text}, but for the warped 5D geometry. In dashed grey we illustrate the small-$r$ asymptotic expectation $\varphi \propto r^{\sqrt{k}/(Fg_5)}$. Here, $f_a / \mpl = 10^{-2}$, and $(\tilde{v}_h,\tilde{v}_v, M_5 /k, g_5^2 k) = (2,1,2, 1)$. (See App.~\ref{app:warped_profiles} for details.)
    }
    \label{fig:radion_profile_warped}
\end{figure}

Let us now repeat the procedure of Sec.~\ref{sec:flat_extra_dim} to study the radion profiles numerically. In App.~\ref{app:warped_profiles} we rederive the radion equation of motion in the warped geometry without recourse to approximations, and verify that it reduces to the flat geometry result~\eqref{eq:rho_EOM} in the limit of zero warp factor. We then solve the equation of motion numerically. As in the flat geometry case, we assume the radion is stabilized by the Goldberger-Wise mechanism. To obtain the radion potential for the warped geometry, we repeat the procedure of the previous section; however, now we add interaction terms for the bulk scalar $\Psi$ on each boundary $i$ orthogonal to the fifth dimension, with $i=h$ for the UV boundary at $\phi =0$ and $i=v$ for the IR boundary at $\phi =\pi$. In particular, we include the actions~\cite{Goldberger:1999uk, Goldberger:1999un}
\begin{equation}
S_i=-\int \D^4 x \sqrt{-g_i} \lambda_i\left(\Psi^2-v_i^2\right)^2 \,,
\end{equation}
with $g_i$ the induced metric on boundary $i$.
The potential obtained is \cite{Goldberger:1999un}
\es{eq:GW_warped}{
V(\varphi)=\frac{k^3}{144 M_5^6} \varphi^4\left(v_v-v_h(\varphi / F)^\epsilon\right)^2 \,,
}
which is valid to leading order in $\epsilon \equiv M_\Psi^2 / 4 k^2 \ll 1$ when the $\lambda_i$ is large (in units of $M_\Psi^{-2}$) on each brane such that $\Psi$ takes the values $v_h (v_i)$ at $\phi=0$ ($\phi=\pi$). 
We impose $v_h> v_v$ such that \eqref{eq:GW_warped} has a minimum at
\es{}{
\frac{\langle\varphi\rangle}{F}=\left(\frac{v_v}{v_h}\right)^{1 / \epsilon}\,.
}
The resulting string profile is shown in  Fig.~\ref{fig:radion_profile_warped} for $f_a/\mpl = 10^{-2}$ and a fiducial choice of parameters specified in App.~\ref{app:warped_profiles}.  We confirm that $\mu_\mathrm{in} \lesssim f_a^2$. However, unlike in the flat geometry, the majority of the tension is not contained in a region exponentially smaller than $\mpl$, suggesting that in this case the string tension may be reliably computed in the EFT.
This might be expected from the dual CFT perspective where both the radion and the axion are composite degrees of freedom with the heavier degrees of freedom reliably integrated out.

\section{String tension from wrapped branes}
\label{sec:string_theory}

We argued in the previous section that the core of the cosmic strings that we study are not resolved in a controlled EFT, neither in four nor in higher spacetime dimensions. Instead, the only (known) controlled description of those objects is found in (super)string theory, as extended fundamental string theory objects wrapped on some compactification cycle. In this section, we explore the top-down string theory perspective on the tension of a given axion string. Namely, given some compactification schemes yielding a 4D axion, we identify the brane which carries the relevant charge in 10D, and compute the 4D tension of the resulting string. We will see that those tensions are much larger than in field theory setups, except when the brane sits at the bottom of a warped throat.  The results, obtained by ignoring the backreaction of the brane on the background fields, match the EFT expectations for the warped and unwarped string tensions detailed in the previous section. That the BPS condition allows one to track objects over weak and strong backreactions is a crucial property of D-brane physics.

We do not attempt to scan full-fledged string compactifications; instead, we extract generic scalings for the two paradigmatic scenarios of large and warped extra dimensions.  Also, although we focus here on ten-dimensional string theory, similar estimates can be performed in eleven-dimensional M-theory, where the strings correspond to wrapped $\mathrm{M}2$ or $\mathrm{M}5$ branes. We stress that the cases we study do not capture all possibilities, instead, they serve as examples that saturate the bounds found in~\eqref{eq:upperBoundTension}-\eqref{eq:lowerBoundTension}. For instance, mildly warped M-theory setups presented in~\cite{Svrcek:2006yi} yield axion string tensions that interpolate between the two extremal values.  We also
do not discuss open string sector axions that arise from matter fields, as they would behave similarly to the normal field theory axion scenario.

Let us start by describing how the axion string uplifts to higher-dimensional objects as one resolves distances smaller than the KK scale. An axion string satisfies
\es{}{
\int_{\gamma} \D a=2\pi f_a n \,,
}
for a curve $\gamma$ encircling the string and an integer winding number $n$. One says that the string is magnetically charged, with charge $n$, under the dimensionless axion $\vartheta\equiv a/f_a$. The meaning of this is the following.
In general, in a $d$-dimensional spacetime and given a $p+1$-form field $C_{p+1}$, a $p$-brane with worldvolume $\Sigma_{p+1}$ can be electrically charged under $C_{p+1}$ or magnetically under the dual $C_{d-p-3}$ defined via $\star \D C_{p+1}= \D C_{d-p-3}$, with $\star$ the Hodge star operator. The electric coupling is captured by the action
\es{eq:brane_action}{
-T_p \int_{\Sigma_{p+1}} \!\!\!\!{\mathrm{vol}_{p+1}} 
+ Q_p\int_{\Sigma_{p+1}}  \!\!\!\!{C_{p+1}} \,,
}
with $Q_p$ the electric charge of the brane, $T_p$ the brane tension, and $\mathrm{vol}_{p+1}$ the $p+1$-form volume element of the brane worldvolume. The charge can be extracted through the integral $\int_{S_{d-p-2}} \star\D C_{p+1}$, where $S_{d-p-2}$ is a surface enclosing the brane. The equation of motion of $C_{p+1}$, whose kinetic-term action reads
\es{}{
-\frac{1}{2} \int \D C_{p+1} \wedge \star \D C_{p+1} \,,
}
 is $\D \star \D C_{p+1} = Q_p \delta(\Sigma_{p+1})$, where $\delta(\Sigma_{p+1})$ is the Poincar\'e dual of $\Sigma_{p+1}$.\footnote{de Rham duality ensures that each non-trivial $p$-cycle $\Sigma_p$ is associated to a non-trivial $(d - p)$-cohomology class. Informally, we may represent the latter by a $(d - p)$ form $\delta(\Sigma_p)$, with indices along the $(d - p)$ directions transverse to the $p$-cycle and coefficient given by a Dirac delta function with support on $\Sigma_p$. }
Therefore, using Stokes theorem for a surface $S_{d-p-1}$ spanning $S_{d-p-2}$,
 \es{}{
\int_{S_{d-p-2}} \star\D C_{p+1} &= \int_{S_{d-p-1}} \D(\star\D C_{p+1}) \\
&= Q_p\int_{S_{d-p-1}} \delta(\Sigma_{p+1})= Q_p \,}
The magnetic charge is, in turn, measured by $\frac{1}{2\pi}\int_{S_{p+2}} \D C_{p+1}$ for a surface $S_{p+2}$ enclosing a $(d-p-4)$-dimensional magnetic brane. Specializing now to the axion string, we may introduce the $2$-form $C_2$ dual to the axion in 4D; {\it i.e.}, $\D_{4D}a = \star\D_{4D}C_2 $, and we see that the electric charge of the string under $C_2$ is the topological axion charge of the string, $Q_2 = 2\pi n f_a$. 

Let us now reinterpret the axion string in the context of compactifications. At some scale, the relevant spacetime dimension is some integer $d\geq 4$, and the relevant field is either a fundamental axion or a $p$-form field $C_p$ of higher degree, from which the axion emerges in the low-energy 4D spacetime. In the latter case, one writes
\es{}{
\frac{a}{f_a} \propto \int_{\Omega} C_p \,,
}
for a $p$-cycle $\Omega$. It is then clear from the above discussion that the axion string uplifts to an object of dimension $(d-p-3)$ magnetically charged under $C_p$,
\es{}{
2\pi n \propto \int_{\gamma\times\Omega} \D C_p \,.
} 
From the 4D perspective, this object appears as a string.

If all KK scales are lower than the string scale, one can track the decompactifications until the generating EFT in 10D, which approximates the full string theory at low energies, breaks down. In that EFT one finds the original $p$-form that gives rise to a given axion in 4D. Nevertheless, the core of the appropriate magnetic solitons corresponding to a 4D axion string remains singular and their tension cannot be reliably computed in the EFT. Fortunately, the UV string theory contains excitations which are extended objects carrying electric or magnetic charge under the massless $p$-form fields of the EFT. For Ramond-Ramond (RR) forms, they correspond to D-branes, while they correspond to the fundamental string or its magnetic dual, the NS5 brane, for the Neveu-Schwarz-Neveu-Schwarz (NSNS) 2-form. $\mathrm{D}p$ branes exist for $p$ odd in type IIB theory, $p$ even in IIA, and $p = 1, 5$ in type I (and do not exist in heterotic theories). Those objects can be reliably studied, and their tension extracted, in a weakly coupled limit of the full string theory.

In string theory, there are various ways to realize axion strings by wrapping $p$-branes on a $(p-1)$-cycle $\Omega$ (of volume $V_{\Omega}$) of the compact manifold and dimensionally reducing the appropriate $(p+1)$-forms.
Taken in isolation in the 10D bulk (without background branes), these extended objects are BPS, so that their fundamental tension $T$ and charge $Q$ as defined in \eqref{eq:brane_action} satisfy
\es{}{
T^2 = \frac{Q^2}{2\kappa_{10}^2} \,,
}
where $\kappa_{10}^{-2}=\frac{4\pi}{g_s^2} M_s^8$, with $M_s = 1/l_s=1/(2\pi\sqrt{\alpha'})$ the string scale and $g_s$ the string coupling. 
In our conventions \cite{Svrcek:2006yi}, the $p$-forms have mass dimension four in 10D. For a $\mathrm{D}p$-brane, $T_{\mathrm{D}p} =2\pi M_s^{p+1}/g_s$, while $T_F=2\pi M_s^2$ and $T_\text{NS5}=\frac{2\pi}{g_s^2}M_s^6$ for the F-string and NS5-brane, respectively. String theory constructions also feature non-BPS branes, whose tension is larger than the one of a would-be BPS brane of the same charge. Furthermore, we stress that the brane action of~\eqref{eq:brane_action} only describes the coupling to the graviton and the minimal couplings to form fields, and it lacks couplings to the dilaton or brane fields, as well as non-topological $p$-form couplings. For simplicity, we focus on backgrounds where the fields concerned by these extra couplings (which depend on the brane under scrutiny) vanish, so that we can reliably use the above brane action. 

We now consider a spacetime with topology $M_4 \times X$, for $M_4$ the 4D Minkowski space and $X$ a 6D compact manifold, such that an appropriate 10D string EFT yields a massless axion in 4D. The 4D Planck mass reads
\es{}{
\mpl^2 = \frac{V_X}{\kappa_{10}^2}= \frac{4\pi}{g_s^2} M_s^8 V_X \,,
}
with $V_X$ the volume of $X$. Upon wrapping a $p$-brane on a $(p-2)$-cycle $\Omega$, the resulting string tension is 
\es{}{
\mu = T_{p} V_{\Omega}\,.
} 
For simplicity we will repeatedly assume that $X$ admits a global product structure of the form  $X=\Omega \times \Omega_{\perp}$, with $\Omega_\perp$ the dimensions not spanned by $\Omega$, such that we may write $V_X = V_{\Omega}V_\perp$, with $V_\perp$ the volume of  $\Omega_{\perp}$.

\subsection{Flat extra dimensions}

We first consider flat extra dimensions, namely the case where the geometry is factorizable: the metric of spacetime is block diagonal on $M_4 \times X$, so that the block with $M_4$ indices does not depend on $X$ coordinates, and reciprocally. We also take vanishing fluxes and trivial dilaton profiles. In order to obtain an axion in 4D, one looks for either a KK scalar or a KK $2$-form zero mode of the 10D fields on $X$.

\subsubsection{Compactifying a $(p+1)$-form down to a 4D $2$-form}\label{sec:flat2form}

We start with the case where the 4D axion is dual to a 4D $2$-form $C_2$, obtained from dimensional reduction of a form $C_{p+1}$ of equal or higher degree. For generic $p$, this situation is for instance encountered in type II string theory, in which case an appropriate $\mathrm{D}p$ brane, electrically charged under $C_{p+1}$, corresponds to the axion string. 

We KK-reduce $C_{p+1}$ on a basis of harmonic forms $\omega$ dual to a basis of the $(p-1)$-th homology group of $X$,
\es{}{
C_{p+1}=\sum_{\omega}\omega_{p-1} \wedge C_2^{\omega} \,,
}
where $C_2^{\omega}$ is a function of the 4D coordinates only. For simplicity , let us assume that the rank of the $(p-1)$-th homology group is $1$, so that it is generated by a single homology class $[\Omega]$. This allows us to avoid discussing mixings between axion fields, though we expect the conclusions to hold if we relaxed this assumption. (We investigate the cosmology of strings in the presence of several axions in Sec.~\ref{sec:multiple_axions}.) We then have a single 4D 2-form, 
\es{}{
C_2=\int_{\Omega}C_{p+1} \ ,
}
where we normalize $\omega$ so that $\int_\Omega \omega=1$. We also assume that $X=\Omega\times \Omega_\perp$, but as above we expect the qualitative features of the argument to remain the same if this is not the case. Wrapping a charged $p$-brane on $\Omega$ leads to a stable configuration, whose worldvolume becomes that of a 4D string$\times \Omega$. It inherits the following coupling to the 4D 2-form:
\es{}{
Q_p\int_{\rm string}\left(\int_\Omega C_{p+1}\right)=Q_p\int_{\rm string}C_2 \ .
}
Thus, this wrapped brane is magnetically charged under the axion and corresponds to its string.

Canonically normalizing the 2-form requires a rescaling \cite{Svrcek:2006yi},
\es{}{
C_2\to \left(\int_X \omega \wedge \star \omega\right)^{-1/2}C_2 = \sqrt{\frac{V_\Omega}{V_\perp}}C_2 \,.
}
From this, we read off the axion decay constant from the charge of the minimally charged brane,
\es{eq:fa_string}{
f_a=\frac{Q_2}{2\pi}=\frac{Q_p}{2\pi}\sqrt{\frac{V_\Omega}{V_\perp}}
}
and relate it to its tension,
\es{eq:fa_mu_string}{
\mu = T_pV_\Omega=\sqrt 2 \pi f_a\mpl \,,
}
where in the second equality we assume that the brane is BPS in 10D.
We also find
\es{}{
\frac{f_a}{\mpl}=\frac{g_s^{2-\delta}}{2\sqrt 2\pi}\frac{l_s^{7-p}}{V_\perp} \ , 
}
where $\delta=0,1,2$ for F strings, D-branes and NS5-branes, respectively. Hence, we see that, in the controlled $g_s\ll 1,V_\perp \gg l_s^{7-p}$ limit, the axion string tension is much larger than the usual field theory result, $\mu \sim f_a^2$.  

We stress that the above applies to the limit case $p+1=2$, where the harmonic $(p-1=0)$-form $\omega$ is a constant ($=1$ given our normalizations), and the canonical normalization of $C_2$ in 4d reads
\es{}{
C_2 \to \left(\int_X 1 \wedge \star 1\right)^{-1/2}C_2=\frac{C_2}{\sqrt{V_X}} \ .
}
The corresponding axion string is an unwrapped $1$-brane.

\subsubsection{Compactifying a $(p+1)$-form down to a 4D axion}\label{sec:flatAxion}

We now consider the case, related to the previous one by electromagnetic duality, where the axions come from the 4D scalar zero modes of a $(p+1)$-form,
\es{}{
C_{p+1} = \sum_\omega \omega_{p+1} \frac{a_\omega}{Q_p} \ ,
}
where $Q_p$ is the charge of a $p$-brane electrically charged under $C_{p+1}$, and the axion normalization is chosen so that it is dimensionless given our conventions for $C_{p+1}$ and $\omega_{p+1}$, and so that it has periodicity $a \to a+2\pi$. Therefore, when a single homology class $[\Omega]$ generates the $(p+1)$-th homology group and $X=\Omega\times \Omega_\perp$, the axion decay constant can be read from its kinetic term,
\es{}{
f_a^2 = \frac{1}{Q_p^2} 
\int_X \omega \wedge \star \omega =
\frac{V_X}{V_\Omega^2 Q_p^2} \,.
\label{eq:fullyWrappedRRfa}
}
We then infer
\es{}{
\frac{f_a}{\mpl}=\frac{g_s^{\delta}}{2\sqrt 2\pi}\frac{l_s^{p+1}}{V_\Omega} \,.
}
The axion string of $a_\omega$ is charged under the 2-form $C_2^{a_\omega}$ dual to the axion in 4D. To highlight the UV origin of this 2-form, we write
\es{}{
\star \D C_{p+1} = \sum_\omega \star \D\left(\omega_{p+1} \frac{a_\omega}{Q_p}\right) &= \sum_\omega \star\left(\omega_{p+1} \frac{\D_{4D}a_\omega}{Q_p}\right) \\
&= \sum_\omega\star_X\omega_{p+1} \frac{\D_{4D}C^{a_\omega}_2}{f_{a_\omega}Q_p} \,,
}
where we use the relation $\D_{10D} = \D_{4D} + \D_X$, that $\omega$ is harmonic on $X$, that, for a $y_1$-form $\omega_1$ on $Y_1$ and a $y_2$-form $\omega_2$ on $Y_2$, $\star_{Y_1\times Y_2}\omega_1\wedge\omega_2=(-1)^{y_2(\text{dim}Y_1-y_1)}\star_{Y_1}\omega_1\wedge \star_{Y_2}\omega_2$,
and finally that the axion is not canonically normalized in $4$D while $p$-forms are in our conventions; hence, $\D a_\omega= f_{a_\omega}^{-1}\star \D C_2^{a_\omega}$. Introducing the dual $C_{7-p}$ of $C_{p+1}$ such that $\star \D C_{p+1}=\D C_{7-p}$, one sees that $C_2^{a_\omega}$ is related to one of the KK zero modes of $C_{7-p}$, 
\es{}{
\D C_{7-p} = \D \left (\sum \omega_{5-p} C^\omega_2 \right) = \sum \omega_{5-p} \D_{4D} C^\omega_2 \,,
}
and that we should identify $C_2^\omega=\sqrt{\int_X \omega_{5-p}\star \omega_{5-p}}C_2^{a_\omega}$ and $\omega_{5-p} \propto 
(\star_X \omega_p)/(f_{a_\omega}Q_p)$. This tells us that the string magnetically charged under the axion is obtained by wrapping the $(6-p)$-brane which has magnetic charge under $C_{p+1}$ on a cycle belonging to the class $[\Omega^\star]$ de Rham dual to $\star_X \omega_{p+1}$.  (This is a $(5-p)$-homology class of $X$, so we do get a string. Also, it makes sense to apply de Rham duality to $\star_X\omega_{p+1}$, which is closed since $\omega_{p+1}$ is harmonic.) Under our assumption that $X=\Omega \times \Omega_\perp$ and that there is a unique class of $(p+1)$-cycles (hence of $(5-p)$-cycles), we find that $[\Omega^*]=[\Omega_\perp]$. The stable brane is that wrapped on $\Omega_\perp$. Dirac quantization condition for a minimally charged brane implies
\es{}{Q_pQ_{6-p} = 2\pi \,,
}
so that the string tension is again
\es{}{\mu = T_{6-p}V_{\Omega_\perp}= \sqrt{2}\pi f_aM_{\rm pl} \,,
}
in agreement with the result found in~\eqref{eq:fa_mu_string}.

\subsection{Warped compactifications}

We now turn to warped compactifications, where the entries of the metric with $M_4$ indices depend on the coordinates on $X$. For simplicity, let us take a spacetime $M_4\times I_1 \times X_5$, where $X_5$ is some 5d manifold of coordinates $z^m$ and $I_1=S_1/\mathbb{Z}_2$ is an orbifolded circle with coordinate $\phi$ and radius $\rho$, equipped with a metric of the form
\es{}{
\D s^2_{10D}=e^{-2k\rho\phi}g_{4D,\mu\nu}\D x^\mu \D x^\nu+\rho^2\D\phi^2+\D s^2_{X_5} \,,
}
so that the compactification on $X_5$ is not warped. We also assume a trivial dilaton profile. An example would be a slice of the ${\rm AdS}_5\times X_5$ geometry obtained from the near geometry of a stack of $\mathrm{D}3$-branes, where $X_5$ could, for instance, be $S_5$ for branes in flat space, or $T_{1,1}$ with topology $S^2 \times S^3$ for branes at a conical singularity. We can then perform the KK reduction on $X_5$ by following the steps in the previous section. In order to get an axion in 4D, one can focus on KK scalars, vectors, or $2$-forms zero modes of the 10D fields on $X_5$.

\subsubsection{Compactifying a $(p+1)$-form down to a $2$-form on a slice of AdS$_5$}

We start again with the case where the 4D axion is dual to a 4D $2$-form $C_2$, and we assume that this 4D $2$-form descends from a $2$-form on AdS$_5$, itself descending from a $(p+1)$-form in 10D. This case also captures that of Section~\ref{sec:EFTSolutionWarped}, where the 4D axion descends from a wrapped gauge field on AdS$_5$: $2$-forms and vectors are electric-magnetic duals of one another in 5D. Axion strings correspond to 1-branes extending in 4D and point-like on $I_1\times X_5$. They have tension $\mu_{\rm 5D}$ in 5D and, following the logic of Section~\ref{sec:flat2form}, we obtain their charge
\es{}{
Q_{2,{\rm 5D}}= 
\sqrt 2\frac{\mu_{\rm 5D}}{M_5^{3/2}} \,.
}
The charge is measured with respect to the canonically normalized $2$-form in 5D, and $M_5$ is the 5D Planck mass: $M_5^3 = (4\pi/g_s^2) M_s^8 V_X$. On this warped space, we can show that a flat KK profile is consistent for the $2$-form, as long as appropriate boundary conditions are chosen.
Therefore the canonically normalized $2$-form in 4D is obtained through the rescaling
\es{}{
C_2\to 
\sqrt{\frac{k}{e^{2k\rho\pi}-1}}C_2 \,,
}
and the 4D charge of the $\mathrm{D}1$-brane is 
\es{}{
Q_2=\sqrt{\frac{k}{e^{2k\rho\pi}-1}}Q_{2,{\rm 5D}}=
\sqrt{\frac{2k}{e^{2k\rho\pi}-1}}\frac{\mu_{\rm 5D}}{M_5^{3/2}} \,.
}
Using now the values of the 4D tensions and Planck masses,
\es{}{
\mu=\mu_{\rm 5D}e^{-2k\rho\phi_\text{brane}} \ , \quad \mpl^2=\frac{1-e^{-2k\pi\rho}}{k}M_5^3 \ ,
}
we obtain
\es{}{
f_a=\frac{Q_2}{2\pi}=
\frac{e^{k\rho(2\phi_\text{brane}-\pi)}}{\sqrt 2 \pi} \frac{\mu}{\mpl} \ .
}
Due to its position-dependent tension, the brane will be driven to the IR brane\footnote{The warping is often supported by fluxes, for instance by an RR 4-form flux in the case of a stack of D$3$-branes at the orbifold fixed point. This flux can cancel the force felt by other probe branes, such as another D$3$ brane displaced in the bulk. Here, we focus on branes which are not affected by the flux, such as D$1$-branes produced after D$3$-brane inflation.} ($\phi_\text{brane}=\pi$), where we see that its tension is warped down with respect to the value found for flat extra dimensions, $f_a \mpl$. Instead, we now find $\mu\sim f_a^2$, in line with the results of Sec.~\ref{sec:EFTSolutionWarped}. 
This result is in line with the expectation that string theory in warped backgrounds has field theory duals, allowing one to describe the axion strings at the IR brane. 

\subsubsection{Compactifying a $(p+1)$-form down to an axion on a slice of AdS$_5$}

We now turn to the case where the 4D axion is simply the zero mode of a scalar on AdS$_5$. Axion strings correspond to 2-branes in 5D wrapped on $I_1$, which carry magnetic axion charge in 5D. From the logic of Section~\ref{sec:flatAxion}, we obtain
\es{}{\mu_{\rm 5D} =\sqrt{2}\pi (f_{a,{\rm 5D}}M_{5})^{3/2} \ ,
}
where the axion decay constant in 5D $f_{a,{\rm 5D}}$ can be read from the 5D axion kinetic term: $f_{a,{\rm 5D}}^3(\partial a)^2/2$. 
The wrapped brane then has a tension
\es{}{\mu =\frac{1-e^{-2k\rho\pi}}{k}\mu_{\rm 5D} \ .
}
The KK profile for the axion zero mode is flat so that the 4D axion decay constant reads
\es{}{f_a 
=\sqrt{\frac{1-e^{-2k\rho\pi}}{k}}f_{a,{\rm 5D}}^{3/2}=\frac{\mu}{\sqrt 2 \pi\mpl} \ .
}
As the brane extends out of the warped region, we recover the flat space relation.
\newline

\section{Axion cosmology in field theory versus string theory UV completions}\label{sec:axion_cosmo}

In the previous sections we describe the structure of field theory and string theory axion strings, but just because the strings exist mathematically does not mean that the strings form dynamically in a cosmological context.  We begin by reviewing the argument for dynamical string formation in the field theory axion scenario, before arguing against axion string formation in extra-dimensional constructions under the standard cosmological paradigm of a {\it generic} scalar-field-driven inflationary epoch followed by reheating. On the other hand, we then discuss how string theory axion strings may form at the end of inflation in the context of D-brane inflation. 

\subsection{No string theory axion string formed after reheating}

We denote the reheat temperature after inflation by $T_{\rm RH}$. (Our logic also extends to the case where the maximum temperature $T_{\rm max}$ reached during reheating is much larger than $T_{\rm RH}$, upon replacing $T_{\rm RH}$ by $T_{\rm max}$.) Let us assume for simplicity that the Universe is radiation-dominated below $T_{\rm RH}$ until matter-radiation equality.
In the minimal PQ theory of \eqref{eq:minimalPQ}
, the PQ scalar $\Phi$ acquires a thermal mass $m_{\rm therm}$ which restores the PQ symmetry for $T \gtrsim f_a$. Thus, if $T_{\rm RH} \gtrsim f_a$ the PQ symmetry is restored; in the unbroken phase $\langle \Phi \rangle = 0$ but $\Phi$ acquires non-trivial thermal fluctuations about this mean value through interactions with the SM bath.  The axion field, which is the phase of $\Phi$, thus has random and uncorrelated values over spatial scales much larger than $1/T \ll H^{-1}$. 

When the UV PQ theory has a $\mathrm{U}(1)_{\rm PQ}$ global symmetry, the theory undergoes a second-order phase transition to the broken phase where $\langle \Phi \rangle \neq 0$ for $T \lesssim f_a$. Thus, by the Kibble-Zurek mechanism~\cite{Kibble:1976sj,Zurek:1985qw}, global strings develop, which are characterized by closed curves encompassing strings where the axion field has a full $2 \pi f_a$ field excursion. A key point of string formation is that in the high-temperature theory, there are no preferred values for the axion field; it takes on random, uncorrelated average values over causally disconnected Hubble patches.  The radial mode is massive and non-relativistic at temperatures below $f_a$; this mode freezes out to its VEV except at the location of string cores. 
Let us stress that, although we focused the discussion on the minimal model of \eqref{eq:minimalPQ}, the conclusion only depends on the fact that there exists a PQ-preserving point in field space that the Universe selects at high enough temperatures, so that it also holds in more elaborate PQ UV completions.

In contrast, the extra dimension UV completion does not have a symmetry-breaking phase transition as $T$ crosses $f_a$, even if $T_{\rm RH} \gg f_a$.\footnote{As noted previously, in scenarios where $T_{\rm RH}$ is larger than the warped down string scale, there can be a Hagedorn phase transition producing F-strings~\cite{ENGLERT1988423, Polchinski:2004ia, Frey:2005jk}. Here, we consider scenarios where the warped down string scale is larger than the scale of inflation $V_{\rm inf}^{1/4}$, so that a Hagedorn phase transition does not occur; if strings are generated by a Hagedorn phase transition then the cosmology would proceed as in Sec.~\ref{sec:network}.} The first point to note is that in the flat extra dimension case, the radion is decoupled from the thermal plasma at temperatures below $T_{\rm RH}$, and cannot adjust to smooth out the core of a would-be axion string.  Indeed, the radion is a gravitational degree of freedom and thus it scatters with the SM plasma in the early Universe with a scattering rate $\Gamma \sim T^3 / M_{\rm pl}^2$. Thus the radion is decoupled from the SM plasma at temperatures $T \lesssim M_{\rm pl}$.  On the other hand, the reheat temperature is constrained by the BICEP2/Keck Array upper limit on the tensor-to-scalar ratio to be less than $T_{\rm RH} \lesssim 10^{16}$ GeV~\cite{Planck:2018jri}.  Thus, post-reheating the radion was never thermalized and instead was frozen at its homogeneous initial misalignment value until Hubble dropped below its mass. After this point the radion redshifts like matter and decays to SM final states.

Let us return to the EFT description of flat extra-dimensional strings given in Sec.~\ref{sec:flat_extra_dim}.
Consider the scenario where $T_{\rm RH} \gg f_a \sim 1 /b$, such that the 5D gauge field $A_M$ is in thermal equilibrium with the SM plasma.  During inflation the field component $A_5$ acquires a homogeneous VEV $\langle A_5 \rangle$ that is related to the specific field value of $A_5$ at the point of space we inflated from.  
This VEV translates to an axion misalignement angle $\langle \vartheta \rangle$, where $\vartheta$ is the dimensionless axion field defined in~\eqref{eq:theta}; the misalignment angle takes on a random value between $0$ and $2 \pi$ with equal probability.

Post reheating, the field $A_5$ has non-trivial thermal fluctuations, but its VEV is preserved by thermal fluctuations.\footnote{In contrast, in the field theory case post-reheating for $T \gg f_a$ the axion, which is the phase of the complex PQ scalar field, does not have normally-distributed thermal fluctuations.}  When $T$ drops below $f_a$ there is no phase transition, but instead the 5D KK modes of $A_M$ become massive, decouple from the SM plasma, and then decay into lighter states.  Since the thermal fluctuations in $\vartheta$ do not create singular field configurations, the axion field $\vartheta$ relaxes to its inflationary-selected VEV $\vartheta_i \equiv \langle \vartheta \rangle$ during the subsequent expansion of the Universe, such that at the QCD phase transition the cosmology is that of a constant initial misalignment angle $\vartheta_i$. 

In the case of extra dimension axions, there is in fact no difference between $T_{\rm RH} > f_a$ and $T_{\rm RH} < f_a$. If $T_{\rm RH} < f_a$ then the cosmology is, as in the field theory case, that of a constant initial misalignment angle $\vartheta_i$.  Importantly, as discussed above, the case $T_{\rm RH} > f_a$ is equivalent to this cosmology; the dynamics is set by a homogeneous, constant, initial misalignment angle $\vartheta_i$. Not only does this imply that axion strings do not form, except in special inflationary scenarios such as D-brane inflation discussed in the next subsection, it also implies that isocurvature constraints arising from quantum fluctuations of the axion (or $A_5$) during inflation are always relevant constraints for the string theory axion. In contrast, for the PQ axion these constraints are only relevant if $T_{\rm RH} < f_a$, as in the other case, the isocurvature perturbations are erased by the initial conditions of the axion field.  Upper limits on the isocurvature perturbations from Planck measurements constrain the Hubble parameter $H_{\rm inf}$ during inflation to be less than~\cite{Planck:2018jri} 
\es{}{
H_{\rm inf} \lesssim 8.6 \times 10^6 \, \, {\rm GeV} \, \left( {f_a \over 10^{11} \, \, {\rm GeV}} \right)^{0.408} \,.
}
Note that at present the strongest upper limit on $H_{\rm inf}$ is $H_{\rm inf} \lesssim 6 \times 10^{13}$ GeV, and this upper limit should only improve by a factor of $\sim$5 in the future~\cite{CMB-S4:2016ple}; this implies, in particular, that a string theory QCD axion is incompatible with a near-term detection of primordial CMB B-mode polarization from inflation, except in special inflationary scenarios such as D-brane inflation that do produce axion strings.  

It is worth commenting on the expected axion mass in the case without axion strings from the misalignment angle alone. We assume a radiation-dominated cosmology below the temperature at which the axion field begins to oscillate and compute the DM abundance for different values of $f_a$ and the initial misalignment angle $\vartheta_i$ using the code package \texttt{MiMeS}~\cite{Karamitros:2021nxi}.
\texttt{MiMeS} solves the axion equations of motion assuming a radiation-dominated universe with the axion susceptibility presented in~\cite{Borsanyi:2016ksw}. (See~\cite{Karamitros:2021nxi} for more details.) For the 68\% (95\%) confidence intervals on the mass prediction we assume $\vartheta_i=[0.16,0.84]\pi$ ($\vartheta_i=[0.025,0.975]\pi$) and find for each $\vartheta_i$ the correct $f_a$ at which $\Omega_a \equiv \Omega_\text{DM} = 0.12h^{-2}$~\cite{Planck:2018vyg}. 
The axion mass $m_a$ that produces the correct DM abundance is predicted to lie in the range $(1.86,\ 53.3)$ $\mu$eV ($(0.0777,\ 118)$ $\mu$eV) at 68\% (95\%) confidence. 
Of course, there may be anthropic reasons why, for a smaller value of $m_a$, the initial misalignment angle of our observable Universe is selected to be near zero in order to give conditions necessary for life~\cite{Hertzberg:2008wr,Tegmark:2005dy}.

\subsection{Axion strings in D-brane inflation}

There is a well-studied non-thermal 
production mechanism for axiverse cosmic strings: strings can form at the end of D-brane inflation~\cite{Sarangi:2002yt,Jones:2003da,Dvali:2003zj}. 
D-brane inflation\footnote{Other inflation models involving D-branes have been proposed, for instance, the closely-related setup with branes at angles~\cite{Garcia-Bellido:2001lbk} or the D3-D7 system~\cite{Dasgupta:2002ew}. Cosmic strings can also be formed in these models~\cite{Kallosh:2003ux,Urrestilla:2004eh,Binetruy:2004hh,Dasgupta:2004dw,Gwyn:2010rj}.} \cite{Dvali:1998pa,Dvali:2001fw} identifies the inflaton with the modulus encoding the separation between a brane and an anti-brane, extended in the four non-compact spacetime dimensions, and possibly in compact ones. 
With a suitable metric and fluxes, an isolated brane does not move in the compact dimension by itself, but in the presence of another brane, there can be an attractive force between them, mediated by long-distance closed string exchange.
If that potential is flat enough, it can support a sufficiently long period of inflation.

At large brane separation, the open strings stretched between them are very massive and do not influence the dynamics. However, at small separations, the lightest mode becomes tachyonic and induces brane-anti-brane annihilation. 
As in the Kibble-Zurek mechanism, this tachyon condenses to a vacuum manifold whose topology is consistent with the formation of cosmic strings \cite{Sen:1999mg,Dvali:2002fi,Dvali:2003zh,Blanco-Pillado:2005oxi}. Some of these strings are D-branes of lower dimension, which can be identified given their RR charges. 
In more detail, each of the brane-anti-brane pair is associated with a $U(1)$ gauge theory and the tachyon degree of freedom couples to one combination of the $U(1)\times U(1)$ theory.
Tachyon condensation then breaks that $U(1)$ and gives rise to cosmic strings, similar to a field theory scenario.
In the core of the string, a non-zero field strength survives and induces a coupling to lower degree RR forms through the Wess-Zumino couplings of the annihilating branes. For the most-studied $\mathrm{D}3$-$\overline{\mathrm{D}3}$ brane system, the strings would be $\mathrm{D}1$-branes extended along one non-compact space dimension. 
As they turn into $\mathrm{D}1$-branes via S-duality, it is also expected that F-strings are formed in the process, as well as any bound state of F- and D-strings of charge $(p,q)$. 
The more general case of D$(3+p)$-$\overline{{\rm D}(3+p)}$ brane inflation leading to ${\rm D}(1+p)$-branes wrapped down to 4D strings has also been discussed \cite{Dvali:2003zj,Jones:2003da}.

The above mechanism of cosmic string formation however relies on two important aspects.
First, the inflaton potential needs to be flat enough to support a sufficient number of $e$-foldings and be consistent with the observations.
Second, for our purpose, we also need to understand whether the cosmic string actually has an axion associated with it.
The first aspect requires a precise computation in a string-theoretic EFT.
Brane-antibrane systems generically do not have a flat-space potential compatible with inflation, although fine-tuning allows one to evade this \cite{Burgess:2001fx,Garcia-Bellido:2001lbk}. 
However, these early works assumed an unspecified mechanism for moduli stabilization, which was then shown to be a crucial part of the discussion in~\cite{Kachru:2003sx} (KKLMMT).
In particular, KKLMMT considered a warped geometry which led to an exponential flattening of the inflaton potential, along with a stabilization of all the moduli -- a significant improvement over earlier constructions.
However, it was found that {\it generic} stabilization mechanisms could still make the inflaton too heavy to support inflation, and some degree of fine-tuning (roughly a percent) might be needed.

The second aspect of producing cosmic axion strings is however more challenging, at least in the context of the KKLMMT construction.
This construction involves a $Z_2$ orientifold which removes the zero mode of the axion fields that couples to the D- and F-strings~\cite{Copeland:2003bj}.
As a result, while the cosmic strings can still be metastable, they would not source axions.
It may be possible to find other, similar constructions to KKLMMT where the axion field does survive in the low-energy spectrum, though the fact that in this canonical D-brane inflation picture there is no axion in the low-energy EFT may also be taken as an additional argument against cosmic axion strings in string theory. 
A similar projection of the NSNS and RR 2-forms out of the spectrum is found in the ${\rm D}3$-${\rm D}7$ inflation model~\cite{Dasgupta:2002ew}.
In the following sections, however, we will assume that an axion zero mode does survive in the low energy EFT in order to discuss the cosmological dynamics of an axion cosmic superstring network generated from, {\it e.g.}, D-brane inflation. On the other hand, the construction of fully-controlled D-brane inflation models whose cosmic superstring networks source axions in the EFT would be an interesting direction for future work.

\section{String network evolution}
\label{sec:network}
As we discuss in Sec.~\ref{sec:axion_cosmo}, string theory axion strings, unlike field theory axion strings, require special inflationary conditions to form, such as forming through D-brane annihilation at the end of brane inflation. In this section, we suppose that string theory axion strings do form in the early Universe, and we discuss the evolution of the resulting string network and the radiation it produces. 

As we show in Sec.~\ref{sec:string_theory}, axion strings in string theory may be interpreted as wrapped D-branes or $\mathrm{F}$-strings that magnetically source the axion.  Such cosmic superstrings have been discussed extensively in the literature as possible sources of GWs, primordial density perturbations, micro-lensing signals, and other early Universe signatures~\cite{Sakellariadou:2009ev, Copeland:2009ga, Danos:2009vv, Ellis:2023tsl}.  Here, we consider the possibility that these superstrings also source axions, and we discuss how the axions modify the superstring network and the resulting radiation.

Let us consider a network of cosmic superstrings that are magnetically charged under an axion $a$ with decay constant $f_a$; as shown in Sec.~\ref{sec:string_theory}, in the absence of strong warping in the extra dimensions,  the tension of the strings may be written as
\es{eq:mu}{
\mu = \kappa f_a M_{\rm pl} \,,
}
where $\kappa$ is a number of order unity. (The string theory calculations of Sec.~\ref{sec:string_theory} suggest $\kappa = \sqrt{2} \pi$,  but here we keep this coefficient more general.)  Note that D-brane inflation may (or may not) form multiple types of strings, such as a network of F- and D-type strings~\cite{Copeland:2003bj}.  Such mixed string networks\footnote{These are commonly termed $(p,q)$ string networks in the superstring literature.} have string junctions, since F-strings can end on D-strings; the evolution of such mixed string networks is more complicated, and we do not consider this possibility further in this work for simplicity. Rather, we suppose that the only strings present cosmologically are those that source the axion of interest $a$.

\subsection{Network evolution: no axion emission}

Let us briefly summarize the evolution of cosmic superstring networks, characterized by a tension $\mu$, without accounting for radiation loss in the form of axions.  Afterwards, we discuss how to incorporate axion emission. In this section, we assume the string network, for flat-extra-dimension strings, evolved entirely during the radiation-dominated era.  The standard assumption for cosmic string networks is that they approach a scaling solution via string intersections with the scaling solution maintaining energy conservation through GW radiation.  

We denote the string re-connection probability as $P$; this is the probability that if two strings intersect they will intercommute (exchange ends). The probability that the strings pass through each other unaffected is $1-P$. The probability $P$ should depend both on the relative velocity between the strings and their relative angle, but it is common to characterize the network by the network-averaged quantity $P$. For field theory local and global strings $P \approx 1$,\footnote{$P$ may also be artificially reduced in field theory models: {\it e.g.}, for $N$ decoupled copies of Abelian-Higgs model, we have $P=1/N$.} but for D-brane string networks it is expected that $P\sim (0.1,1)$~\cite{Jackson:2004zg}. ($P$ may be even smaller if the strings are able to ``miss'' each other in one of the extra dimensions.) 
Suppose that we start with a network consisting of long strings, with lengths much larger than a horizon size. As the strings evolve according to their Nambu-Goto equations of motion, they can intersect, forming for example closed string loops. Left in isolation a closed string loop will disappear by GW radiation.  In particular, a string loop $L$ of length $\ell$ emits energy with rate
\es{eq:dot_EGW}{
{\D E_{\rm GW} \over \D t} = r_G[L] G \mu^2 \,,
}
where $r_G[L]$ is a dimensionless constant that depends on the shape (but not the size) of $L$ and where $G$ is Newton's constant.  For typical string loops expected in cosmological context $r_G[L] \approx 50$~\cite{Vachaspati:1984gt,Allen:1991bk,Allen:1994iq,Blanco-Pillado:2017oxo}.  In this work we assume, for simplicity, that all $r_G[L]$ take the same value $r_G[L] \approx r_G \approx 50$.  This implies that a loop in isolation changes length with time according to $\ell(t) = \ell_0 - r_G G \mu t$, with $\ell_0$ the initial length, implying the strings with length smaller than roughly $r_G G \mu t$ at given cosmological time $t$ will decay within a Hubble time. 

Modern simulations~\cite{Blanco-Pillado:2011egf,Blanco-Pillado:2013qja} in addition to older works and analytic arguments~\cite{Bennett:1987vf,Albrecht:1989mk,Allen:1990tv} (see~\cite{Auclair:2019wcv} for a review of different approaches) suggest that regardless of the initial string network properties the string network reaches a scaling solution.  One way of characterizing the scaling solution is through the differential number density of sub-horizon-size loops of length $\ell$ per unit length, $n(\ell,t)$.  In the scaling regime, the number density at any time may be related to a universal function $n(x)$ through $n(\ell,t) = t^{-4} n(x)$ with $x = \ell / t$. 
The velocity-dependent one-scale (VOS) model~\cite{Martins:1996jp,Martins:2000cs,Martins:2003vd,Sousa:2013aaa} is a semi-analytic approach to describing the string network whereby one models the production of loops through a loop-production function that splits loops off of long strings. The VOS equations of motion then relate the average distance between long strings and the average string velocity; solving the VOS equations of motion with an ansatz for the loop-production function, and imposing energy conservation, which then leads to solutions for the loop number density.  VOS-derived number densities suggest that (for $P = 1$)
\es{eq:n_x}{
n(x) \approx {\alpha \over  (x + r_G G \mu)^{5/2}} \,,
}
for $x \lesssim 0.1$ and $\alpha \approx 0.18$; these values agree with those in modern numerical simulations such as~\cite{Blanco-Pillado:2013qja}, and so we adopt them throughout this work. 
On the other hand, some simulations such as~\cite{Ringeval:2005kr,Lorenz:2010sm} predict more small-scale loops relative to~\eqref{eq:n_x}, but we do not consider these results here (i) because they do not so clearly obey energy conservation, and (ii) because such results would produce more axion and GW radiation relative to our fiducial choice in~\eqref{eq:n_x}. (Hence, our results are more conservative from an observational perspective.) 

Generically, one expects that the energy density in the string network during the scaling regime grows as the reconnection probability $P$ decreases, since the network is, {\it e.g.}, able to less efficiently lose energy to small loops.  However, dedicated simulations with $P <1$~\cite{Avgoustidis:2005nv} suggest that the string energy density is roughly constant over the range $0.1 \lesssim P \lesssim 1$, which is the range of probabilities we are primarily interested in for D-brane networks, and so we adopt the $P = 1$ results above in the analyses that follow.    

Let us now compute the energy production rate $\Gamma_{\rm GW}$ of energy per unit time per unit volume by string loops within the scaling regime. We may write this rate as 
\es{eq:Gamma_GW_rate}{
\Gamma_{\rm GW} = r_G G \mu^2 \int \D \ell n(\ell,t) 
&\approx (16 \alpha/ 3) H^3 \mu  (r_G G  \mu)^{-1/2} \\
&\approx {16 \sqrt{8 \pi \kappa} \alpha \over 3 \sqrt{r_G}} H^3 \sqrt{ f_a {M_{\rm pl}^3}} \,,
}
where in the last line we use the expression for $\mu$ in~\eqref{eq:mu}. (Note that~\eqref{eq:Gamma_GW_rate} only accounts for emission from loops with $x \lesssim 0.1$.) 
In addition to the energy production rate into GWs it is also important to know the frequency spectrum of emitted GWs.

An individual string loop of length $\ell$ may emit GW radiation with a complicated spectrum at frequencies above roughly the first harmonic frequency $\omega_1 = 4 \pi / \ell$, depending on the loop shape and in particular the distribution of cusps and kinks on the loop. Simulations suggest, however, that averaged over an ensemble of loops of length $\ell$ one may approximate $\D \dot E_{\rm GW}^\ell/ \D \omega \propto \omega^{-4/3}$ for $\omega > \omega_1$~\cite{Blanco-Pillado:2017oxo}, where $\D \dot E_{\rm GW}^\ell/ \D \omega$ denotes the energy loss per unit time per unit frequency of a loop of length $\ell$.  Here, however, we follow~\cite{Auclair:2019wcv} and let, for loops of length $\ell$,
\es{eq:dGamma}{
{\D \dot E_{\rm GW}^\ell \over \D \omega} = \dot E_{\rm GW} \left\{
\begin{array}{cc}
0 & \quad \omega < \omega_1 \\
 \omega_1^{q - 1} {q - 1 \over \omega^q}  & \quad \omega > \omega_1 \,,
\end{array}
\right.
}
with $2> q > 1$.  Recall that $\dot E_{\rm GW}$ is given in~\eqref{eq:dot_EGW} and is independent of the string length $\ell$.  For $\omega \gg \omega_1$ one expects, for example, $q = 4/3$ ($q = 5/3$) if the emission is dominated by cusps (kinks)~\cite{Auclair:2019wcv}.  We may then compute the network-averaged emission rate per unit time per unit volume per unit frequency by
\es{eq:dGamma_exact}{
{\D \Gamma_{\rm GW} \over \D \omega} &= \int \D\ell n(\ell,t) {\D \dot E_{\rm GW}^\ell \over \D\omega} \,.
}
To make progress analytically we approximate the loop number density in~\eqref{eq:n_x} to be
\es{}{
n(x) \approx {2 \over 5}\left\{ 
\begin{array}{cc}
{\alpha \over x_{\rm UV}^{5/2} } & x < x_{\rm UV} \\
{\alpha \over x^{5/2}} & x > x_{\rm UV} \,,
\end{array}
\right. \qquad x_{\rm UV} \equiv r_G G \mu \,.
}
Note that the overall factor of $2/5$ is chosen such that the total number of loops, integrated over $\ell$, matches the value found using~\eqref{eq:n_x}.
Then, we may perform the integral in~\eqref{eq:dGamma_exact} to compute
\begin{widetext}
\es{eq:dGamma_approx}{
{\D \Gamma_{\rm GW} \over \D \omega} &= r_G G \mu^2 \left\{
\begin{array}{cc}
{4 \alpha \over 5 \pi \sqrt{x_{\rm UV}}}{q - 1 \over 2 q + 1}  H^2 \sqrt{ {\omega \over \omega_0} } \,, & \omega< \omega_0\,, \\
{2 \alpha \over 5 \pi \sqrt{x_{\rm UV}}} {q - 1 \over 2-q} H^2 \left( {5 \over 1 + 2 q} \left(  {\omega_0 \over \omega}\right)^q - \left( {\omega_0 \over \omega} \right)^2\right) \,, & \omega > \omega_0\,,
\end{array}
\right.
\qquad \omega_0 \equiv {8 \pi \over x_{\rm UV}} H \,.
}
\end{widetext}
Note that integrating~\eqref{eq:dGamma_approx} over $\omega$ returns the full emission rate in~\eqref{eq:Gamma_GW_rate}.

The emission rate in~\eqref{eq:dGamma_approx} is the instantaneous GW emission rate, but more often we are interested in the full energy density frequency spectrum $\partial \rho_{\rm GW} / \partial \omega$ at some time $t$. The energy density spectrum may be related to the instantaneous emission spectrum by (see, {\it e.g.},~\cite{Gorghetto:2021fsn})
\es{}{
{\partial \rho_{\rm GW} \over \partial\omega} = \int_0^t \D t' \left( {R(t') \over R(t)} \right)^3 {\D \Gamma_{\rm GW}(t',\omega') \over \D \omega'} \,,
\label{eq:drhoGW_domega}
}
where $\omega' \equiv \omega R(t) / R(t')$ with $R$ the scale factor.  Substituting~\eqref{eq:dGamma_approx} into the expression above and neglecting any possible change in the relativistic degrees of freedom for simplicity yields the result 
\begin{widetext}
\es{eq:differential_GW_spectrum}{
{\partial \rho_{\rm GW}(\omega,t) \over \partial\omega} = \Gamma_{\rm GW} {x_{\rm UV} \over 5 \cdot 8 \pi \cdot H^2} 
\left\{
\begin{array}{cc}
{4 (q-1) \over 2 q +1} \sqrt{ {\omega \over \omega_0} } \,, & \omega < \omega_0 \\
{1 \over 2-q} \left[ 3(q - 1) \left( {\omega_0 \over \omega} \right)^2 - {15 \over 2q + 1}\left( {\omega_0 \over \omega} \right)^q + 5 (2 - q) \left( {\omega_0 \over \omega} \right) \right] \,, & \omega > \omega_0 \,.
\end{array}
\right.
} 
\end{widetext}
Integrating the equation above over $\omega$ shows that most of the energy is contained in the high-$\omega$ tail, which gives the logarithmically divergent contribution to the energy density
\es{}{
\rho_{\rm GW}(t) = \Gamma_{\rm GW} H^{-1} \log(f_a / H) + \cdots \,,
}
where we take the UV cut-off to be $f_a$ and neglect sub-leading, finite terms.

\subsection{Network evolution: including axion emission}

We now include the effects of axion emission on the network evolution and compute the resulting axion energy density and frequency spectrum.  We conjecture that a string $L$ of length $\ell$ loses energy to axion radiation at a rate
\es{}{
{\D E_a \over \D t} = r_a[L] f_a^2 \,,
}
where $r_a[L]$ is a shape parameter that does not depend on the string length. Such a formula is valid for global axion strings, and from the perspective of the axion the extra-dimensional strings look like global strings; in particular, the axion field winds around the string cores in the same way it does for global strings. In the context of global string simulations the network-averaged quantity $r_a[L]$ has been measured to be $r_a[L] \approx r_a \approx {\mathcal O}( 10)$~\cite{Hagmann:1990mj,Davis:1986xc,Davis:1985pt,Vilenkin:1986ku}.  Referring to~\eqref{eq:dot_EGW}, we see that the energy loss of string loops to axions is parametrically the same as that to GWs (see also~\cite{Firouzjahi_2008}):
\es{}{
{\dot E_{\rm GW} \over \dot E_{a}} = {r_G \over r_a} {\kappa^2 \over 8 \pi} \,. 
}
Thus, the string loop density takes the same form as~\eqref{eq:n_x} but with the replacement
\es{eq:r_G_sub}{
r_G \to r \equiv r_G +r_a {8 \pi \over \kappa^2} \,.
}
Numerically, we expect $r \sim 50 - 100$.
The GW emission rate is then nearly the same as that given in~\eqref{eq:Gamma_GW_rate} except that  $r$ and $r_G$ enter separately:
\es{eq:Gamma_GW_rate_axion}{
\Gamma_{\rm GW} &= {16 r_G \sqrt{8 \pi \kappa} \alpha \over 3 r^{3/2} } H^3 \sqrt{ f_a {M_{\rm pl}^3}} \,.
}
The axion emission rate is closely related to that of GWs:
\es{eq:Gamma_}{
\Gamma_a = {16 r_a (8 \pi)^{3/2} \alpha \over 3 r^{3/2} \kappa^{3/2}} H^3 \sqrt{f_a M_{\rm pl}^3} \,.
}
With the change given in~\eqref{eq:Gamma_GW_rate_axion}, the GW energy density spectrum is the same as that given in~\eqref{eq:differential_GW_spectrum}.
Similarly, the axion energy density is, to leading order in large $\log(f_a / H)$,
\es{}{
\rho_a(t) = \Gamma_a H^{-1} \log(f_a / H) + \cdots \,.
\label{eq:rho_a_ST}
}

The differential axion spectrum calculation is analogous to that performed previously for GWs.  If the differential axion spectrum $\D \dot E_a^\ell / \D \omega$ takes the form of~\eqref{eq:dGamma}, for $2 > q > 1$, then the result in~\eqref{eq:differential_GW_spectrum} applies also for the axion spectrum, given the substitution in~\eqref{eq:r_G_sub}.  Let us assume this is the case for the moment, though we revisit this point later in the next paragraph.  For the QCD axion DM calculation that we present in the next section, we need the axion number density $n_a(t)$, which may be calculated through 
\es{}{
n_a(t) = \int {\D \omega \over \omega} {\partial \rho_a(\omega,t) \over \partial \omega} \,.
\label{eq:na_def}
}
Performing this integral using~\eqref{eq:differential_GW_spectrum} we calculate
\es{}{
n_a(t) &= \Gamma_a {3 (q - 1) x_{\rm UV} \over  16 \pi q  H^2} \\
&=2 \sqrt{ {2 \over \pi}} {q - 1 \over q}{r_a \over \sqrt{r} \sqrt{\kappa} } \alpha H f_a \sqrt{f_a M_{\rm pl}} \,.
\label{eq:na_eval_qg1}
}

Note that some global string simulations suggest that $q=1$ ({\it e.g.},~\cite{Buschmann:2021sdq}), in which case the formula above does not apply.  
In this case, repeating the above procedure we compute
\es{}{
n_a^{q=1}(t) \approx {4 \over 3} \sqrt{ {2 \over \pi}} {r_a \over \sqrt{r} \sqrt{\kappa} } \alpha H f_a {\sqrt{f_a M_{\rm pl}} \over \log(f_a / H)} \,,
\label{eq:axion_n_a_q_eq_1}
}
to leading order in large $\log(f_a / H)$.  Note also that by computing the ratio $\rho_a / n_a$ we may infer that at time $t$ the typical axion energy is $\omega \sim H \mpl / f_a$.

\section{The QCD axion DM mass for string theory axion strings}\label{sec:implications}
In the field theory UV completion for the QCD axion, where the PQ symmetry is broken after inflation with $N_{\rm dw} = 1$, such that the axion DM abundance is predominantly produced by string radiation, it has been most recently estimated that the correct DM abundance is obtained for an axion mass in the range $m_a \in (40,180)$ $\mu$eV~\cite{Buschmann:2021sdq} (but see~\cite{Gorghetto:2020qws}).  
In this section, we imagine that the QCD axion DM abundance is produced by a network of string theory axion strings, and we compute the axion mass that gives the correct DM abundance assuming both unwarped and warped axion strings.\footnote{In App.~\ref{sec:alp_network} we consider string theory axion strings that source axion-like particles and not the QCD axion; we show they are constrained by contributions to $N_{\rm eff}$.}  We show that the unwarped QCD axion string result is in strong tension with observational upper limits on the QCD axion mass, while the warped result is the same, with a few caveats, as that from field theory UV completions.

\subsection{Axion strings from warped compactifications}
Let us first review how to estimate $m_a$ in the standard field theory axion string scenario, with $N_{\rm dw} = 1$.  In this case, the string network evolves as under the $m_a = 0$ scenario until shortly before the QCD phase transition, when $m_a(T) \approx 3 H(T)$. Note that during the QCD phase transition $m_a$ rises rapidly with inverse temperature: we use the approximation (for $T \gg \Lambda$)~\cite{Wantz:2009it}
\es{eq:ma_growth}{
f_a^2 m_a^2(T) = \alpha_a \Lambda^4 \left( {\Lambda \over T} \right)^n \,,
}
with $\alpha_a \approx 4.6 \times 10^{-7}$, $n \approx 8.16$, and $\Lambda = 400$ MeV.  Note that for $T \ll \Lambda$ the axion mass asymptotes to its zero-temperature value~\cite{Gorghetto:2018ocs}:
\es{eq:ma_zero_T}{
f_a^2 m_a^2 = \Lambda_1^4 \,, \qquad \Lambda_1 \approx 75.4 \, \, {\rm MeV} \,. 
}
We define the time (temperature) when $3 H(T) = m_a(T)$ as $t_*$ ($T_*$). 

The network itself does not begin to collapse at $t_*$; rather, the network collapses within approximately a Hubble time of the time $t_{\rm coll}$, 
which may be estimated by setting the energy density in domain walls to the energy density in strings. Let us imagine a generic axion string loop at the time $t_{\rm coll}$, which we take to be circular with radius $r_{\rm loop} = H^{-1} / \sqrt{\xi}$, with $\xi \sim {\mathcal O}(1-10)$ the number of strings per Hubble patch~\cite{Safdi:2022xkm}.  The energy associated with the axion string forming the circumference of this loop is approximately 
\es{eq:E_string}{
E_{\rm string} \approx  2 \pi r \mu_{\rm eff}\approx  2 \pi \mu_{\rm eff} / (H\sqrt{\xi}) \,. 
}
On the other hand, for $N_{\rm dw} = 1$ the string loop bounds a single domain wall configuration, with the domain wall stretched over an area approximately $\pi r_{\rm loop}^2$ with surface tension, due to the axion configuration in the vicinity of the domain wall, $\sigma \approx 8 m_a f_a^2$~\cite{Amin:2010jq}.  The energy associated with the domain wall is 
\es{eq:E_dw}{
E_{\rm dw} \approx \sigma \pi r_{\rm loop}^2 \approx 8 m_a f_a^2 \pi / (H^2\xi) \,.
}
Equating $E_{\rm string} \approx E_{\rm dw}$ then leads to the collapse time estimate 
\es{eq:warp_coll}{
m_a(T_{\rm coll}) &\approx {\pi \over 4} \sqrt{\xi} \log(m_s / H) H(T_{\rm coll}) \\
&\approx 200 H(T_{\rm coll}) \,,
}
where in the second line above we take the benchmark values at the QCD phase transition for the PQ string $\xi \sim 17$ and $\log(m_s / H) \sim 70$~\cite{Buschmann:2021sdq}.  Note that this implies $T_{*} / T_{\rm coll} \approx 2$.  

Axions are no longer efficiently created for $T < T_{\rm coll}$, since the string network collapses within a Hubble time of $T_{\rm coll}$. In fact, we argue that -- by number density -- most axions are created by $T_*$. The reason is that while a significant amount of energy is deposited into axions between $T_*$ and $T_{\rm coll}$, that energy is deposited into fewer axions, by number, given the rapid increase of $m_a$ with time, as we explain further below.

Let us suppose that the axion-string network emits axions with an instantaneous differential emission spectrum $\propto k^{-q}$ for some index $q$ for $H \ll k \ll f_a$; Ref.~\cite{Buschmann:2021sdq} measured $q = 1.02 \pm 0.04$. Larger $q$ values produce more DM, so to be conservative let us assume $q = 1.06$.  In this case, the axion number density at $T_*$ is estimated to be
\es{eq:n_aT}{
n_a(T_*) \approx {8 \pi f_a^2 H(T_*) \over \delta} \sqrt{\xi_*} \log_* \,,
}
with $\xi_* \approx 17$ and $\log_* \equiv \log(f_a / H(T_*)) \approx 70$ the values at $T_*$ and with $\delta \approx 113$ measured in simulations~\cite{Buschmann:2021sdq}.  Assuming number density conservation for $T < T_*$ one may then redshift the result in~\eqref{eq:n_aT} to, {\it e.g.}, matter-radiation equality or today to compute the DM abundance. To achieve the correct DM abundance, one finds the string-induced DM abundance $\Omega_a^{\rm str}$ to be~\cite{Buschmann:2021sdq}
\es{eq:Omega_a_str}{
\Omega_a^{\rm str} \approx 0.12 h^{-2} \, \left( {f_a \over 4.1 \cdot 10^{10} \, {\rm GeV}} \right)^{1.17} {113 \over \delta } \sqrt{{\xi_* \over 17}} {\log_* \over 70} \,.
}
The contribution to $\Omega_a^{\rm str}$ from axions produced between $T_*$ and $T_{\rm coll}$ was estimated in~\cite{Buschmann:2021sdq} to raise $\Omega_a^{\rm str}$ by, at most, a factor of $3/2$.

The result in~\eqref{eq:Omega_a_str} was computed for global axion strings, suggesting $m_a \sim (40,180)$ $\mu$eV in that case to obtain the observed DM abundance~\cite{Buschmann:2021sdq}. On the other hand, we claim that this result also applies to string theory axion string networks where the axion arises from a strongly warped cycle, since in this case -- as shown in Secs.~\ref{sec:axion_strings} and~\ref{sec:string_theory} -- the tension of the axion string is the same as that found in field theory axion models.  The tension of the axion strings dictates the evolution of the network, with one possible exception. In the field theory axion models the intercommutation of axion strings is purely deterministic ({\it i.e.}, $P = 1$). On the other hand, string theory axion strings may have $P < 1$ due to either the quantum nature of the string cores (F-strings or wrapped D-branes) or due to the string missing each other in the additional extra dimensions.  We suspect that $P < 1$ raises the number of strings per Hubble $\xi$ relative to the $P = 1$ scenario, though the $P < 1$ case has never been simulated for warped axion strings where the energy loss is dominated by axion emission instead of GW emission. Increasing $\xi$, as seen in~\eqref{eq:Omega_a_str}, would raise the DM abundance at fixed $f_a$, thus requiring a larger $m_a$ to obtain the correct DM abundance. However, without dedicated simulations of the $P < 1$ scenario we are unable to quantify this effect, though it would be an interesting direction for future work.

\subsection{Axion strings from unwarped compactifications}
\label{sec:nonwarped_axion_strings}

We now consider the scenario where the axion arises from an unwarped cycle so that the axion string tension is that given in~\eqref{eq:mu}. 
First, we repeat the calculation for the collapse time of the string network for the case of string theory axion strings.
Towards that end, it is useful to calculate the expected string length in Hubble units $\langle \ell / H^{-1} \rangle$ for string loops, which we may do using~\eqref{eq:n_x} to find: 
\es{}{
H\langle \ell \rangle = r G \mu = {r \kappa \over 8 \pi} {f_a \over M_{\rm pl}} \,.
}
Then, equating the string tension with the domain wall tension we see that the network collapses at a temperature
\es{}{
T_{\rm coll} \approx {\Lambda_{1}^2 \over \sqrt{m_a M_{\rm pl}}} \,,
\label{eq:Tcoll_ST}
}
where we use the zero-temperature relation for $m_a f_a$ in~\eqref{eq:ma_zero_T}. 
This relation is justified since for $m_a \sim 10$ $\mu$eV the collapse temperature is $T_{\rm coll} \sim {\mathcal O}(0.1 \, \, {\rm MeV})$, with the collapse temperature only being lower for higher axion masses.  On the other hand, the network does not efficiently produce axions at temperatures below that where the typical axion momentum, $k \sim H M_{\rm pl} / f_a$, drops below $m_a$. We define this temperature as $T_{\rm NR}$, and it is given approximately by $T_{\rm NR} \sim \Lambda_{1}$.  We may then estimate the contribution to the DM abundance from the string network prior to the epoch defined by $T_{\rm NR}$ by
\es{}{
\Omega_a^{\rm str} &\approx {m_a n_a(T_{\rm NR}) \over T_{\rm MRE}^4} \left( {T_{\rm MRE} \over T_{\rm NR} } \right)^3 \\
&\sim {\Lambda_1^2 \over \sqrt{m_a M_{\rm pl}} T_{\rm MRE} } \\
&\sim 10^4 \sqrt{ {10 \, \, \mu{\rm eV} \over m_a }} \,,
\label{eq:abundance_relativistic}
} 
with $T_{\rm MRE} \sim 1\,\, {\rm eV}$ the temperature of matter-radiation equality.  Thus, we conclude that string theory axion strings overproduce the DM abundance for any allowable QCD axion mass $m_a$, given that neutron star and stellar cooling constraints $m_a \lesssim 10 - 30$~meV, depending on the UV completion~\cite{Buschmann:2021juv}. (The QCD axion mass would have to be around 100 keV to produce the correct DM abundance.) In Appendix \ref{app:precise_abundance} we calculate the abundance more carefully by including ${\mathcal O}(1)$ numerical factors and accounting for axions emitted up to the time of collapse with energies $\omega \ge m_a$, and we find that the above conclusion is unchanged.  Note also that while the calculation above is strictly only valid for $P = 1$, going to $P < 1$ only increases the number of strings per Hubble and thus exacerbates the problem of overproducing the DM abundance.

Even a period of early matter domination (EMD) is not sufficient to dilute the axion abundance.
If an EMD occurs before the network collapses and before the axions become non-relativistic, the network will adjust to a scaling solution during the EMD era.
Therefore, axions are not diluted when the EMD ends, in contrast to other forms of radiation not sourced by the network.
On the other hand, if an EMD occurs after the network collapses then the situation is different.
First, we consider the case where EMD starts immediately below $T_{\rm NR}$, to estimate the maximal amount of dilution.
Since EMD has to end by $T_{\rm RH}\sim 4$~MeV to ensure successful big bang nucleosynthesis (BBN), the axion abundance can be diluted maximally by a factor of $R_{\rm NR}/R_{\rm RH} \sim (T_{\rm RH}/T_{\rm NR})^{4/3} \sim 10^{-2}$, which is not sufficient, given~\eqref{eq:abundance_relativistic}.
Here $R_{\rm NR}$ and $R_{\rm RH}$ are scale factors at $T_{\rm NR}\sim \Lambda_1$ and $T_{\rm RH} \sim 4$~MeV, the end of the assumed EMD period, respectively.
In Appendix~\ref{app:EMD} we give a more detailed estimate of entropy dilution from an EMD period, reaching the same conclusion.

To further dilute the QCD axion abundance we need the network to collapse earlier.  Along these lines, in the following section, we show that the QCD axion may constitute the DM if it is accompanied by a heavier axion, such that a linear combination of the two axions is sourced by the string network.
The heavier axion causes an early collapse of the network, raising $T_{\rm coll}$ above $T_{\rm NR}$, as we elaborate below.

\section{Axiverse string cosmology}
\label{sec:multiple_axions}

We now examine how the QCD axion abundance is altered if the strings source a linear combination of axion mass eigenstates, as is likely to happen in axiverse constructions (see, {\it e.g.},~\cite{Bachlechner:2017zpb,Ho:2018qur,Hu:2020cga,Chadha-Day:2021uyt,Foster:2022ajl,Gavela:2023tzu,Murai:2023xjn,Cyncynates:2023esj} for field theory examples of multi-axion scenarios). We consider a 4D EFT of the axiverse with $n_a$ axions and $n_I$ instantons given by
\es{}{
& \mathcal{L} \supset -\frac{K_{ij}}{2}\partial_\mu a_i\partial^\mu a_j-\Lambda_n^4\left[1-\cos \left( Q_{ni}\frac{a_i}{f_{a_i}} +\delta_n \right)\right] \,,
\label{eq:axiverse_EFT}
}
where summation of repeated indices is understood, with $i,j\in \{1,...,n_a\}$ and $n\in \{1,...,n_I\}$. $K$ is a non-degenerate symmetric kinetic matrix, and $f_{a_i}$ is the decay constant of axion $a_i$; {\it i.e.}, $a_i=a_i+2\pi f_{a_i}$. We also define $\vartheta_i = a_i/f_{a_i}$ with periodicity $2\pi$. $Q$ is the matrix of instanton charges associated to instanton scales $\Lambda_n$. In this field basis of $2\pi$-periodic scalars, the charges $Q_{ni}$ are integers. We take $\Lambda_{i} < \Lambda_{i+1}$ for each $i$, and below we focus on the case where QCD generates the weakest instanton.\footnote{If this is not the case, the QCD axion DM abundance may be further diluted.} 
Note that $\Lambda_1$ is defined for QCD in~\eqref{eq:ma_zero_T}.
We assume that there are no temperature-dependent instanton potentials other than that from QCD.\footnote{Note, however, that for instanton potentials coming from D-branes wrapping cycles in the compactified dimensions, a  temperature-dependent potential could arise if the dilaton VEV or volume modulus changes with temperature.}
For the QCD instanton potential, we assume the temperature dependence~\eqref{eq:ma_growth}
until $T \approx 152$~MeV, when the temperature-dependent growth saturates the zero-temperature result; for $T \lesssim 152$~MeV we assume a constant value for $m_a f_a$ equal to $\Lambda_1^2$.
This is a rough approximation since the parametrization in~\eqref{eq:ma_growth} is technically only valid for $T\gg 400 \, \mathrm{MeV}$, but this approximation is sufficient for the accuracy of the calculations in this section.

We denote the basis of mass eigenstates by $\varphi_i$, and the associated masses $M_{\varphi_i}$. We suppose that additional bare masses are negligible, which is expected for axiverse axions. We further assume that $n_a \geq n_I$ such that, barring certain alignments of the instanton charges, through a field redefinition we may set $\delta_i =0$ for all $i$, in which case we maintain the solution to the strong-{\it CP} problem. 
Lastly, we suppose that all strings in the network source the axion $\vartheta_1$ with winding number $1$, whereas the other $\vartheta_i$ for $i>1$ have winding number $0$, while $\vartheta_1$ is not a mass eigenstate. 
Namely, we assume that the strings produced after inflation source only the axion $\vartheta_1$. This would be expected if the string cores are resolved by the same kind of higher-dimensional D-branes wrapped on the same cycle of the internal manifold, as may happen after D$(3+p)$-$\overline{{\rm D}(3+p)}$-brane inflation with $p>0$.

\subsection{Cosmology with two axions: generalities}

We now focus on the case $n_I= n_a =2$, choosing the instanton charges so that $N_\mathrm{DW}=1$ when circling around the string under consideration.\footnote{In App.~\ref{sec:Ndw} we derive a general relation for the instanton charges in order to give $N_{\rm DW}=1$ for an arbitrary number of axions.} We show how the abundance of QCD axions emitted by the network can avoid the over-closure issue encountered in the single axion story.  The intuition is that the heavy axion state can cause the string network to collapse earlier, thus reducing the abundance of the lighter QCD axion state.
This can be seen via an estimate similar to~\eqref{eq:abundance_relativistic} which shows the abundance scales as $1/m_a^{1/2}$.
 
The axions associated with each instanton are 
\es{eq:thetadefs}{
 \vartheta_{\text {QCD}} &\equiv Q_{11} \vartheta_1+Q_{12} \vartheta_2 \ , \\
\vartheta_{\Lambda_2} &\equiv Q_{21} \vartheta_1+Q_{22} \vartheta_2 \ .
}
Let $(a_1', a_2')$ denote the axion basis after canonically normalizing the axions. Writing the Cholesky decomposition of the kinetic mixing matrix as $K = V^\top V$ and denoting the mass matrix in the basis $(a_1,a_2)$ by $M$, it follows that the mass matrix in the basis $(a_1',a_2')$ is $(V^{-1})^\top M V^{-1}$. For concreteness, let us take $K = \left(\begin{array}{cc}
1 & \epsilon \\
\epsilon & 1
\end{array}\right)$. 
To avoid fields with negative norms and tachyonic instabilities we must impose $|\epsilon| < 1$. Then $V^{-1}=\left(\begin{array}{cc}
1 & -\frac{\epsilon}{\sqrt{1-\epsilon^2}} \\
0 & \frac{1}{\sqrt{1-\epsilon^2}}
\end{array}\right)$, and by Taylor expanding \eqref{eq:axiverse_EFT}, we obtain the mass matrix of the canonically normalized axions
\es{eq:mass_matrix}{
\mathcal{L} \supset -\frac{1}{2}\left(\begin{array}{l}
a_1' \\
a_2'
\end{array}\right)^{\top}
\left(\begin{array}{cc}
M_{a_1'}^2 & M_{a_1' a_2'}^2\\
M_{a_1' a_2'}^2 & M_{a_2'}^2
\end{array}\right)
\left(\begin{array}{l}
a_1' \\
a_2'
\end{array}\right) \,,
}
with
\es{}{
 M_{a_1'}^2&=M_{a_1}^2\\
 M_{a_2'}^2&=\frac{M_{a_1}^2\epsilon^2 -2\epsilon  M_{a_1 a_2}^2+M_{a_2}^2}{1-\epsilon^2}\\
 M_{a_1' a_2'}^2&=\frac{-\epsilon M_{a_1}^2 +M_{a_1 a_2}^2}{\sqrt{1-\epsilon^2}} \,,
}
where
\es{}{
 M_{a_1}^2&=\frac{Q_{11}^2\Lambda_1^4+Q_{21}^2\Lambda_2^4}{f_{a_1}^2} \\
 M_{a_2}^2&=\frac{Q_{12}^2\Lambda_1^4+Q_{22}^2\Lambda_2^4}{f_{a_2}^2} \\
 M_{a_1 a_2}^2&=\frac{Q_{11} Q_{12} \Lambda_1^4+Q_{21} Q_{22} \Lambda_2^4}{ f_{a_1} f_{a_2}} \,.
}
Diagonalizing \eqref{eq:mass_matrix}, we see that the mass eigenbasis $(\varphi_1,\varphi_2)$ is related to the basis $(a_1',a_2')$ by a rotation
\es{}{
\left(\begin{array}{l}
\varphi_2 \\
\varphi_1
\end{array}\right)=\mathcal{R}(\theta)\left(\begin{array}{l}
a_1' \\
a_2'
\end{array}\right)  \,,\qquad \mathcal{R}(\theta)\equiv \left(\begin{array}{cc}
\cos \theta & \sin \theta \\
-\sin \theta & \cos \theta
\end{array}\right)\,,
\label{eq:mass_eigenbasis}
}
where
\es{}{
\cos 2 \theta&=\frac{M_{a_1'}^2-M_{a_2'}^2}{\Delta^2} \\ \sin 2 \theta&=\frac{2 M_{a_1' a_2'}^2}{\Delta^2} 
\\\left(\Delta^2\right)^2& =\left(M_{a_1'}^2-M_{a_2'}^2\right)^2+4\left(M_{a_1' a_2'}^2\right)^2 \,,
}
such that the masses of the eigenstates are 
\es{}{
M_{\varphi_1}^2& =\frac{1}{2}\left(M_{a_1'}^2+M_{a_2'}^2-\Delta^2 \right) \\
M_{\varphi_2}^2& =\frac{1}{2}\left( M_{a_1'}^2+M_{a_2'}^2+\Delta^2\right) \ .
} 

In order to compute the abundances of $\varphi_1$ and $\varphi_2$ axions emitted by the network, we need to first understand how the axion energy emitted by a string loop is split between them.  
We focus on $a_1$ for this purpose since we are considering a scenario where only $a_1$ is sourced by the network.
The matrix sending the basis $(\varphi_2,\varphi_1)$ back to the original basis $(a_1,a_2)$ is $P\equiv V^{-1}\mathcal{R}(-\theta)$. 
It follows that the 2-form dual to $a_1$, which couples to the string worldsheet $X^{\mu}(\tau,\sigma)$, can be written 
$C=P_{12}C_1+P_{11}C_2$
with $\D_{4d}\varphi_i =  \star\D_{4d}C_i$. Let us assume that the mass of each eigenstate is negligible relative to $f_{a_1}$, which holds until the $\Lambda_2$ instanton becomes relevant and quickly stops the emission of the heavy axion. Then the action for the string and the fields $(\varphi_1, \varphi_2)$ is  
\es{}{
S=&-\mu \int \D \tau \D \sigma \sqrt{-\gamma}-\frac{1}{6} \int \D^4 x F_i^{\mu \nu \rho} F_{i,\mu \nu \rho} \\&-\frac{2\pi f_{a_1}}{\sqrt{2}} \int \D \tau \D \sigma \epsilon^{a b} \partial_a X^\mu \partial_b X^\nu C_{\mu \nu}(X) \,,
} 
with $\gamma_{a b}=\partial_a X^\mu \partial_b X_\mu$, $\mu$ the string tension, and $F_i$ the field strength of $C_i$. After fixing the Lorentz gauge $\partial^\mu C_{i,\mu\nu}=0$ and the string worldsheet diffeomorphism invariance, the equation of motion for the dual axion fields is \cite{Gorghetto:2021fsn}
\es{}{
\partial_\alpha \partial^\alpha C_i^{\mu \nu}=g_i \int \D \sigma\left(\dot{X}^\mu X^{\prime \nu}-\dot{X}^\nu X^{\prime \mu}\right) \delta^3(\vec{x}-\vec{X}) \,,
}
with 
$g_i=\sqrt{2}\pi f_{a_1}P_{1i}$. 
Therefore $C_i \propto g_i$, so $\D E_{\varphi_i}/\D t$, which can be obtained from the stress-energy tensor of $\varphi_i$ sourced by the string, scales like $g_i^2$. The emission rates from a given string loop are thus $\D E_{\varphi_{i}}/\D t = r_a g_i^2/\sqrt{g_1^2+g_2^2}$. Explicitly,
\es{eq:emission_km}{
P_{12}&=-\sin(\theta)-\frac{\epsilon}{\sqrt{1-\epsilon^2}}\cos(\theta) \,,\\
P_{11}&=\cos(\theta)-\frac{\epsilon}{\sqrt{1-\epsilon^2}}\sin(\theta) \,.
}
We see that the presence of kinetic mixing $\epsilon$ can alter the emission of $\varphi_1$ and $\varphi_2$.
In the fine-tuned cases where $\epsilon=-\sin(\theta)$ ($\epsilon=\cos(\theta))$, the network does not emit the lighter (heavier) mass eigenstate at all.
However, barring such extreme cases, for typical values of $\epsilon$ and $\theta$, we can capture the generic properties of the cosmological evolution by working with $\epsilon = 0$, which we will do in the following.
The presence of kinetic mixing, along the lines of~\cite{Gendler:2023kjt}, can be taken into account by redoing the following analysis starting with~\eqref{eq:emission_km}. 

We also assume that $\Lambda_2 \gg \Lambda_1 = 75 \, \mathrm{MeV}$,
and that $r_f \equiv f_{a_1}/f_{a_2}$ is $\mathcal{O}\left((\Lambda_2/\Lambda_1)^2\right)$.
Then we find
\es{eq:emission_rate_charge_dep}{
\frac{\D E_{\varphi_{1}}}{\D t}&=r_a\sin^2\theta \approx  r_a \frac{Q_{22}^2r_f^2}{Q_{21}^2+Q_{22}^2r_f^2} \,,\\
\frac{\D E_{\varphi_{2}}}{\D t}&=r_a\cos^2\theta \approx  r_a \frac{Q_{21}^2}{Q_{21}^2+Q_{22}^2r_f^2} 
\,,
}
such that, unless the instanton charges are tuned or there is a hierarchy amongst the $f_{a_i}$, the rates of energy emitted into each mass eigenstate are comparable. Further, the mass-decay constant relations are
\es{eq:2axionsQCDmass}{
f_{a_1}M_{\varphi_1} &\approx \Lambda_1^{2} R_Q \,, \\
f_{a_1}M_{\varphi_2} &\approx \Lambda_2^2 \sqrt{Q_{21}^2 + r_f^2Q_{22}^2}
\,,
}
where we define
\es{eq:RQdef}{R_Q \equiv \frac{|Q_{11}Q_{22}-Q_{21}Q_{12}|r_f}{\sqrt{Q_{21}^2+Q_{22}^2r_f^2}} 
\,.
}
Lastly, let us comment on the periodicity of the (canonically normalized) QCD axion. At energies below $\Lambda_2$, setting $\vartheta_{\Lambda_2}=0$ in \eqref{eq:thetadefs} gives $\vartheta_1=(-Q_{22}/Q_{21})\vartheta_2$, such that the periodicity of $\vartheta_1$ is $\Delta \vartheta_1 = -2\pi Q_{22}/\mathrm{gcd}(Q_{22},Q_{21})$, with $\mathrm{gcd}$ the greatest common divisor. Hence, the periodicity of $\vartheta_\mathrm{QCD}$ is 
\es{eq:QCDaxion_periodicity}{
\Delta \vartheta_\mathrm{QCD} = -\left(Q_{11}-\frac{Q_{21} Q_{12}}{Q_{22}}\right)  \frac{2 \pi Q_{22}}{\operatorname{gcd}\left(Q_{21}, Q_{22}\right)} \,.
}
Therefore, defining the domain wall (DW) number as
\es{eq:NDWQCD}{
N^\mathrm{QCD}_\mathrm{DW}=\frac{|Q_{11}Q_{22}-Q_{21}Q_{12}|}{\mathrm{gcd}(Q_{21},Q_{22})} \,,
}
it follows that $\vartheta_\mathrm{QCD}' \equiv \vartheta_\mathrm{QCD}/N^\mathrm{QCD}_\mathrm{DW}$ is $2\pi$ periodic. Writing the Lagrangian \eqref{eq:axiverse_EFT} in terms of  $\vartheta_\mathrm{QCD}'$ we obtain 
\es{eq:2axion_Lagrangian}{
\mathcal{L} \supset \frac{1}{2}f_{a_\mathrm{QCD}}^2 (\partial \vartheta_\mathrm{QCD}')^2 - \Lambda_1^4 \left[1-\cos\left(N^\mathrm{QCD}_\mathrm{DW}\vartheta_\mathrm{QCD}'\right)\right] \,, 
}
where we define the decay constant of $\vartheta_\mathrm{QCD}'$ as
\es{eq:fqcd}{
f_{a_\mathrm{Q C D}} \equiv  f_{a_1} \frac{\sqrt{Q_{22}^2 + r_f^{-2}Q_{21}^2}}{|\operatorname{gcd}\left(Q_{21}, Q_{22}\right)|} \,.
}
As we now discuss, the DW number of the QCD axion is generally different from the DW number which determines whether the string network collapses due to DW formation.  

When $T$ drops below $\Lambda_2$, the first relation of \eqref{eq:emission_rate_charge_dep} remains valid, but we can approximate $\D E_{\varphi_2}/\D t \approx 0$.
Below $T\approx \Lambda_2$, domain walls develop and potentially make the string network collapse. Whether that happens immediately or at much lower temperatures depends on $Q_{21}$, which is the DW number associated to the part of the potential generated by the strongest instanton.
In particular, for $T\approx \Lambda_2$, only the $\Lambda_2^4(1-\cos(Q_{2i}\vartheta_i))$ part of the potential in~\eqref{eq:axiverse_EFT} is active and $Q_{21}$ determines the DW structure; if $Q_{21}\neq 1$ the network will be stable until $T\approx \Lambda_\text{QCD}$.\footnote{Note that if the lightest axion was an ultralight axion-like particle instead of the QCD axion, the network would persist until $T \sim \Lambda_1$,  at which point $N_\mathrm{DW}$ would be given by \eqref{eq:NDW}.} However, we focus on the case where the network decays early, so as to reduce the QCD axion abundance, and take $Q_{21}=1$ in what follows.\footnote{To enlarge the parameter space in terms of instanton charges, notice that the same result could be achieved with $Q_{21}\neq 1$ if there is more than one instanton stronger than QCD. The only criterion is that the potential has $N_\text{DW}=1$ at $T\approx \Lambda_2$.}

Let us denote the DW tension by $\sigma$.  In the limit $\Lambda_2 \gg \Lambda_1$ of primary interest we can identify $\sigma = 8f_{a_1}\Lambda_2^2$ from the single-axion calculation~\cite{Hiramatsu:2012sc}.
The domain wall tension causes the network to collapse, as we discuss further below, shutting off the emission of the lighter QCD axions.

\subsection{Cosmology with two axions: no warping}
\label{sec:cosmo_NW}

When there is no warping the string tension $\mu$ is parametrically $\sim f_a\mpl$ as given in~\eqref{eq:mu}. Therefore, assuming the entire network evolution occurs during a radiation-dominated era, the temperature at which the domain wall tension drives the network to collapse is
\es{eq:T_coll_multi_axion_v2}{
T_{\text {coll }} \approx \frac{1}{\left(32 \pi^3\right)^\frac{1}{2}} \left(\frac{90}{g_{*,\mathrm{coll}}}\right)^{\frac{1}{4}} \sqrt{\frac{r\sigma}{\mpl}} \,,
}
with $g_*(T)$ the number of relativistic degrees of freedom at temperature $T$, which we evaluate numerically using formulae from \cite{Saikawa:2018rcs}, and where $r$ is defined in~\eqref{eq:r_G_sub}.

Let the $\varphi_1$ states become non-relativistic at temperature $T_\mathrm{NR, \varphi_1}$. We focus on the case   $T_\mathrm{coll} > T_\mathrm{NR, \varphi_1}$, such that the network collapses before the lighter mass-eigenstate $\varphi_1$ becomes non-relativistic.\footnote{If this is not the case, it is straightforward to show that the QCD axion DM abundance would still be overproduced.}
Thus, we require 
\es{eq:Lambda2_lower_bound}{
\Lambda_2 > \frac{\sqrt{\kappa R_Q}}{4}\left(\frac{g_{*,\mathrm{coll}}}{g_{*,\mathrm{QCD}}}\right)^{\frac{1}{4}}\frac{1}{\log\left(\frac{f_{a_1}}{H_\mathrm{coll}}\right)}\Lambda_\mathrm{QCD} \sqrt{\frac{\mpl}{f_{a_1}}} \,.
}
At the time of collapse, the $\varphi_1$ have characteristic energy $\omega_\mathrm{coll} \sim H(T_\mathrm{coll}) \mpl / f_{a_1}$,
which then redshifts as 
\es{eq:TNRphi1relation}{\omega(T < T_\mathrm{coll}) = \frac{R_\mathrm{coll}}{R}\omega_\mathrm{coll}\,.
} 
From entropy conservation ($T \propto g_{*,s}(T)^{-\frac{1}{3}}/R)$, it follows that
the temperature at which the emitted axions become non-relativistic is
\es{}{
T_\mathrm{NR, \varphi_1}  &\sim \frac{\kappa\sqrt{\pi r}}{128\pi^2}\frac{1}{\log(\mpl/\Lambda_2)^2}
\left(\frac{g_{*,s,\mathrm{coll}}} {g_{*,s,\mathrm{NR},\varphi_1}}\right)^{\frac{1}{3}}\\
&\times \left(\frac{90}{g_{*,\mathrm{coll}}}\right)^{\frac{1}{4}}
\frac{\sqrt{f_{a_1}\mpl}}{\Lambda_2}M_{\varphi_1} \ .
\label{eq:TNRphi1_v2}
}
Let us assume that $T_{\mathrm{NR,\varphi_1}}<100 \, \mathrm{MeV}$, such that we may set $M_\mathrm{\varphi_1}$ to its zero-temperature value. 
 We also assume, for simplicity, that the strings radiate axions with a conformal spectrum $\D E / \D k \propto 1/k$.
 Then, the present-day abundance is
\begin{widetext}
\es{eq:Omega_varphi1_today}{
\Omega_{\varphi_1, \mathrm{str}} &\sim \frac{\rho_{\varphi_1}(T_\mathrm{coll})}{\rho_{c,0}} \left(\frac{R_\mathrm{coll}}{R_\mathrm{NR, \varphi_1}}  \right )^4 \left(\frac{R_\mathrm{NR, \varphi_1}}{R_\mathrm{0}} \right )^3  \\ 
&\sim \frac{2\sqrt{2}\pi^{2}}{135} \frac{g_{*,s,0}g_{*,\mathrm{coll}}}{g_{*,s,\mathrm{coll}}}\left(\frac{90}{g_{*,\mathrm{coll}}}\right)^\frac{1}{4}\frac{r_a\alpha}{r\sqrt{\kappa}}\frac{1}{\log(\mpl/\Lambda_2)}
 \frac{\Lambda_1^2}{\Lambda_2 } \frac{T_0^3}{3H_0^2 \mpl^2} \frac{r_f^2Q_{22}^2}{(1+r_f^2Q_{22}^2)}R_Q \\
&\sim \Omega_c \frac{1.8\cdot 10^{3} \,\mathrm{GeV}}{\Lambda_2} \frac{g_{*,\mathrm{coll}}}{g_{*,s,\mathrm{coll}}}\left(\frac{90}{g_{*,\mathrm{coll}}}\right)^\frac{1}{4}\left(\frac{r_a}{10}\right)\left(\frac{\alpha}{0.18}\right)   \left(\frac{r} {100}\right)^{-1} \left(\frac{\kappa}{\pi\sqrt{2}}\right)^{-\frac{1}{2}}\frac{30}{\log(\mpl/\Lambda_2)}
 \frac{r_f^2Q_{22}^2}{(1+r_f^2Q_{22}^2)}R_Q
\,,
}
\end{widetext}
where we use \eqref{eq:rho_a_ST},  \eqref{eq:2axionsQCDmass}, and \eqref{eq:emission_rate_charge_dep}. Here $\rho_{c,0}=3H_0^2 \mpl^2$ denotes the critical density at the present day. Above, we use $H_0=1.4 \times 10 ^{-42} \, \mathrm{GeV}$,  $T_0 = 2.4 \times 10^{-13} \, \mathrm{GeV}$, and $g_{*,s,0}=3.36$.
For $\Lambda_2$ sufficiently large such that the DM abundance is not exceeded and $f_{a_1} > 10^{9} \, \mathrm{GeV}$, our assumption that  $T_{\mathrm{NR,\varphi_1}}<100 \, \mathrm{MeV}$ is indeed valid.

Note that~\eqref{eq:Omega_varphi1_today} suggests we may achieve the correct DM abundance (or a sub-leading DM abundance) from strings for $\Lambda_2 \gtrsim 10^3$ GeV.  In the limit $f_{a_1} < f_{a_2}$ the DM abundance is further suppressed by the factor $r_f^2 = f_{a_1}^2 / f_{a_2}^2$, allowing for smaller $\Lambda_2$ without overproducing the DM abundance.  With that said, there are a few caveats to the result in~\eqref{eq:Omega_varphi1_today}, which we expand upon below: (i) the QCD axion $\varphi_1$ has a misalignment contribution to its DM abundance, which should be accounted for; and (ii) the relic abundance of $\varphi_2$ may give rise to an EMD period, which then leads to a more complicated relation than~\eqref{eq:Omega_varphi1_today}.  In the numerical results, we illustrate below we allow for the initial misalignment angle of $\varphi_1$, which we denote by $\varphi_{i,i}$, to be in the range $\varphi_{i,1} \in [0.025, 0.975] \pi$, and we compute the misalignment contribution to the DM abundance using \texttt{MiMeS}~\cite{Karamitros:2021nxi} (see Sec.~\ref{sec:axion_cosmo} for details). 

We now discuss the period of EMD that can be brought upon by the relic population of cold $\varphi_2$'s.
The $\varphi_2$ become non-relativistic before the network collapses, and, if stable, their relic density is set by $T_{\mathrm{NR},\varphi_2} \sim \sqrt{M_{\varphi_2} f_{a_1}}\sim \Lambda_2$.
Assuming entropy conservation, the $\varphi_2$ density at temperature $T$ is given by
\es{}{
\rho_{\varphi_2}(T) &\sim \rho_{\varphi_2}(T_{\mathrm{NR}, \varphi_2}) \left(\frac{R_{\mathrm{NR}, \varphi_2}}{R} \right)^3 \\
&\sim \frac{16\sqrt{2}\pi^2}{135}\Lambda_2 T^3\left(\frac{f_{a_1}}{\mpl}\right)^{\frac{1}{2}} \frac{r_a\alpha}{r\kappa} \\
&\times  \frac{g_{*,\mathrm{NR},\varphi_2} g_{*,s}(T)}{g_{*,s,\mathrm{NR},\varphi_2}} \left(\frac{90}{g_{*,\mathrm{NR},\varphi_2}}\right)^\frac{1}{4}(1+r_f^2Q_{22}^2)^{-\frac{3}{4}} 
\,.
\label{eq:rhovarphi2}
} 
Therefore, using~\eqref{eq:rhovarphi2} we can obtain the temperature $T_{\rm dom}$ where $\varphi_2$ axions would dominate the total energy density,
\es{eq:T_dom_v2}{
T_\mathrm{dom} &\sim \frac{32\sqrt{2}}{9}\frac{r_a\alpha}{r\kappa}\frac{g_{*,s,\mathrm{dom}}g_{*,\mathrm{\Lambda_2}}}{g_{*,\mathrm{dom}}g_{*,s,\mathrm{\Lambda_2}}}\left(\frac{90}{g_{*,\mathrm{\Lambda_2}}}\right)^\frac{1}{4} \\
&\times \Lambda_2 \sqrt{\frac{f_{a_1}}{\mpl}} (1+r_f^2Q_{22}^2)^{-\frac{3}{4}} \,.
}
At $T=T_\mathrm{dom}$ the Universe enters a period of EMD, which is allowed as long as $\varphi_2$  decays sufficiently quickly to respect observational constraints from BBN. In this case, the QCD axion DM abundance is diluted by the entropy injected by the decay of $\varphi_2$ axions.  The details of this scenario are worked in App.~\ref{sec:EMD}, where we self-consistently account for the decays of $\varphi_2$ to photons and to gluons, when kinematically accessible.  

\begin{figure}[!tb]
\begin{center}
\includegraphics[width=0.9\linewidth]{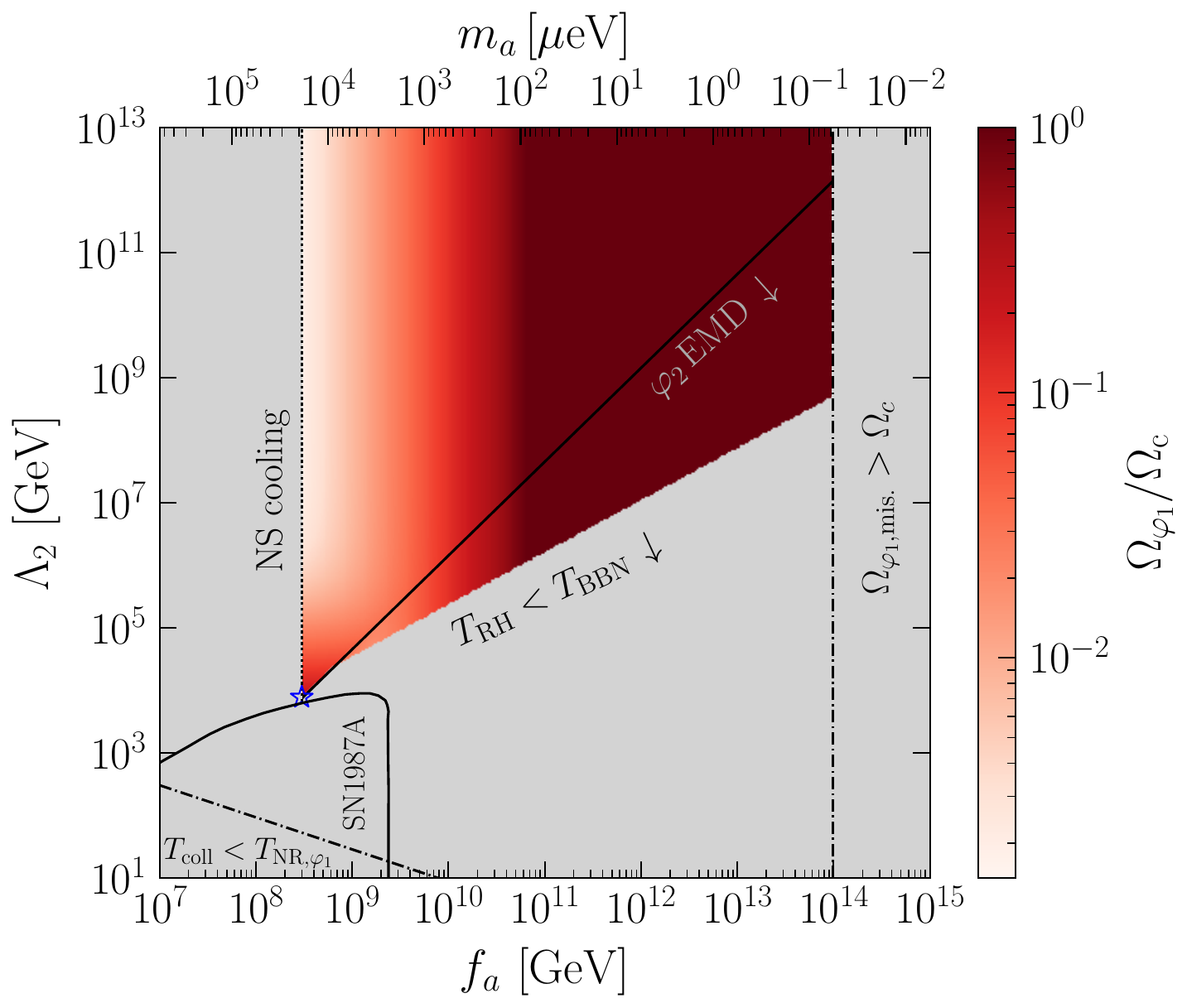} 
\caption{The space of $(f_{a_1}\equiv f_a, \Lambda_2)$ 
for the two-axion scenario for our fiducial choice of dimensionless parameters. Excluded parameter space is shaded gray. Within the viable parameter space, the fraction of the DM abundance sourced by the string network and misalignment is indicated according to the color bar. We allow for an initial misalignment angle $\theta_i$ in $[0.025, 0.975]\pi$.  Thus, axions may constitute the DM for $f_{a} \gtrsim 5 \times 10^{10} \mathrm{GeV}$, where the abundance is set almost entirely by misalignment of the lighter eigenstate.  For smaller $f_{a}$ the QCD axion may only constitute a sub-dominant fraction of the DM $(\lesssim 25 \%)$, as in the lower left corner of the viable region, where the string network provides the dominant contribution. (Going beyond our canonical choice of dimensionless parameters, however, for example, by allowing $f_{a_1} \neq f_{a_2}$, the correct DM abundance may be achieved.)
In the region below the solid black line the $\varphi_2$ eventually dominates the energy density, initiating an EMD era, and must subsequently decay before BBN. The blue star indicates the choice of parameters for which we show the GW signal in Fig.~\ref{fig:domega_df}.}
\label{fig:2axion_unwarped_param_space_rf_1.pdf}
\end{center}
\end{figure}

For illustrative purposes, let us consider the scenario where $f_{a_1} = f_{a_2} \equiv f_a$.  We also make the choice, relevant for determining the duration of the EMD epoch, that the heavy axion couples to electromagnetism with coupling (see App.~\ref{sec:EMD}) $g_{a_2\gamma\gamma} = \alpha_\mathrm{EM}C_{a_2\gamma}/(2\pi f_a)$, where $\alpha_\mathrm{EM}$ is the fine structure constant, and we take $C_{ {a_2} \gamma} = 1$. 
In Fig.~\ref{fig:2axion_unwarped_param_space_rf_1.pdf} we illustrate the parameter space for this model as spanned by $f_a$ and $\Lambda_2$, making the additional assumptions that $\kappa = \pi \sqrt{2}$, $r_G=87$, $r_a=10$, $r=100$,  $\alpha = 0.18$, and $(Q_{11},Q_{12},Q_{21},Q_{22})=(2,1,1,2)$, in addition to varying the initial misalignment angle of the QCD axion within the range previously mentioned. Note that within the allowed parameter space
neutron star cooling provides a lower bound on \mbox{$f_{a_1}$ at the level $f_{a} \gtrsim 3 \times 10^{8} \, \mathrm{GeV}$}, up to minor model dependence depending on the UV completion~\cite{Buschmann:2021juv}.
There is an additional constraint on the gluon coupling of massive axion-like particles from observations of Supernova 1987A, which is relevant for the $\varphi_2$ parameter space. We use the bound of~\cite{Ertas:2020xcc}, which found $4\pi c_{\mathrm{g}} / \Lambda \gtrsim 4\times 10^{-8} \, \mathrm{TeV}^{-1}$ for masses up to $\sim 200 \, \mathrm{MeV}$ (see App.~\ref{sec:EMD} for the conventions of these couplings).  We also exclude the regions of parameter space where the reheat temperature from EMD would be below the temperature of BBN and where the misalignment contribution to the DM abundance would overproduce the observed abundance, within the amount we allow ourselves to tune the initial misalignment angle. 

Combining all the constraints mentioned above, we deduce that the QCD axion may only compose a subdominant ($\lesssim 25 \%$) fraction of the DM (the starred point in Fig.~\ref{fig:2axion_unwarped_param_space_rf_1.pdf}), unless the present-day axion abundance is set almost entirely by misalignment, in which case it may be all of the DM but the strings do not play an important role. Allowing for a small hierarchy between the decay constants $f_{a_1}/f_{a_2} \gtrsim 10$, it is possible to have the QCD axion constitute the DM with the dominant contribution being from the string network for $f_{a_1}\sim 10^9 \, \mathrm{GeV}$.  Interestingly, as we discuss further in Sec.~\ref{sec:GW}, even if the string-generated axions are only a subfraction of the DM abundance, as is the case by the starred point in parameter space in Fig.~\ref{fig:2axion_unwarped_param_space_rf_1.pdf}, the string network may still give rise to a GW signal within reach of future observations.

Lastly, let us comment on how the phenomenology discussed in this section generalizes to $n_a > 2$ axions. We assume that $Q_{n_a1}=1$, such the collapse of the network is set by the heaviest instanton and that there is no kinetic mixing or large hierarchies amongst the $f_{a_i}$. Then, there are at least two important differences with respect to the case $n_a=2$. First, unless the instanton charges are tuned, we naively expect that the energy emitted from a shrinking loop of string is split roughly equally into each mass eigenstate.
Thus, the abundance of QCD axions emitted by the string network is suppressed by a factor $\propto$$1/n_a$, making it harder for the QCD axions generated from the string to account for the full DM abundance in the limit $n_a \gg 1$. Secondly, the abundance may be further diluted due to a period of EMD in several steps if, after the heaviest mass eigenstate decays, some of the lighter mass-eigenstates also come to dominate the energy density. 

In App.~\ref{sec:cosmo_W} we discuss the scenario of a warped axion string that produces two or more mass-eigenstate axions. In this case, the heavier axion may also reduce the relic abundance, though the effect is less dramatic since the QCD axion does not overproduce the relic abundance in the single axion case.

\subsection{Gravitational wave signal from unwarped axion strings}
\label{sec:GW}

\begin{figure}[!tb]
\begin{center}
\includegraphics[width=0.9\linewidth]{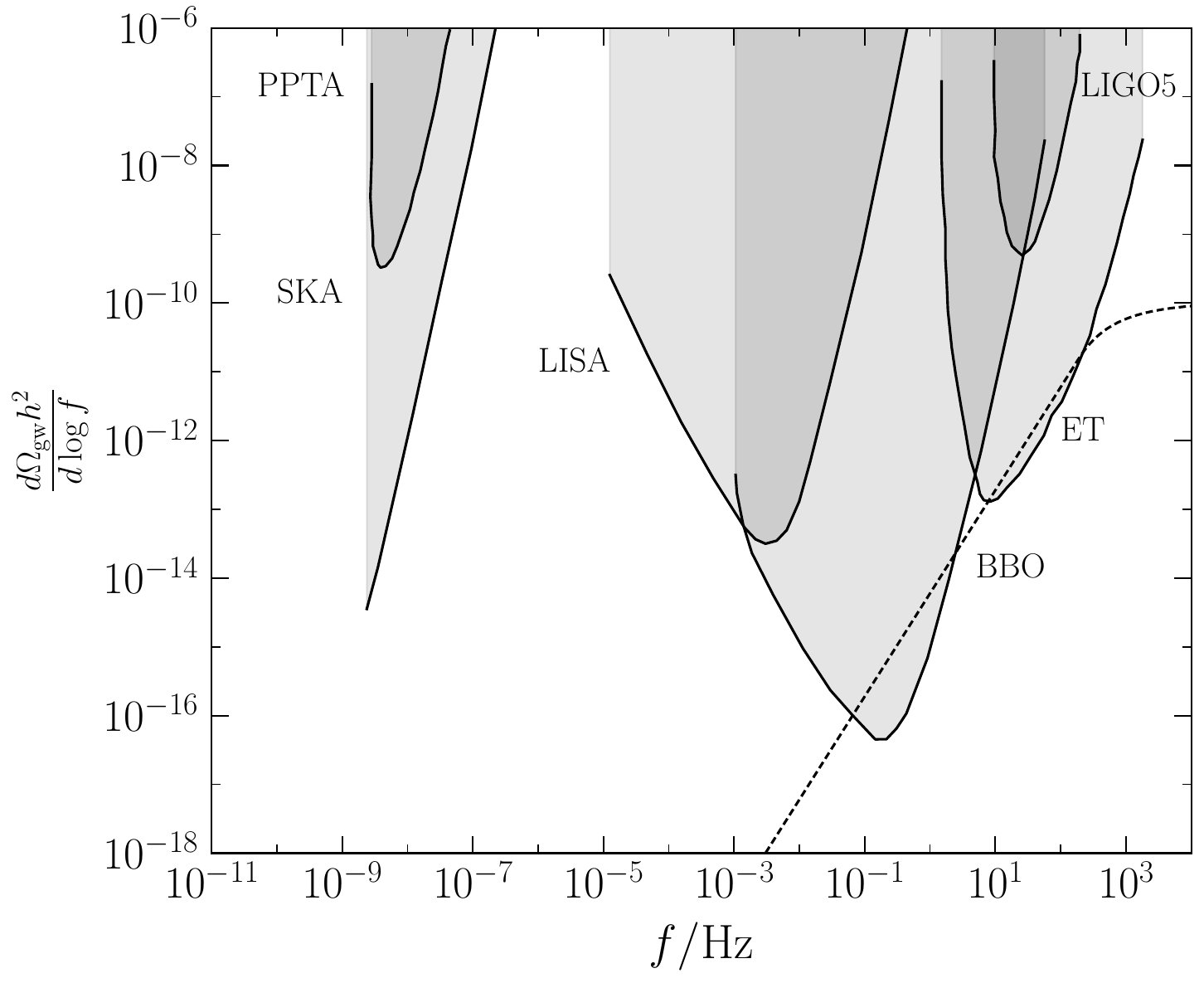} 
\caption{The present-day GW spectrum for the two-axion scenario, compared to existing and projected constraints from GW observatories (shaded). Here $f_{a} = 3 \times 10^8 \, \mathrm{GeV}$, $\Lambda_2 = 7.8\times 10^{3}\, \mathrm{GeV}$, corresponding to the QCD axion constituting $25\%$ of the DM. This point in parameter space corresponds to the blue star in Fig.~\ref{fig:2axion_unwarped_param_space_rf_1.pdf}, and the choice of dimensionless parameters is the same as in Fig.~\ref{fig:2axion_unwarped_param_space_rf_1.pdf}.}
\label{fig:domega_df}
\end{center}
\end{figure}

We now compute the GW signal from the network in the two-axion, unwarped scenario. We further restrict to the part of parameter space where there is no period of EMD. 
From \eqref{eq:differential_GW_spectrum}, the GW spectrum at the time of network collapse is 
\es{eq:rhoGW_coll}{
\frac{\D \rho_g}{\D \log \omega} = \left({32 r_G \sqrt{2 \pi \kappa} \alpha \over 15 r^{3/2} }\right)H_\mathrm{coll}^2 \sqrt{f_{a_1}\mpl^3} \left(x f\left(x\right)\right),
} 
where $f(x)$ is shorthand for the bracketed term in~\eqref{eq:differential_GW_spectrum} and $x=\omega/\omega_{0,{\rm coll}}$.  The GWs redshift freely after the network collapses, so that the present-day GW spectrum is, using~\eqref{eq:T_coll_multi_axion_v2},
\begin{widetext}
    \es{eq:Omega_GW_multi_axion}{
\frac{h^2 \D\Omega_\mathrm{GW,0}}{\D\log(\omega)} \equiv  \frac{h^2}{\rho_{c,0}}\frac{\D \rho_g}{\D \log \omega}\left[t_0, \omega \right] 
&=  \frac{h^2}{\rho_{c,0}}\left(\frac{R_\mathrm{coll}}{R_0}\right)^4\frac{\D \rho_g}{\D \log \omega}\left[t_{\mathrm{coll}}, {\omega  R_0 \over R_{\mathrm{coll}}}\right] \\
&=  \frac{16(2\pi^5)^\frac{1}{2}}{2025} \frac{g_{*,\mathrm{coll}}g_{*,s,0}^\frac{4}{3}}
{g_{*,s,\mathrm{coll}}^\frac{4}{3}}h^2 \frac{T_0^4}{H_0^2\mpl^2}\left({r_G \sqrt{\kappa} \alpha \over  r^{3/2} }\right) \times \sqrt{\frac{f_{a_1}}{\mpl}}\left[x f\left(x\right) \right] |_{x = \frac{R_0\omega}{R_\mathrm{coll}\omega_{0,\mathrm{coll}}}} \\
&\approx  1.4 \cdot 10^{-4} g_{*,\mathrm{coll}}^{-\frac{1}{3}} \left({r_G \sqrt{\kappa} \alpha \over  r^{3/2} }\right)\sqrt{\frac{f_{a_1}}{\mpl}} \times \left[x f\left(x\right) \right] |_{x = \frac{R_0\omega}{R_\mathrm{coll}\omega_{0,\mathrm{coll}}}} \,.
}
\end{widetext}
Note in~\eqref{eq:Omega_GW_multi_axion} $\omega$ denotes the present day  frequency whereas in~\eqref{eq:rhoGW_coll} $\omega$ denotes the frequency at the collapse time, and the two are related by an appropriate redshift factor.
Lastly, we need to evaluate
\es{eq:x_val}{
\frac{R_0\omega}
{R_\mathrm{coll}\omega_{0,\mathrm{coll}}}
= \frac{45^{1\over 4}\sqrt{r}\kappa}{16\cdot 2^{3\over 4}\pi^\frac{3}{2}} \frac{\omega}{T_0}  \frac{g_{*,s,\mathrm{coll}}^\frac{1}{3}}{g_{*,\mathrm{coll}}^\frac{1}{4}g_{*,s,0}^\frac{1}{3}}\frac{\sqrt{f_{a_1}\mpl}}{\Lambda_2} \,.
}
In Fig.~\ref{fig:domega_df} we show the result of~\eqref{eq:Omega_GW_multi_axion} for the case where the QCD axion constitutes $25\%$ of the DM, for our fiducial choice of parameters (the starred point in parameter space in Fig.~\ref{fig:2axion_unwarped_param_space_rf_1.pdf}).  We also assume the GW emission is cusp-dominated ($q=5/3$). In this case the GW signal is within reach of proposed future probes such as the Einstein Telescope (ET) \cite{Punturo:2010zz} and the Big Bang Observer (BBO) \cite{Yagi:2011wg}, but not detectable by LISA \cite{LISA:2017pwj}, Square Kilometer Array (SKA) \cite{Janssen:2014dka}, Parkes Pulsing Timing Array (PPTA) \cite{Shannon:2015ect}, or LIGO's O5 observing run (LIGO5) \cite{LIGOScientific:2014qfs}. 

\section{Discussion}
\label{sec:discussion}

The so-called ``post-inflationary scenario'' has been treated as a leading and predictive picture of axion cosmology for decades, where PQ symmetry breaking below the reheating scale after inflation gives rise to a network of axion strings. At the same time, axions emerging in the context of string theory compactifications have been pursued in part due to their ability to more naturally address the PQ quality problem. However, as we show these two pictures are at odds.

QCD axion strings from string theory axions do not generically form. The strings must be formed at high temperatures, in the context of the UV theory, and while field theory axion strings form generically, we show that string theory axion strings require special inflationary conditions such as D-brane annihilation at the end of D-brane inflation.  Moreover, even in the case of D-brane inflation in the canonical model of KKLMMT~\cite{Kachru:2003sx}, the resulting cosmic superstring network does not source axions, since the axion is projected out of the EFT by an orientifold. 

Even if QCD axion strings from string theory axions did form (for example, at the end of D-brane inflation), we show that they overproduce the DM abundance, unless the axion arises from a strongly warped cycle. One possible way out of the above conundrum, which we discuss, is if string theory produces an axiverse, and a heavier axion state collapses the axion string network at an earlier time before the QCD phase transition. We show that the relic abundance of QCD axions produced prior to the network collapse can make up a sizeable fraction of the DM in this case. Moreover, strong GW signals are possible, due to the enhanced tension of the axion strings, and this could be observable in future GW observatories.

Our work relies heavily on analytic estimates for the axion production rates from cosmic superstrings. On the other hand, dedicated simulations of cosmic superstrings that self-consistently account for axion production have never been performed and would greatly increase the accuracy and robustness of the conclusions of this work.  Such simulations would need to self-consistently evolve both the axion field sourced by the D- or F-strings in addition to the strings themselves, accounting for the possibility of a non-trivial interconnection probability for the string cores.  Moreover, string theory may give rise to more complicated networks~\cite{Copeland:2003bj} containing both F- and D-strings, as well as other branes wrapped on the internal manifold. Those various strings could end on each other, forming, {\it e.g.}, junctions, which were not considered in this work. Some of them could develop instabilities, {\it e.g.}, due to background branes. Here, we have considered the simplest cosmic superstring scenario that is most analogous to the field theory axion string network, but we have only scratched the surface of possibilities. Exploring the more general parameter space for axion superstring networks would be an interesting direction for future work.

This work is of direct experimental relevance because it de-motivates, though does not exclude, the QCD axion mass prediction for the post-inflationary axion scenario since this scenario does not generically arise in string theory QCD axion constructions. In field theory axion constructions this cosmological scenario predicts $m_a \in (40,180)$ $\mu$eV to achieve the observed DM abundance~\cite{Buschmann:2021sdq}.  String theory QCD axions most generically achieve the relic abundance through the misalignment mechanism and not from string production, in which case the QCD axion mass is predicted to lie in the range $(1.9,53)$ $\mu$eV, without allowing for more than 1$\sigma$ tuning of the initial misalignment angle.  On the other hand, given the unknowns related to the UV completion of the axion theory, inflation, and possible anthropic selection effects, a broad experimental search program for QCD axions is necessary over the full possible mass range~\cite{Adams:2022pbo}.

\subsubsection*{\textbf{Acknowledgements}}

{\it
We thank Zackaria Chacko, Thibaut Coudarchet, Nathaniel Craig, Emilian Dudas, Joshua Foster, James Halverson, Anson Hook, Andrew Long, Lisa Randall, Matt Reece, and Raman Sundrum for helpful conversations.
J.B., Q.B., and B.R.S. are supported in part by the DOE Early Career Grant DESC0019225. M.B. was in part supported by the DOE under Award Number DESC0007968 and acknowledges funding from the European Research Council (ERC) under the European Union’s Horizon 2020 research and innovation programme (Grant agreement No. 864035). S.K. is supported partially by the NSF grant PHY-2210498 and the Simons Foundation. Q.B. and S.K. thank Aspen Center for Physics (supported by NSF grant PHY-2210452) for hospitality while this work was in progress.\\
}

\appendix

\section{More precise QCD axion DM abundance calculation}
\label{app:precise_abundance}

Here we compute the abundance of axions emitted by strings up until the time of network collapse, taking into account that the emitted axions are nonrelativistic at late times. 
As the precise form of the axion emission spectrum at energies $\omega < m_a$ is not known, we approximate that emission is cut off completely in this regime, i.e,
\es{}{
{\partial \rho_{\rm GW} \over \partial\omega} \approx \int_0^t dt' \left( {R(t') \over R(t)} \right)^3 {d \Gamma_{\rm GW}(t',\omega') \over d \omega'} \Theta(\omega' - m_a(t))\,,
}
with $\Theta$ the Heaveside step function. In the following let us assume that $q>1$ to be conservative,
and for simplicity, we ignore the time dependence of $m_a$. Then \eqref{eq:drhoGW_domega} is unchanged for $\omega > m_a$, and for $\omega < m_a$, we have the modification
\begin{widetext}
\es{eq:differential_GW_spectrum_v2}{
{\partial \rho_{\rm GW}(\omega,t) \over \partial\omega} = \Gamma_{\rm GW} {x_{\rm UV} \over 5 \cdot 8 \pi \cdot H^2} 
\left\{
\begin{array}{cc}
{4 (q-1) \over 2 q +1} \sqrt{ {\omega \over \omega_0} } \left(\frac{\omega}{m_a}\right)^{\frac{3}{2}}\,, & \omega < \sqrt{m_a\omega_0} \\
{1 \over 2-q} \left[ 3(q - 1) \left( {\omega_0 \over \omega} \right)^2\left(\frac{m_a}{\omega}\right) - {15 \over 2q + 1}\left( {\omega_0 \over \omega} \right)^q \left(\frac{m_a}{\omega}\right)^{q-1} + 5 (2 - q) \left( {\omega_0 \over \omega} \right) \right] \,, & \omega > \sqrt{m_a\omega_0} \,,
\end{array}
\right.
} 
\end{widetext}
which reduces to \eqref{eq:drhoGW_domega}  for $m_a = \omega$. Then we obtain, after the appropriate replacements for the axion energy emission rate,
\begin{widetext}
\es{}{
n_a(t) &= \Gamma_{a} {x_{\rm UV} \over 5 \cdot 8 \pi \cdot H^2} \mathcal{F}\left(\frac{\omega_0}{m_a}\right)\\
\mathcal{F}\left(\frac{\omega_0}{m_a}\right) &\equiv \frac{10(q-1)}{2q-1} \left(\frac{\omega_0}{m_a}\right)^{\frac{1}{2}} - \frac{30(q-1)}{(2-q)(2q+1)(2q-1)q} \left(\frac{\omega_0}{m_a}\right)^{q} + \frac{q-1}{2(2-q)} \left(\frac{\omega_0}{m_a}\right)^{2}  \,.
}
\end{widetext}
Hence, writing $n_a = C f_a\sqrt{f_a \mpl} \mathcal{F} H$, with the prefactor $C$ given as in \eqref{eq:na_eval_qg1}, we obtain
\es{}{
\Omega_a^{\rm str} &\approx {m_a n_a(T_\mathrm{coll}) \over T_{\rm MRE}^4} \left( {T_{\rm MRE} \over T_\mathrm{coll}} \right)^3 \\
&=C\left(\frac{r}{2\pi}\right)^{-\frac{1}{2}}\mathcal{F}\left(\frac{\omega_0(t_\mathrm{coll})}{m_a}\right) \frac{\Lambda_1}{T_{\rm MRE}} \,.
}
By the time $t_\mathrm{coll}$ of network collapse, we have 
\es{}{
\frac{m_a}{\omega_0(t_\mathrm{coll})} = \frac{\kappa}{32\pi} \frac{m_a \mpl}{\Lambda_1^2} =  \frac{\kappa}{32\pi} \cdot 10^4 \frac{m_a}{\mu \mathrm{eV}}
\label{eq:ma_over_omega0} \,.
}
Let us assume that $m_a \gtrsim 1 \, \mathrm{neV}$ such that the argument of $\mathcal{F}$ is much less than unity. Keeping the leading order term in $\mathcal{F}$, we have finally 
\es{}{
\Omega_a^{\rm str} &\approx 10^{6} 
 \cdot \frac{64}{r}\left(\frac{\kappa}{\sqrt{2\pi}} \right)^{-1} \frac{(q-1)}{q}\frac{10(q-1)}{2q-1}  r_a\alpha \sqrt{\frac{1 \mu \mathrm{eV}}{m_a}} \,.
}
Hence, the DM abundance is overproduced for any allowed $m_a$.

\section{QCD axion DM abundance with early matter domination}
\label{app:EMD}
Here we show that even with a period of early matter domination (EMD), the QCD axion DM abundance from string axion string networks is still overproduced. We denote the reheat temperature by $T_\mathrm{RH}$ and assume instantaneous reheating. BBN constrains $T_\mathrm{RH}> 4 \,\mathrm{MeV}$~\cite{Kawasaki:2000en}\cite{Hannestad:2004px}. 
If the string network evolves for long enough during EMD then it attains a scaling regime with $\xi$ constant, and the formulae concerning the emission of axions and GWs are unchanged up to order one dimensionless prefactors. 
For simplicity, we assume that the axion emission becomes inefficient during the EMD era. The ratio of Hubble rates between the time when the axions become non-relativistic and the time of reheating is  $H_\mathrm{NR}/H_\mathrm{RH} \sim (\Lambda_1 / T_\mathrm{RH})^2$. Hence this scenario is consistent only if $T_\mathrm{RH} < \Lambda_1$. Then, assuming the axion is emitted with spectral index $q=1$, we have
\es{EMD_single_axion_precise}{
\Omega_a^{\rm str} &\sim {\rho_a(T_{\rm NR}) \over \rho_{c,0}} \left(  R_{\rm NR} \over {R_{\rm RH}} \right)^3 \left(  R_{\rm RH} \over {R_{\rm 0}} \right)^3 \\ 
&\sim 3.4 \Omega_c \left(\frac{T_\mathrm{RH}}{4 \, \mathrm{MeV}} \right)\left(\frac{f_a}{10^8 \, \mathrm{GeV}}\right)^{\frac{1}{2}} \left(\frac{r_a}{10}\right)\left(\frac{\alpha}{0.18}\right)   
\\  
&\times \left(\frac{r} {100}\right)^{-\frac{3}{2}} \left(\frac{\kappa}{\pi\sqrt{2}}\right)^{-\frac{3}{2}} \log\left(  \frac{4.3 \cdot 10^{28} f_a}{10^8 \, \mathrm{GeV}}\right)
\,,
}
where we use that $(  R_{\rm NR} / {R_{\rm RH}} )^3 =(H_{\rm RH}/ H_{\rm NR})^2$ and  $T_\mathrm{NR} \sim \Lambda_1$. Hence even if $T_\mathrm{RH} = 5 \, \mathrm{MeV}$, the DM is overproduced unless $f_a \lesssim 10^{8}\, \mathrm{GeV}$, which is excluded by neutron star cooling ~\cite{Buschmann:2021juv}.

\section{ALP string networks}
\label{sec:alp_network}
Here we consider the case of non-warped compactifications where the axion sourced by the string network is an ALP  which is not the QCD axion. In this scenario~\eqref{eq:T_coll_multi_axion_v2} implies that the collapse temperature is $T_\mathrm{coll} \sim f_a \sqrt{m_a/\mpl}$, 
such that for  $m_a  \lesssim 10^{-3} \,\mathrm{ eV} \left(10^{12} \,\mathrm{GeV}/f_a\right)^2 $
the string network may survive until the time of BBN, in which case $f_a$ is constrained by the contribution to $N_\mathrm{eff}$ from relativistic axions emitted by the network. 
Using the energy density of string theory axions~\eqref{eq:rho_a_ST} and the bound from~\cite{Gorghetto:2021fsn} (see also~\cite{Benabou:2023ghl}), we find the constraint $f_a \lesssim 7.4 \times 10^{13} \, \mathrm{GeV}$, for $r_a = 10$, $\alpha=0.18$, $c_1=0.25$, $\kappa = \pi\sqrt{2}$, $r=100$, and $\log(f_a/H) = 70$.

\section{String profiles in the warped 5D geometry}
\label{app:warped_profiles}

Here we rederive the radion equation of motion in the 5D warped geometry without recourse to approximations and verify that it reduces to the flat geometry result in the limit of zero warp factor. With the same notation as in main text,~\eqref{eq:S_U1_warped} and~\eqref{eq:rad_action_warp} give the combined action 
\es{}{
	S_G + S_{\mathrm{U}(1)} =&~ \frac{2M_5^3}{k}\int \D^4x\sqrt{-g}\left(1-(\varphi/F)^2\right)R^{(4)} \\
	&-\frac{1}{2}\int \D^4x\sqrt{-g}\partial_\mu\varphi \partial^\mu \varphi \\
	&- \frac{1}{g_5^2}\int \D^4x\sqrt{-g} \frac{2k\pi^2}{(F/\varphi)^2-1}(\partial_\mu A_5) (\partial^\mu A_5)\, .
}
Here $F = \sqrt{24M_5^3/k}$.
We can do a Weyl transformation to reduce the Ricci term $R^{(4)}$ in its canonical form.
To that end, we define $g_{\mu\nu} = h(\varphi) \bar{g}_{\mu\nu}$, where $ h(\varphi) = 1 / (1-(\varphi/F)^2)$.
Then we can derive
\es{}{
 (1-(\varphi/F)^2) \sqrt{-g} R^{(4)} = \sqrt{-\bar{g}}\left(\bar{R}-\frac{3}{2}\frac{h'^2}{h^2}\bar{g}^{\mu\nu}\partial_\mu\varphi \partial_\nu\varphi\right).
}
The part of the action containing $A_5$ and $\varphi$ becomes after simplification,
\begin{equation}
\begin{aligned}
	S_G + S_{\mathrm{U}(1)} \approx &~-\frac{1}{2}\int \D^4x\sqrt{-\bar{g}} \frac{(\partial_\mu\varphi)^2}{(1-(\varphi/F)^2)^2}   \\
	&~-\int \D^4x\sqrt{-\bar{g}} V(\varphi) 
 \\&~-\frac{1}{g_5^2}\int \D^4x\sqrt{-\bar{g}} \frac{2k\pi^2 (\partial_\mu A_5)^2}{((F/\varphi)^2-1)(1-(\varphi/F)^2)}.
\end{aligned}
\label{eq:warped_action_precise}
\end{equation}
As a check, we can take the limit of $k\rightarrow 0$ using $\varphi/F = \exp(-k\pi \rho)$ (ignoring the potential momentarily),
\begin{equation}
\begin{aligned}
	S_G + S_{\mathrm{U}(1)} = &~-\frac{1}{2}\int\D^4x \sqrt{-\bar{g}} \frac{6M_5^3}{k\rho^2}(\partial_\mu\rho)^2 \\
	&~-\frac{1}{2kg_5^2}\int\D^4x \sqrt{-\bar{g}} \frac{1}{\rho^2} (\partial_\mu A_5)^2 \,.
\end{aligned}
\end{equation}
The $1/k$ appears in the gauge boson kinetic term because of the Weyl rescaling, $ h(\varphi) = 1 / (1-(\varphi/F)^2)$, which becomes $1/(2k\pi\rho)$ in the limit of small $k$.
Therefore, to reduce to the case of flat extra dimension, we should identify, $\langle\rho\rangle \equiv 1/(2k\pi)$, and use $1/g_4^2 = (2\pi \langle \rho \rangle)/g_5^2$.
We also use the relation between the 5D and 4D Planck scales, $\mpl^2 \approx 8M_5^3\pi \langle \rho \rangle \equiv 4M_5^3/k$.
Then the above reduces to 
\es{eq:warped_matching_flat}{
	S_G + S_{\mathrm{U}(1)} =  -\frac{3\mpl^2}{4}\int\D^4x \sqrt{-\bar{g}} \frac{(\partial_\mu\rho)^2}{\rho^2} \\
	-\int\D^4x \sqrt{-\bar{g}} \frac{(\partial_\mu A_5)^2}{2g_4^2\rho^2}.
}
\vspace{1em}
This matches with the 4D action for a flat extra dimension~\eqref{eq:Srhoa}.
Let us denote $\psi = \varphi/F$. Specializing to the string ansatz $\psi = \psi(r)$, $A_5 = \theta/(2\pi)$, the equation of motion for $\psi$, from~\eqref{eq:warped_action_precise}, is 
\es{eq:warped_eom_precise}{
    \psi''+\frac{1}{r}\psi' + \frac{1}{1-\psi^2}\left[(\psi')^2(4-2\psi) - \frac{k}{F^2g_5^2}\frac{1}{r^2}\psi(1+\psi^2)\right] \\-\frac{1}{F^2}(1-\psi^2)^2 \frac{\partial V}{\partial \psi}=0 \,,
}
which reduces to~\eqref{eq:radion_eom_warp} in the limit $\psi \ll 1$.
To solve this numerically, we define as in the flat geometry $\tilde{\rho} = \rho/b$, $\tilde{r} = r/b$, and solve \eqref{eq:warped_eom_precise} in the form
\begin{widetext}
\es{eq:eom_numerics_warped}{
    \tilde{\rho}''+\frac{1}{\tilde{r}}\tilde{\rho}'  - bk(\tilde{\rho}')^2 \left[\pi +\frac{\pi\psi(4-2\psi)}{1-\psi^2}\right]  +\left(\frac{k}{F}\right)^2 \frac{1}{kg_5^2}\frac{1}{\pi bk}\frac{1+\psi^2}{1-\psi^2}\frac{1}{\tilde{r}^2} 
    - \left(\frac{k}{F}\right)^2\frac{bk}{\pi ^2k^5}\frac{(1-\psi^2)^2}{\psi^2}\frac{\partial V}{\partial \rho} = 0
     \,,
}
\end{widetext}
where here the primed quantities are with respect to $\tilde{r}$. Analogously to the flat geometry, the small $r$ asymptotic form is $\rho = -\frac{1}{\sqrt{k}F\pi g_5} \log(cr)$, and at large $r$  we have for $\rho \approx b + \beta /r^2$ for some $\beta$. Hence, the boundary conditions may be chosen as $\tilde{\rho}' = - \frac{k}{F}\frac{1}{\sqrt{k}g_5 (\pi bk)}  \frac{1}{\tilde{r}}$ at $\tilde{r}=\tilde{r}_\mathrm{min} $ and $\tilde{\rho} = 1 -\tilde{\rho}' \tilde{r}/2$ at $\tilde{r} = \tilde{r}_\mathrm{max}$. 
Note that the Goldberger-Wise potential \eqref{eq:GW_warped} gives 
\es{eq:GW_warped_app}{
\frac{\partial V}{\partial \rho}=-\frac{\pi k^4F^4M_\Psi^3}{72 M_5^6} \psi^3 \left(\tilde{v}_v-\tilde{v}_h \psi^\epsilon \right)\left(2\tilde{v}_v-\tilde{v}_h (2+\epsilon)\psi^\epsilon \right) \,,
}
where we define the dimensionless quantities $\tilde{v}_v = v_v/M_\Psi^{\frac{3}{2}}$, $\tilde{v}_h = v_h/M_\Psi^{\frac{3}{2}}$.
Our fiducial choice of parameters is $(\tilde{v}_h,\tilde{v}_v, M_5 /k, g_5^2 k) = (2,1,2, 1)$.  
For a given value of $f_a/\mpl$, we determine $bk$ from ~\eqref{eq:fa_warped}. In particular, for small $f_a/\mpl$, our choice of parameters gives $f_a/\mpl = \frac{1}{g_5\sqrt{k}}\frac{k}{\mpl} \frac{\langle\varphi\rangle}{F} \sim \frac{1}{63} e^{-\pi b k}$.

\section{Domain wall number in axiverse scenario}
\label{sec:Ndw}

In this Appendix, we compute the domain wall number $N_{\rm dw}$ in the axiverse construction discussed in Sec.~\ref{sec:multiple_axions}.  Here, we allow for an arbitrary number of axions. The condition which allows the string/DW network to eventually collapse; {\it i.e.}, the condition which implies that the domain wall number $N_\mathrm{DW}$ of the zero-temperature potential is equal to one. 
To begin, note that $N_\mathrm{DW}$ is the number of local minima of the instanton potential as $\vartheta_1$ varies from $0$ to $2\pi$. 
Denoting the vector $\boldsymbol{\vartheta} = (\vartheta_1,...,\vartheta_{n_a})^\top$ , a local minimum occurs if and only if  
\es{}{
\mathbf{0} = \frac{\partial V}{\partial \boldsymbol{\vartheta}} = Q^\top \left[\begin{array}{c}
\Lambda_1^4\sin\left((Q\vartheta)_1 \right) \\
\vdots \\
\Lambda_{n_I}^4\sin((Q\vartheta)_{n_I})
\end{array}\right] \,,
\label{eq:dV_dtheta}
}
and the Hessian $\left(\frac{\partial V}{\partial \vartheta_i \partial \vartheta_j}\right) _{ij} =Q^\top \text{diag}\left[\Lambda_i^4\cos((Q\vartheta)_i)\right] Q $ is positive-definite. 

Let us take the case $n_I = n_a$ for simplicity, and suppose that $Q$ is invertible, which is expected in axiverse constructions \cite{Heidenreich:2020pkc}. Then ~\eqref{eq:dV_dtheta} is equivalent to $\cos((Q\vartheta)_i) = \pm 1$ for all $i$. The positive-definiteness of the Hessian then implies that $\cos((Q\vartheta)_i) =1$ for all $i$ (assuming all the $\Lambda_i$ are nonzero). 
That is, there must exist some $\mathbf{N} \in \mathbb{Z}^{n_a}$ for which $\boldsymbol{\vartheta}/(2\pi) = Q^{-1}\mathbf{N}$. We may express $Q^{-1} = \frac{1}{\operatorname{det}(Q)}{C^\top}$ where $C$ is the matrix of cofactors, which has integer entries. In particular, 
\es{}{
\frac{\vartheta_1}{2\pi} = \frac{1}{\operatorname{det}(Q)} \left(\left(C^{\top}\right)_{1,1} ,\cdots \left(C^{\top}\right)_{1,n_a}\right)\mathbf{N} \,.
}
By Bézout's identity, for any multiple $k$ of the greatest common divisor (gcd) of the elements of the first row  of $C^\top$, there exists an $\mathbf{N}$ such that the dot product above is equal to $k$. Thus
\es{eq:NDW}{
N_{\text {DW }}=\left|\frac{\operatorname{det}(Q)}{\operatorname{gcd}\left(\left(C^{\top}\right)_{1,1} \ldots\left(C^{\top}{ }\right)_{1, n_a}\right)}\right| \,.
}
In particular, we have $N_\mathrm{DW} \leq|\operatorname{det} Q|$, such that $N_\mathrm{DW}=1$ is ensured if $|\operatorname{det} Q| =1$. Note that for $n_a=2$, \eqref{eq:NDW} reduces to $N_\mathrm{DW}=|(Q_{11}Q_{22}-Q_{12}Q_{21})/\mathrm{gcd}(Q_{22},Q_{12})|$.

\section{Early matter domination from heavy axions in unwarped scenario}
\label{sec:EMD}

The $\varphi_2$ states could decay to SM fields, in particular to SM gauge fields, or to other axions.\footnote{There could be additional decay channels if there are 
dark sector D-branes, which are generically expected, and which could host dark $\mathrm{U}(1)$’s or confining gauge groups.} In the case of strong hierarchies between the instanton scales, Ref.~\cite{Gendler:2023kjt} showed that decays into axions are suppressed relative to the photon channel, 
which has decay rate
\es{eq:axion_to_photon_decay}{ 
\Gamma_{\varphi_2 \to \gamma\gamma} = \frac{M_{\varphi_2}^3 g^2_{a_2\gamma\gamma}}{64\pi} \,,
}
where the axion-photon coupling is given by $g_{a_2\gamma\gamma} = \alpha_\mathrm{EM}C_{a_2\gamma}/(2\pi f_{a_2})$, where $\alpha_\mathrm{EM}$ is the fine structure constant,
and $C_{a_2\gamma}$ is assumed to be $\mathcal{O}(1)$.
If the SM lives on a brane that overlaps with the cycle on which the $p$-form is wrapped down to an axion, then {\it e.g.}, axion-fermion couplings may be generated, such that decays into electrons may also be relevant. However, as this is highly model-dependent we do not consider this possibility here. On the other hand, decays into QCD states are relevant (and less model-dependent) as there is generically the axion-gluon coupling 
$\mathcal{L}\supset- \frac{\alpha_s}{8\pi}\vartheta_{\mathrm{QCD} } G^{a,\mu \nu} \tilde{G}_{a,\mu \nu}$,
with $\alpha_s$
the dimensionless $\mathrm{SU}(3)$ gauge coupling.
Writing $\vartheta_\mathrm{QCD}$ in the mass-eigenstate basis using  \eqref{eq:thetadefs}, the $\varphi_2$ component has coefficient $c_g /f_{a_2}$, with
\es{}{
c_g &\equiv {r_f}^{-1}Q_{11}\cos(\theta)+Q_{12}\sin(\theta)\\
 &= \frac{Q_{11}Q_{21}{r_f}^{-1}+Q_{12}Q_{22}r_f}{\sqrt{Q_{21}^2+Q_{22}^2r_f^2}} \,.
}
Thus $\varphi_2$ couples to gluons via \es{eq:gluon_coupling_varphi_2}{\mathcal{L} \supset- \frac{4\pi\alpha_s c_g}{\Lambda}\varphi_2 G^{a,\mu \nu} \tilde{G}_{a,\mu \nu}\,,
} 
with $\Lambda= 32\pi^2 f_{a_2}$.
The total decay rate to gluons is calculated at one-loop in perturbative QCD as \cite{Aloni:2018vki}
\es{eq:gluon_decay}{
\Gamma_{\varphi_2 \to gg} \approx \frac{32 \pi \alpha_s^2 c_g^2 M_{\varphi_2}^3}{\Lambda^2}\left[1+\frac{83 \alpha_s}{4 \pi}\right] \Theta(M_{\varphi_2} -\Lambda_{p\mathrm{QCD}}) \,,
}
where $\alpha_s$ is evaluated at the renormalization group scale $M_{\varphi_2}$ and we impose that the rate vanishes outside of the regime of validity of perturbative QCD, which we may take as $\Lambda_{p\mathrm{QCD}}\approx 1.8$~GeV~\cite{Aloni:2018vki}.
For the parameter space of our interest (specifically, above $T_{\rm RH} < T_{\rm BBN}$ line in Fig.~\ref{fig:2axion_unwarped_param_space_rf_1.pdf}), we find $M_{\varphi_2} > \Lambda_{\mathrm{pQCD}}$ and therefore we can parameterize the width as $\Gamma_{\varphi_2,\mathrm{tot}}=\beta^2 M_{\varphi_2}^3/(64\pi f_{a_2}^2)$, where 
\es{eq:betadef}{
\beta^2 \equiv \left(\frac{\alpha_\mathrm{EM}C_{a_2\gamma}}{2\pi}\right)^2 + \frac{\alpha_s^2c_g^2}{32\pi^3}\left(1+\frac{83 \alpha_s}{4 \pi}\right) \,.
}
Approximating that the $\varphi_2$ axions decay instantaneously at the time $t_\mathrm{dec} \equiv \Gamma_{\varphi_2, \mathrm{tot}}^{-1}$, the reheat temperature   $T_\mathrm{RH}$ is determined by continuity of the energy density,  
\es{}{\rho_\mathrm{tot}(T_\mathrm{RH})=\frac{\pi^2}{30} g_*\left(T_{\mathrm{RH}}\right) T_{\mathrm{RH}}^4=\frac{4}{3} M_{\mathrm{pl}}^2 \Gamma_{\varphi_2, \mathrm{tot}}^2 \,,
} 
giving
\es{eq:Tdec}{
T_\mathrm{RH} = 
\beta\frac{\Lambda_2^3}{f_{a_2}} \left(\frac{2}{3}\frac{\mpl }{64\pi f_{a_1}^3} \frac{\sqrt{90}}{\pi\sqrt{g_{*,\mathrm{dec}}}}\right)^{\frac{1}{2}} \left(1 + r_f^2 Q_{22}^2\right)^\frac{3}{4} \,.
}
To proceed with the calculation of the abundance, let us approximate that the EMD era
ends instantaneously at the reheat temperature $T_\mathrm{RH}$. From  \eqref{eq:T_coll_multi_axion_v2} and \eqref{eq:T_dom_v2} we find that necessarily the network collapses before the $\varphi_2$ states may dominate, such that at $T=T_\mathrm{coll}$ the Universe is still radiation-dominated. 
We enter an EMD era if $\varphi_2$ decays after it dominates the energy density which is achieved if $\Gamma_{\varphi_2,{\rm tot}} \lesssim H_{\rm dom}$, equivalently,
\es{}{
\Lambda_2 &\lesssim 
\frac{16\cdot2^\frac{1}{4} \sqrt{\pi}}{3\sqrt{\beta}}\left(\frac{r_a \alpha}{r\kappa}\right)^\frac{1}{2}    \frac{f_{a_1}^\frac{3}{2} }{\sqrt{r_f \mpl}} 
\left(\frac{g_{*,\mathrm{dec}}}{g_{*,\Lambda_2}}\right)^{\frac{1}{8}}\\
&\times 
\left(\frac{g_{*,\mathrm{dom}}g_{*,s,\mathrm{\Lambda_2}}}{g_{*,s,\mathrm{dom}}g_{*,\mathrm{\Lambda_2}}}\right)^{\frac{1}{2}}
(1+r_f^2Q_{22}^2)^{-\frac{3}{4}}  \,.
}
Assuming an EMD era occurs, the final $\varphi_1$ abundance depends on whether the $\varphi_1$ axions become non-relativistic during or after the EMD era. 
First, let us consider the case where the $\varphi_1$ axions become non-relativistic during the EMD era.
From \eqref{eq:TNRphi1relation} and using that $H^2 \propto R^{-3}$ during the EMD era, it follows that the Hubble scale $H_{\mathrm{NR},\varphi_1}$ when $\varphi_1$ becomes non-relativistic obeys the relation, 
\es{eq:EMDrelation}{
M_\mathrm{\varphi_1} \sim \omega_\mathrm{coll} \frac{R_\mathrm{coll}}{R_\mathrm{dom}} \left(\frac{H_\mathrm{dom}}{H_{\mathrm{NR},\varphi_1}}\right)^{-\frac{2}{3}}\,.
}
The present-day $\varphi_1$ can then be computed as, 
\es{}{
\Omega_{\varphi_1,  \mathrm{str}} \sim &~ {\frac{\rho_{\varphi_1}(T_\mathrm{coll})}{\rho_{c,0}}} \left(\frac{R_\mathrm{coll}}{R_\mathrm{dom}}  \right )^4{\left(\frac{R_\mathrm{dom}}{R_\mathrm{NR, \varphi_1}}  \right )^4} \\ 
&\times {\left(\frac{R_\mathrm{NR, \varphi_1}}{R_\mathrm{RH}}  \right )^3}  \!\!{\left(\frac{R_\mathrm{RH}}{R_\mathrm{0}} \right )^3}  \\ 
\sim &~ \frac{\pi\sqrt{2\kappa}}{576\cdot 10^\frac{3}{4}}\frac{\Lambda_2 \Lambda_1^2 \mpl}{f_{a_1}^2f_{a_2}} \frac{T_0^3}{H_0^2\mpl^2}\left(\frac{g_\mathrm{coll} g_\mathrm{dec}}{g_{\Lambda_2}}\right)^\frac{3}{4}  \frac{\beta}{\log\left(\frac{\mpl}{\Lambda_2}\right)}
 \\
 &\times \left(\frac{g_{s,0}g_{s,\Lambda_2}}{g_{s,\mathrm{coll}} g_{s,\mathrm{dec}}}\right)
 Q_{22}^2 r_f^2 (1 + Q_{22}^2 r_f^2)^{1/2}R_Q\,,
\label{eq:Omega_varphi1_today_EMD_case1}
}
where we use $R \propto H^{-2/3}$ for the factors accounting for redshifting during the EMD era, and the relations \eqref{eq:EMDrelation},  $H_\mathrm{RH} \sim T_\mathrm{RH}^2/\mpl$. 
We find the same result if the $\varphi_1$ becomes non-relativistic after EMD.
Note that the $\Lambda_2$ dependence of ~\eqref{eq:Omega_varphi1_today_EMD_case1} is the result of two competing effects. For smaller $\Lambda_2$ the network collapses later~\eqref{eq:T_coll_multi_axion_v2}, which tends to increase the $\varphi_1$ abundance. 
On the other hand, smaller $\Lambda_2$ also lowers the $T_\mathrm{RH}$, implying the EMD era persists longer and leads to more dilution of the $\varphi_1$ abundance.
The second effect turns out to be more important than the first.
Lastly, we also express the BBN constraint  $T_\mathrm{RH}> 4 \,\mathrm{MeV}$ \cite{Kawasaki:2000en}\cite{Hannestad:2004px} using~\eqref{eq:Tdec} which gives,  
\es{eq:BBNbound}{
\Lambda_2 &> \left( 1.7 \cdot 10^{5}  \, \mathrm{GeV}\right) \left(\frac{f_{a_1}}{10^{9}\, \mathrm{GeV}}\right) ^{\frac{2}{3}} 
\\
&\times \left(\frac{f_{a_1}}{\mpl} \frac{64\pi^2\sqrt{g_{*,\mathrm{dec}}}}{\sqrt{90}}\right)^\frac{1}{6}(\beta r_f)^{-1/3} \left(1 + r_f^2 Q_{22}^2\right)^{-\frac{1}{4}}\,.
}

\section{Axiverse string cosmology: warping}
\label{sec:cosmo_W}

We may repeat analysis in Sec.~\ref{sec:cosmo_NW} for an axion string that sources two axion mass eigenstates but for strings from warped compactifications. We assume the tension is $\mu_\mathrm{eff} = \pi f_{a_1}^2 \log(m_s/H)$ with $m_s \sim f_{a_1}$, ignoring subdominant UV contributions. 

We work with the same axion EFT as in the previous subsection.
The Hubble rate at the time of collapse of the string-domain-wall-network is modified from that previously found in~\eqref{eq:T_coll_multi_axion_v2} and we use~\eqref{eq:warp_coll} instead to write,
\es{eq:Hcoll_warped}{
H(T_\mathrm{coll}) = \frac{4 \Lambda_2^2}{\pi \sqrt{\xi_\mathrm{coll}} f_{a_1} \log(m_s/H_\mathrm{coll})} \,.
}
The abundance of $\varphi_1$ is computed as in the unwarped scenario; however, now we have $\rho_{a_1}(T_\mathrm{coll}) = 4 \xi H^2(T_\mathrm{coll}) \mu_\mathrm{eff} $ 
and the typical axion momentum is 
$\omega \sim \delta_1 \sqrt{\xi} H \log(m_s/H) $, with $\delta_1 = 6.2 \pm 0.4$ measured from field theory simulations  \cite{Buschmann:2021sdq}. 
For numerical evaluation, we take $\delta_1=6.2$, and assume the functional form $\xi = 0.24\log(m_r/H)$, consistent with the simulations of \cite{Buschmann:2021sdq}.

Therefore, assuming that $\varphi_1$ becomes non-relativistic after the network collapses, its abundance from string emission is set by
\es{}{T_\mathrm{NR,\varphi_1} &\sim \frac{M_{\varphi_1}T_\mathrm{coll}}{H_\mathrm{coll}
\delta_1\sqrt{\xi_{\mathrm{coll}}}
\log(m_s/H_\mathrm{coll})}\left(\frac{g_{*,s,\mathrm{coll}}}{g_{*,s,\mathrm{NR},\varphi_1}}\right)^\frac{1}{3}
\\
&=\frac{M_{\varphi_1}}{2\Lambda_2}\frac{1}{
\delta_1\sqrt{\xi_{\mathrm{coll}}}
}
\left(\frac{\sqrt{\xi_\mathrm{coll}}}{\log(m_s/H_\mathrm{coll})}f_{a_1}\mpl\right)^{\frac{1}{2}}\\
&\times \left(\frac{g_{*,s,\mathrm{coll}}}{g_{*,s,\mathrm{NR},\varphi_1}}\right)^\frac{1}{3} \left(\frac{90}{g_{*,\mathrm{coll}}}\right)^\frac{1}{4}\,,
}
Self consistency requires $T_{\mathrm{NR},\varphi_1} < T_\mathrm{coll}$, which is valid 
in the case we are considering, $\Lambda_2 \gg \Lambda_1$. Note that this calculation is unlike in the single-axion case where the QCD axion becomes non-relativistic before the network collapses. 

If $T_{\mathrm{NR},\varphi_1}< 100 \, \mathrm{MeV}$, then we may ignore the temperature dependence of $M_{\varphi_1}$ and we obtain
\es{}{
\Omega_{\varphi_1, \mathrm{str}} &\sim 
\frac{\pi^{3}}{135} \left(\frac{f_{a_1}}{\mpl}\right)^\frac{3}{2}\frac{g_{*,s,0}g_{*,\mathrm{coll}}}{g_{*,s,\mathrm{coll}}}\left(\frac{90}{g_{*,\mathrm{coll}}}\right)^\frac{1}{4}\frac{\Lambda_1^2}{\Lambda_2} \frac{ \xi_\mathrm{coll} ^\frac{5}{4} }{ {
\delta_1\xi_{\mathrm{coll}}^\frac{1}{2}
}}\\
&\times \left( \log \left(\frac{f_{a_1} m_s }{\Lambda_2^2}\right)\right)^\frac{1}{2} \frac{T_0^3}{H_0^2 \mpl^2} \frac{r_f^2Q_{22}^2}{(1+r_f^2Q_{22}^2)}R_Q \\
&\sim 
\Omega_c\frac{6 \cdot 10^{-6} \,\mathrm{GeV}}{\Lambda_2}\left(\frac{f_{a_1}}{10^{10}\, \mathrm{GeV}}\right)^\frac{3}{2}\frac{g_{*,\mathrm{coll}}}{g_{*,s,\mathrm{coll}}}\left(\frac{90}{g_{*,\mathrm{coll}}}\right)^\frac{1}{4}  \\ 
&\times \frac{ \xi_\mathrm{coll} ^\frac{5}{4} }{ 
{
\delta_1\xi_{\mathrm{coll}}^\frac{1}{2}
}} \left(\log\left(\frac{f_{a_1}\, m_s}{\Lambda_2 ^2}\right)\right)^{\frac{1}{2}} \frac{r_f^2Q_{22}^2}{(1+r_f^2Q_{22}^2)}R_Q
\,.
\label{eq:Omega_varphi1_today_warped_answer}
}
However, unlike in the unwarped case, $T_{\mathrm{NR},\varphi_1} > 100 \, \mathrm{MeV}$ holds in a portion of the viable parameter space. In this case we find
\es{}{
\Omega_{\varphi_1, \mathrm{str}} &\sim 
1.1 \cdot 10^{-4} \left(\frac{f_{a_1}}{10^{12} \mathrm{GeV}}\right)^{1.9}\left(\frac{\Lambda_2}{1 \mathrm{GeV}}\right)^{-0.197} \\
&\times
\frac{g_{*,s,\mathrm{NR},\varphi_1}^{0.27}g_{*,\mathrm{coll}}^{0.95}}{g_{*,s,\mathrm{coll}}^{1.27}} \xi_\mathrm{coll}^{1.05} \left( \log \left(\frac{f_{a_1} m_s }{\Lambda_2^2}\right)\right)^{0.9}   \\
&\times \frac{r_f^2Q_{22}^2}{(1+r_f^2Q_{22}^2)}R_Q^{0.2}
\frac{1}{\delta_1^{0.2}{\xi^{0.1}_{\mathrm{coll}}}}
\label{eq:Omega_varphi1_today_warped_answer_mass_growth}\,.
}

On the other hand, the $\varphi_2$ become nonrelativistic before the network collapses, and, if stable, their relic density is set by $T_{\mathrm{NR},\varphi_2} \sim \sqrt{M_{\varphi_2}\mpl} \sim \Lambda_2 \sqrt{\mpl/f_{a_1}}$. 
Thus we have
\es{eq:rhophi2_warped}{
\rho_{\varphi_2,\mathrm{str}}(T) &\sim \rho_{\varphi_2}(T_{\mathrm{NR}, \varphi_2}) \left(\frac{R_{\mathrm{NR}, \varphi_2}}{R} \right)^3 \\
&\sim \
\frac{2 \pi^{5 / 2}}{45} \Lambda_2 T^3 \left(\frac{f_{a_1}}{\mpl}\right)^\frac{3}{2} 
\frac{\xi_{\mathrm{NR},\varphi_2}^\frac{3}{4}}{\sqrt{\delta_1}}
\left(\log \left(\frac{m_s f_{a_1}}{\Lambda_2^2}\right)\right)^\frac{1}{2} \\ &\times \frac{g_{*,\mathrm{NR},\varphi_2}g_{*,s}(T)}{g_{*,s,\mathrm{NR},\varphi_2}}
\left(\frac{90}{g_{*,\mathrm{NR},\varphi_2}}\right)^\frac{1}{4}(1+r_f^2 Q_{22}^2)^{-\frac{3}{4}}
\,,
}
which is comparable to the energy density of the $\varphi_2$ states produced through misalignment for order one initial misalignment angles, since their oscillation temperature satisfies $T_\mathrm{osc} \sim T_{\mathrm{NR},\varphi_2}$, and thus $\rho_{\varphi_2,\mathrm{str}}(T_{\mathrm{NR},\varphi_2})\sim \rho_{\varphi_2.\mathrm{mis.}}(T_{\mathrm{NR},\varphi_2})\sim f_{a_1}^2 H^2(T_{\mathrm{NR},\varphi_2})$. 
The $\varphi_2$ axions would dominate the universe at the temperature 
\es{eq:T_dom_warped}{
T_\mathrm{dom} &\sim
\frac{4 \sqrt{\pi}}{3  } \Lambda_2\left(\frac{f_{a_1}}{\mpl}\right)^{3 / 2}
\frac{\xi_{\mathrm{NR},\varphi_2}^\frac{3}{4}}{\sqrt{\delta_1}}
\left(\log \left(\frac{ f_{a_1} m_s }{ \Lambda_2 ^2}\right)\right)^{\frac{1}{2}} \\ &\times 
\left(\frac{90}{g_{*,\mathrm{NR},\varphi_2}}\right)^\frac{1}{4} \frac{g_{*,s,\mathrm{dom}}g_{*,\mathrm{NR},\varphi_2}}{g_{*,\mathrm{dom}}g_{*,s,\mathrm{NR},\varphi_2}}(1+r_f^2 Q_{22}^2)^{-\frac{3}{4}}  \,,
}
up to order one modifications due to the misalignment contribution, which we account for numerically. The $\varphi_2$ dominate before decaying if 
\es{}{
\Lambda_2 &\lesssim  \left(\frac{32\pi^\frac{3}{2}}{3\beta}\right)^\frac{1}{2}\frac{f_{a_1}^2}{\mpl}\left(\frac{g_{*,\mathrm{dec}}}{g_{*,{\rm NR},\varphi_2}}\right)^{\frac{1}{8}} \\
&\times 
\frac{\xi_{\mathrm{NR},\varphi_2}^\frac{3}{8}}{\delta_1^\frac{1}{4}}
\log \left(\frac{ f_{a_1} m_s }{ \Lambda_2 ^2}\right) ^\frac{1}{4}r_f^{-\frac{1}{2}}(1+r_f^2Q_{22}^2)^{-\frac{3}{4}} \,.
}
Note that the time at which $\varphi_2$ dominates the energy density must either occur before BBN, and for self-consistency, we must have $T_\mathrm{dom} \le T_\mathrm{MRE}$. If there is no EMD period and $\varphi_2$ is stable up until today, then the string contribution to the relic abundance of $\varphi_2$ is 
\es{}{
\Omega_{\varphi_2, \mathrm{str}} &\sim 
0.85 \left(\frac{f_{a_1}}{10^{10}\,\mathrm{GeV}}\right)^\frac{3}{2}\frac{g_{*,\mathrm{NR}.\varphi_2}^\frac{3}{4}}{g_{*,s,\mathrm{NR}.\varphi_2}}\frac{\Lambda_2}{10^3 \, \mathrm{GeV}} \\
&\times \log^{\frac{1}{2}} \left(\frac{m_s f_{a_1}}{\Lambda_2^2}\right)
\frac{\xi_{\mathrm{NR},\varphi_2}^\frac{3}{4}}{\sqrt{\delta_1}}
\frac{1}{(1+r_f^2Q_{22}^2)^\frac{3}{4}} \,.
\label{eq:Omega_varphi1_today_warped_answer}
}

\begin{figure}[!tb]
\begin{center}
\includegraphics[width=0.9\linewidth]{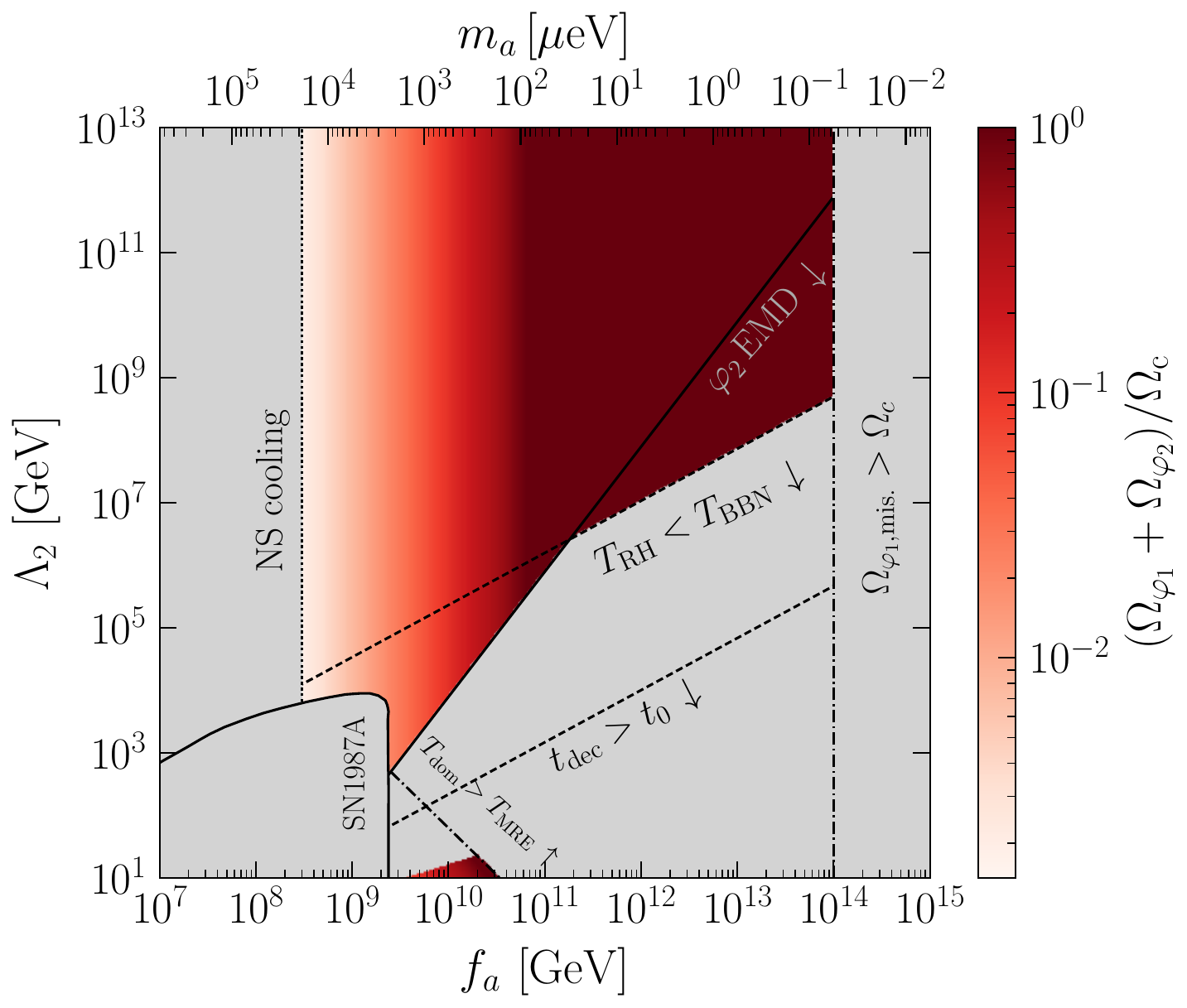} 
\caption{The space of $(f_a, \Lambda_2)$ for the two axion scenario in the warped case,  for our fiducial choice of dimensionless parameters. The color scheme is as in Fig.\ref{fig:2axion_unwarped_param_space_rf_1.pdf}.
}
\label{fig:2axion_warped_param_space.pdf}
\end{center}
\end{figure}

We are left with a small region of parameter space where, at the time of matter-radiation equality, $\varphi_2$ has not yet decayed and is non-relativistic. In this region the mass of the heavier eigenstate is an $\mathrm{eV}$ to $\mathrm{keV}$ and its axion-photon coupling is constrained by observations of the Leo T dwarf galaxy \cite{Wadekar:2021qae}, X-ray observations of galaxies \cite{Cadamuro:2011fd}, and \textit{XMM-Newton} observations of the Milky Way  \cite{Foster_2021} (for a summary of these limits, see \cite{AxionLimits}).
The combination of these constraints excludes almost all of the aforementioned region, leaving only a small sub-region centered around $f_a\sim 10^{10} \,\mathrm{GeV}$, with a weak hierarchy of instanton scales $\Lambda_2/\Lambda_1\sim 130$. The allowed parameter space is summarized in 
Fig.~\ref{fig:2axion_warped_param_space.pdf}. 

\bibliography{Bibliography}

\end{document}